\pdfoutput=1

\documentclass[11pt]{article}

\usepackage[final]{acl}

\usepackage{times}
\usepackage{latexsym}

\usepackage[T1]{fontenc}

\usepackage[utf8]{inputenc}

\usepackage{microtype}

\usepackage{inconsolata}

\usepackage{graphicx}

\usepackage{amsfonts}
\usepackage{amssymb}
\usepackage{textcomp}
\usepackage{verbatim}
\usepackage{pifont}
\usepackage{fontawesome}
\usepackage[clock]{ifsym}
\usepackage{upgreek}
\usepackage{xspace}
\usepackage{upquote}
\usepackage{bbding}

\usepackage{xcolor}
\usepackage{transparent}

\usepackage[most]{tcolorbox}

\usepackage{adjustbox}

\usepackage{xparse}
\usepackage{scalerel}
\usepackage{pgfplots}
\usepackage{pgfplotstable}
\usepgfplotslibrary{groupplots}
\usepgfplotslibrary{statistics}
\usetikzlibrary{matrix}
\pgfplotsset{compat=newest}
\usetikzlibrary{pgfplots.fillbetween,patterns}
\usepackage{tikz}
\usepackage{subcaption}
\pgfplotsset{compat=1.18,
    /pgfplots/xbar legend/.style={
    /pgfplots/legend image code/.code={%
       \draw[##1,/tikz/.cd,yshift=-0.25em]
        (0cm,0cm) rectangle (3pt,0.8em);},
   },
   /pgfplots/ybar legend/.style={
    /pgfplots/legend image code/.code={%
       \draw[##1,/tikz/.cd,yshift=-0.25em]
        (0cm,0cm) rectangle (3pt,0.8em);},
   },
}

\usepackage{array}
\usepackage{multirow}
\usepackage{colortbl}
\usepackage{arydshln}
\usepackage{tabularx}
\usepackage[para,online,flushleft]{threeparttable}
\usepackage{stackengine}
\usepackage{hhline}
\usepackage{diagbox}
\usepackage{booktabs}

\usepackage{bm}
\usepackage{amsfonts}
\usepackage{nicefrac}

\usepackage{mathtools}

\usepackage{enumitem}

\lstdefinestyle{mystyle}{
    backgroundcolor=\color{lightgray!20},
    commentstyle=\color{green!60!black},
    keywordstyle=\color{blue!80!black},
    numberstyle=\tiny\color{gray},
    stringstyle=\color{orange!90!black},
    basicstyle=\ttfamily\footnotesize,
    breakatwhitespace=false,
    breaklines=true,
    captionpos=b,
    keepspaces=true,
    numbers=left,
    numbersep=5pt,
    showspaces=false,
    showstringspaces=false,
    showtabs=false,
    tabsize=2,
}
\NewDocumentCommand\portsIcon{}{\resizebox{!}{3.6mm}{\includegraphics{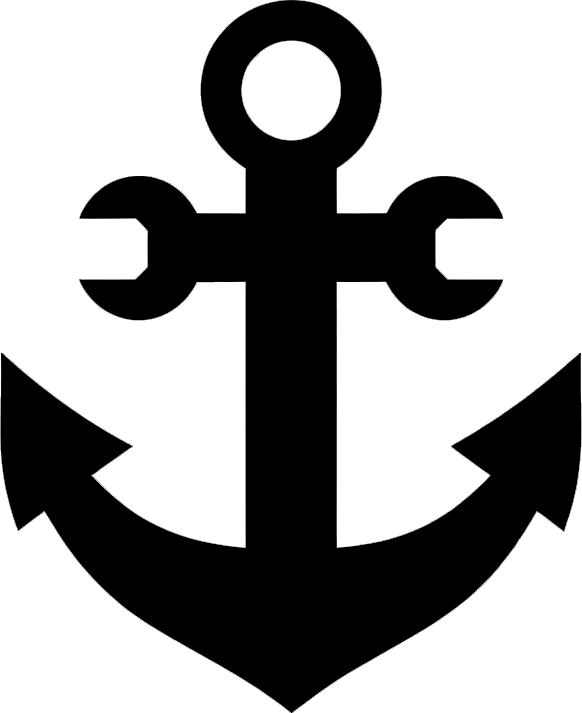}}}

\DeclareMathOperator*{\E}{\mathbb{E}} 
\newcommand{\abs}[1]{\lvert #1 \rvert}

\newcommand{\ports}{\texttt{PORTS}\xspace}
\newcommand{\replug}{\textsc{RePlug LSR}\xspace}
\newcommand{\orpo}{\textsc{ORPO}\xspace}

\newcommand{\bge}{\textsc{BGE}-base\xspace}
\newcommand{\robertaShort}{RoBERTa\xspace}
\newcommand{\bgeShort}{\textsc{BGE}\xspace}
\newcommand{\modernBert}{ModernBERT-base\xspace}

\newcommand{\modernBertShort}{ModernBERT\xspace}

\newcommand{\llamaBig}{\textsc{Llama}3-8B\xspace}
\newcommand{\groqLlama}{\textsc{Llama3-Groq}-8B-Tool-Use\xspace}
\newcommand{\codestral}{\textsc{Codestral}-22B-v0.1\xspace}
\newcommand{\qwen}{\textsc{Qwen}3-4B\xspace}
\newcommand{\llamaSmall}{\textsc{Llama}3.2-3B\xspace}
\newcommand{\gemma}{\textsc{Gemma}3-1B\xspace}

\newcommand{\lossPorts}{$\mathcal{L}_\texttt{PORTS}$\xspace}
\newcommand{\lossReplug}{$\mathcal{L}_\text{replug}$\xspace}

\newenvironment{allintypewriter}{\ttfamily}{\par}

\definecolor{tablehighlight}{RGB}{239, 239, 239}

\definecolor{plotbackground}{HTML}{F4F7FB}

\definecolor{baselinecolor}{HTML}{C2E9F5}
\definecolor{replugcolor}{HTML}{FFDE70}
\definecolor{portscolor}{HTML}{FFAF1A}

\definecolor{seenbackgroundcolor}{HTML}{F4F4F4}
\definecolor{unseenbackgroundcolor}{HTML}{E8E8E8}
\definecolor{codestralcolor}{HTML}{CD1200}
\definecolor{llama3groqcolor}{HTML}{FF8000}
\definecolor{llama3color}{HTML}{FFC72E}
\definecolor{darkerbaselinecolor}{HTML}{7CCFE9}
\definecolor{robertabackgroundcolor}{HTML}{F8F2F4}
\definecolor{robertahighlightcolor}{HTML}{CF9FC3}
\definecolor{bgebackgroundcolor}{HTML}{F4F5FB}
\definecolor{bgehighlightcolor}{HTML}{828DD3}
\definecolor{agglomerativecolor}{HTML}{606EC7}
\definecolor{dbscancolor}{HTML}{187A7B}
\definecolor{kmeanscolor}{HTML}{4BBEAF}
\definecolor{avgcolor}{HTML}{9C4BBE}

\definecolor{bluebar}{HTML}{797DE3}

\title{{\portsIcon}\hspace{1.5mm}PORTS: Preference-Optimized Retrievers for Tool Selection\\with Large Language Models}

\author{Lorenzo Molfetta\ \ \ \ Giacomo Frisoni\ \ \ \ Nicolò Monaldini\ \ \ \ Gianluca Moro \\
        Department of Computer Science and Engineering, University of Bologna \\
        \{lorenzo.molfetta, giacomo.frisoni, gianluca.moro\}@unibo.it,\\ nicolo.monaldini@studio.unibo.it}

\begin{document}

\maketitle

\begin{abstract}
Integrating external tools with Large Language Models (LLMs) has emerged as a promising paradigm for accomplishing complex tasks.
Since LLMs still struggle to effectively manage large tool collections, researchers have begun exploring retrieval-based methods to pre-select the most relevant options, addressing input length and latency constraints.
However, existing retrievers are often misaligned with tool-calling LLMs due to their separate training processes. This paper presents \texttt{PORTS}, a novel odds ratio preference optimization method for training retrievers aimed at tool selection. Using a perplexity-inspired preference signal from a frozen LLM, our approach fine-tunes a retriever to find helpful tools by optimizing the correlation between the selection probabilities and the downstream performances while jointly enforcing a contrastive semantic loss between documentation strings. The versatility of \texttt{PORTS} and its ability to significantly improve tool selection accuracy are demonstrated through extensive experiments on six datasets, two encoder models, and three LLMs with diverse prior knowledge. With low computational demands, our alignment process facilitates generalization to new queries and tools, proving valuable for practical applications with evolving toolsets.\footnote{Code, models, and datasets are publicly available at \href{https://github.com/disi-unibo-nlp/ports}{https://github.com/disi-unibo-nlp/ports}}
\end{abstract}

\def\thefootnote{*}\footnotetext{The definitive, copyrighted, peer-reviewed, and edited version of this article is published in the Proceedings of the 2025 Conference on Empirical Methods in Natural Language Processing, pp. 10007--10030, Suzhou, China, 2025. Association for Computational Linguistics. \url{https://aclanthology.org/2025.emnlp-main.507/}. DOI: \href{https://doi.org/10.18653/v1/2025.emnlp-main.507}{10.18653/v1/2025.emnlp-main.507}.}\def\thefootnote{\arabic{footnote}}

\section{Introduction}

\begin{flushright}
\textit{``The right tool for the right job.''}---\emph{Proverb}
\end{flushright}

Equipping Large Language Models (LLMs) with the capability to dynamically interact with external tools\footnote{Consistent with~\citet{DBLP:journals/corr/abs-2405-17935}, we argue that all external means of augmenting LLMs should be classified as tools.
Accordingly, we regard individual APIs as separate tools.} has garnered significant research attention.
This integration not only improves the problem-solving potential of LLMs, but also dramatically expands their functional scope~\cite{DBLP:conf/nips/Yao0YN22,DBLP:journals/corr/abs-2203-05115}.
When presented with a user query, tool-augmented LLMs can determine when and how to utilize specific tools to generate more accurate and informative responses.
For example, tools can enable LLMs to use a calculator, set calendar events, and access real-time weather information.
As the field continues to evolve, LLMs with tools are expected to play a pivotal role in shaping the future of Natural Language Processing (NLP)~\cite{DBLP:journals/corr/abs-2405-17935}.

\begin{figure}[!t]
    \centering
    \begin{subfigure}[t]{\linewidth}
        \hspace{-2mm}
        \begin{tikzpicture}
            \footnotesize
            \node[draw=none, inner sep=2pt, align=center, text width=8cm] {
            \begin{tabular}{l|lll}
                \multirow{2}{*}{\includegraphics[width=.75cm]{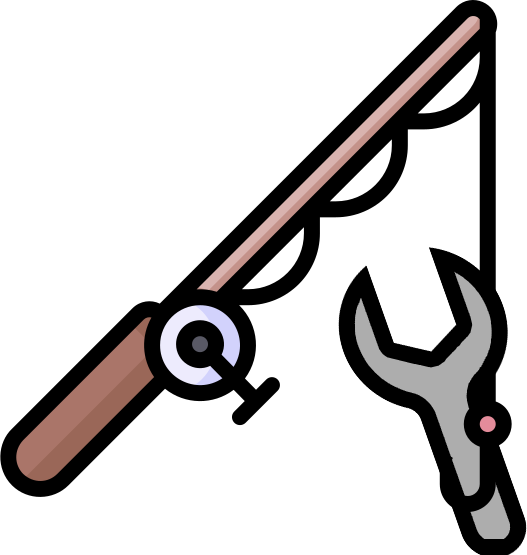}\hspace{.08cm}} & \multicolumn{3}{l}{\textbf{Tool Selection}}\\
                & \hspace{.08cm}\ref{plot:baseline} Baseline & \ref{plot:ports1} \textsc{RePlug} & \ref{plot:ports2} $\texttt{PORTS}\;\text{(Ours)}$\\
            \end{tabular}
            };
        \end{tikzpicture}
    \end{subfigure}\\[-1mm]
    \begin{subfigure}[H]{\linewidth}
        \vspace{3mm}
        \hspace{-1ex}  
        \begin{tikzpicture}[node font=\footnotesize]
        \begin{axis}[
            width=.95\linewidth, height=5.2cm,
            xmajorgrids=true,
            grid style={draw=none}, 
            axis background/.style={fill=plotbackground!50},
            every y tick/.style={/pgfplots/major tick length=0pt}, 
            xbar=2pt, 
            bar width=10pt,
            enlarge y limits=0.12, 
            xtick={0, 10, ..., 80},
            xmin=0, xmax=85,
            xlabel={Avg. Recall (\%)},
            xlabel style={font=\small},
            ytick={23,78},
            yticklabels={BGE, RoBERTa},
            every tick label/.append style={font=\fontsize{8}{8}\selectfont},
            nodes near coords,
            nodes near coords style={font=\small},
            axis on top=true,
            axis line style={draw=black},
        ]

        \addplot [draw=none, fill=baselinecolor] coordinates {(8.8,{100})};\label{plot:baseline}
        \addplot [draw=none, fill=replugcolor] coordinates {(50.8,{100})};\label{plot:ports1}
        \addplot [draw=none, fill=portscolor] coordinates {(57.5,{100})};\label{plot:ports2}

        \addplot [draw=none, fill=baselinecolor] coordinates {(54.4,0)};
        \addplot [draw=none, fill=replugcolor] coordinates {(62.5,0)};
        \addplot [draw=none, fill=portscolor] coordinates {(65.6,0)};

        \node[anchor=east] at (axis cs:7,93) {\includegraphics[height=8pt]{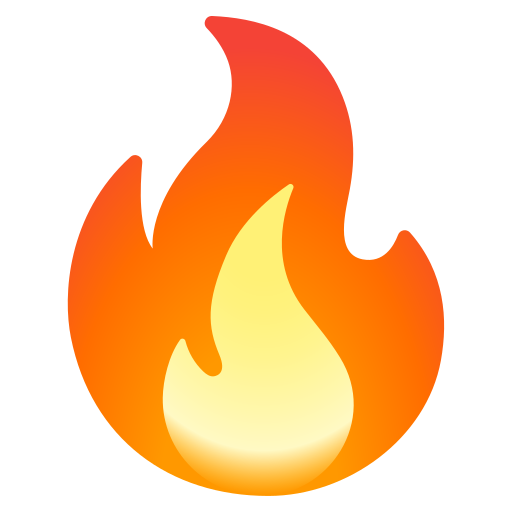}};
        \node[anchor=east] at (axis cs:7,78) {\includegraphics[height=8pt]{fire_icon.png}};
        \node[anchor=east] at (axis cs:7,64) {\includegraphics[height=8pt]{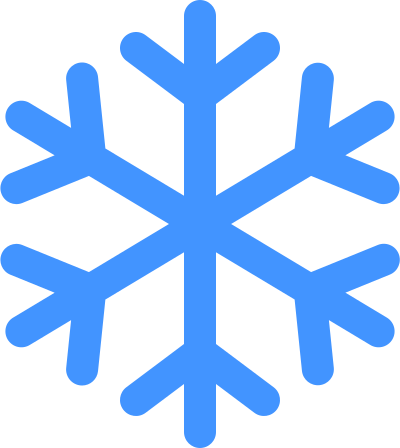}};
        \node[anchor=east] at (axis cs:7,36) {\includegraphics[height=8pt]{fire_icon.png}};
        \node[anchor=east] at (axis cs:7,21) {\includegraphics[height=8pt]{fire_icon.png}};
        \node[anchor=east] at (axis cs:7,7) {\includegraphics[height=8pt]{snowflake_icon.png}};
        
        \end{axis}
        \end{tikzpicture}
    \end{subfigure}
    \caption{\textbf{Results overview.} Comparison between frozen, \textsc{RePlug}-tuned, and \texttt{PORTS}-tuned retrievers. Scores are averaged across Recall@\{1,2,3\} for three LLMs (if trained) and six datasets (test set).}
    \label{fig:abstract}
\end{figure}

Fine-tuning LLMs with tool usage examples is expensive and confines the acquired knowledge to a predefined set of tools~\cite{DBLP:journals/corr/abs-2305-13068,DBLP:conf/nips/YangSLZGLS23}. 
The in-context learning paradigm alleviates these issues, but the limitations in input length and noise for lengthy prompts make it impractical to manage many descriptions or demonstrations directly~\cite{DBLP:journals/tacl/LiuLHPBPL24,DBLP:journals/corr/abs-2405-16089}, introducing efficiency and accuracy challenges in tool-selection tasks, mainly when precise parameter specification and schema typing are paramount.
Furthermore, when faced with hundreds of tool docstrings in the prompt, the language model alone struggles to identify the most suitable one, often resulting in suboptimal performance~\cite{DBLP:journals/corr/abs-2405-17935}, increased computational needs, and high costs.
Recently, the focus has shifted towards the use of retrievers to effectively support LLMs in the tool selection process~\cite{DBLP:conf/iclr/QinLYZYLLCTQZHT24,DBLP:conf/aaai/GaoSZF0RC0R24,DBLP:journals/corr/abs-2312-10332}.
Retrieval-enhanced pipelines filter the top-$K$ most suitable tools for a given query, aiming to reduce noise and enhance the LLM's ability to select the right tool and configure the necessary parameters for calling.

Although several publications have explored retrievers for tool selection, optimization of the retrieval component itself has received little consideration. Clustering approaches have proven successful in multi-dimensional representation spaces~\cite{DBLP:conf/adc/LodiMS10}, but current methods predominantly employ non-parametric indexing techniques~\cite{DBLP:journals/corr/abs-2305-15334} or standard encoder models trained with supervised signals~\cite{DBLP:conf/iclr/QinLYZYLLCTQZHT24}. 
Whereas these methodologies can be effective in isolation, they often falter when integrated into a broader pipeline, primarily due to misalignment between the training criteria used for the retrieval and generation modules.
A significant challenge arises when tools with descriptions similar to the user query are ultimately irrelevant or potentially misleading for the LLM~\cite{DBLP:journals/corr/abs-2311-09210,DBLP:conf/icml/ShiCMSDCSZ23}. Furthermore, tools can exhibit subtle differences, such as variations in the names, numbers, and types of input parameters, which complicate effective selection--an issue that is increasingly prevalent given the rapid proliferation of publicly available tools and Model Context Protocol servers. In such scenarios, conditioning the encoder on the LLM output may provide additional training signals that benefit the selection process.
However, most existing retriever adaptation techniques require the LLM to be trained from scratch~\cite{DBLP:journals/jmlr/IzacardLLHPSDJRG23,DBLP:conf/iclr/Lin0CSL00KSLZY24,DBLP:conf/nips/ChengLCL0023}, which can be prohibitively costly or unfeasible with closed-source solutions characterized by no access to internal representations.
Recent research has explored an alternative approach: training encoder models with the support of LLMs, using them as supervision signals to better align representations with task-specific objectives~\cite{DBLP:journals/corr/abs-2504-13181}. 
We focus on this emerging class of methods, investigating how LLMs can be effectively leveraged and optimized to guide encoder alignment through feedback incorporation for tool retrieval.

In this paper, we propose a new method to train preference-optimized retrievers for tool selection (\ports), aligning them with the needs of the LLM responsible for tool usage.
Our training scheme adapts a pre-trained encoder model with supervision signals from a black-box LLM, preferring retrieving tool docstrings that stimulate the downstream LLM to use the right tool.
Introducing a novel contrastive preference loss enables a more accurate selection process in application domains where multiple tools might adequately serve the task and yield coherent yet inaccurate results.
We conduct experiments on six public datasets. The results are analyzed with various classes of encoders and LLMs.
To gauge generalizability, tests are carried out with in-domain and out-of-domain tools.
We conclude that \ports can effectively increase the tool selection performance of the baseline retriever, with low computational overhead. 
Figure~\ref{fig:abstract} shows the averaged metric gains of our alignment method in retrieving tools seen during training.

\section{Related Work}

\paragraph{Tool Learning} Recent studies in language modeling have explored the use of non-differentiable tools to supplement the knowledge stored in the model weights, offloading tasks to external modules. They broadly fall into two categories. \textit{Tuning-based methods} train models to use one or a few tools in specific domains. Example works include TALM~\cite{DBLP:journals/corr/abs-2205-12255}, Toolformer~\cite{DBLP:conf/nips/SchickDDRLHZCS23}, ToolLLaMA~\cite{DBLP:conf/iclr/QinLYZYLLCTQZHT24}, Gorilla~\cite{DBLP:journals/corr/abs-2305-15334}, ToolkenGPT~\cite{DBLP:conf/nips/HaoLWH23}, and Granite~\cite{DBLP:journals/corr/abs-2407-00121}. These models are trained on datasets where the input text is augmented with tool calls. During inference, when such invocations are identified, the decoding process is paused, the corresponding tool is executed, and the result is incorporated before resuming text generation. Specialized tool-calling models mostly rely on synthetic instruction-tuning data generated from proprietary models such as GPT-4~\cite{DBLP:journals/corr/abs-2303-08774}. Among the few exceptions intended for commercial applications is the NexusRaven series~\cite{nexusraven}. However, LLM fine-tuning is only applicable to open-source models and is generally hindered by expensive data collection and computing infrastructure, as well as poor flexibility in accommodating emergent or updated tools. Conversely, \textit{tuning-free methods} are compatible with all LLMs and capitalize on in-context learning abilities, showing tool descriptions and demonstrations directly in the prompt~\cite{DBLP:conf/nips/0001ST00Z23,DBLP:journals/corr/abs-2306-06624,DBLP:conf/iclr/YaoZYDSN023}. As the tool arsenal grows (e.g., 16.000+~\cite{DBLP:conf/iclr/QinLYZYLLCTQZHT24}), a retriever becomes essential. Retrieval-based and fine-tuning methods can be combined to achieve better performance~\cite{DBLP:conf/aaai/GaoSZF0RC0R24}.

\paragraph{Tool Retrieval} Merging Retrieval-Augmented Generation (RAG) and tool calling enables LLMs to evaluate a small subset of retrieved tools and select the most suitable for response formulation. Tool retrieval approaches can be classified into two main types: term-based and semantic-based. Term-based techniques, exemplified by TF-IDF~\cite{DBLP:journals/jd/Jones04} and BM25~\cite{DBLP:journals/ftir/RobertsonZ09}, rely on exact term matching and utilize sparse representations for both tool docstrings and queries. For instance, Gorilla~\cite{DBLP:journals/corr/abs-2305-15334} implements tool retrieval by combining BM25 with GPT-Index. In contrast, semantic-based methods utilize neural networks to learn the relationships between queries and tool descriptions. CRAFT~\cite{DBLP:conf/iclr/YuanC000J24}, for example, instructs LLMs to generate fictitious tool descriptions conditioned on the input queries, then uses pre-trained SimCSE for similarity computation. 
Few studies focus on training the retriever itself, while iterative refinement with cosine similarity has proven effective for adapting representations~\cite{DBLP:conf/ic3k/DomeniconiMPS14a,DBLP:conf/ic3k/DomeniconiMPS14,DBLP:conf/ic3k/DomeniconiMPP15}. 
TPTU~\cite{DBLP:journals/corr/abs-2311-11315}, ToolLLaMA~\cite{DBLP:conf/iclr/QinLYZYLLCTQZHT24} and Confucius~\cite{DBLP:conf/aaai/GaoSZF0RC0R24} fine-tune a SentenceBERT model using contrastive learning objectives. ProTIP~\cite{DBLP:journals/corr/abs-2312-10332} fine-tunes BERT-base with a contrastive loss optimized for progressive tool retrieval. 
COLT~\cite{DBLP:journals/corr/abs-2405-16089} models collaborative relationships among multiple tools using graphs and implements tool retrieval through cross-view graph contrastive learning. 
ToolRerank~\cite{zheng-etal-2024-toolrerank} addresses the re-ranking stage of tool retrieval, proposing an adaptive and hierarchy-aware method. However, these studies do not consider LLM preferences in specializing the retriever.

\section{Methodology}

\begin{figure*}[!htb]
    \centering
    \includegraphics[width=\linewidth]{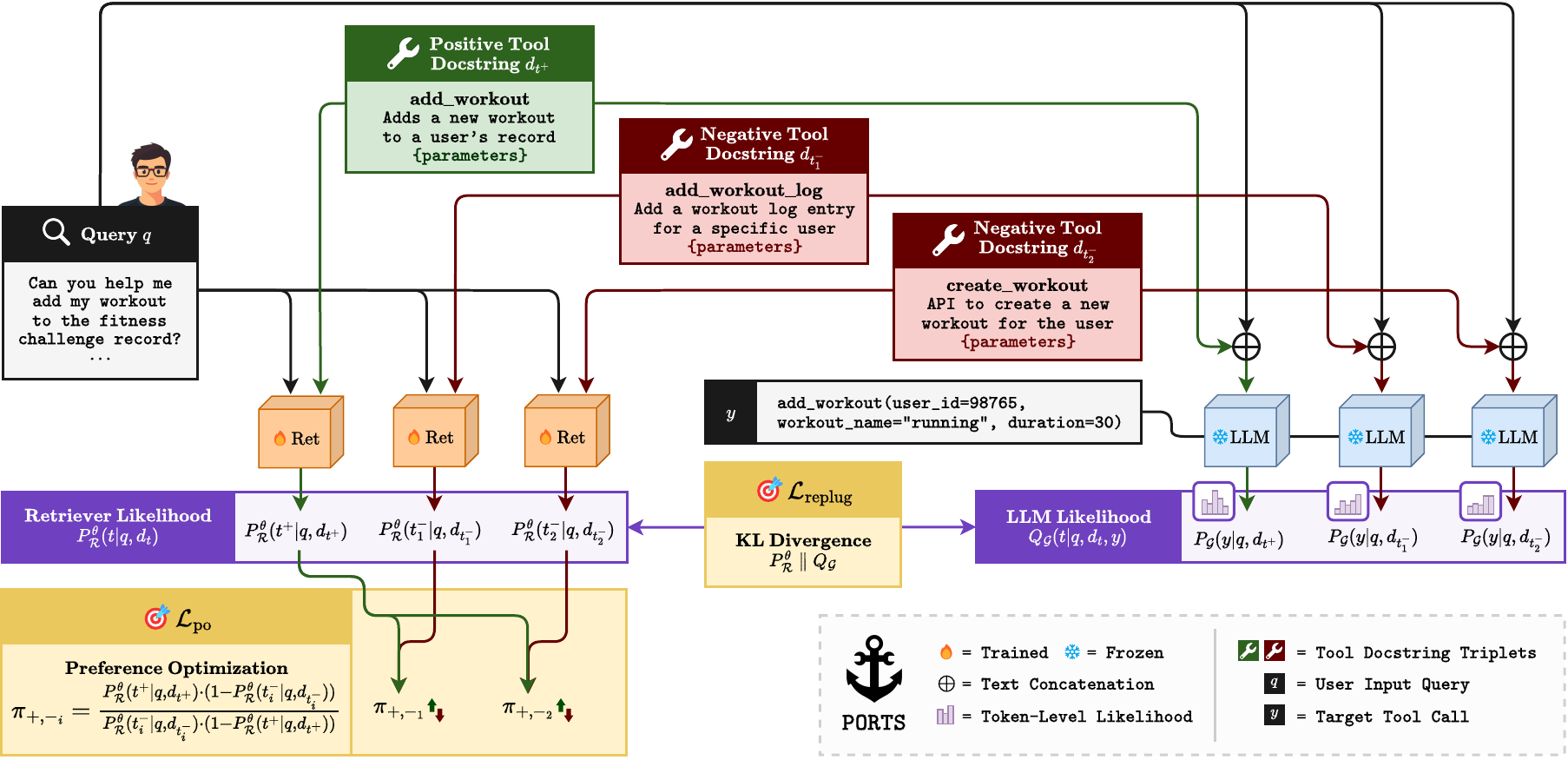}
    \caption{\textbf{\ports training process.} Simplified illustration of the \ports' training recipe with one positive and two negative tool docstrings from APIBench. Input tool documentation triplets are encoded independently and prompted separately to the frozen LLM.  
    The retriever is fine-tuned to align tool selection probabilities with the correct answer likelihood while maximizing the ratio between the odds of selecting the right tool and the wrong ones.}
    \label{fig:overview}
\end{figure*}

In this section, we introduce \ports and elaborate on its theoretical motivations, design, and training losses. 
Figure~\ref{fig:overview} illustrates our architecture.

\subsection{Preliminary}

RAG is a widely used framework for augmenting LLMs with external knowledge sources.
While models like RETRO~\cite{DBLP:conf/icml/BorgeaudMHCRM0L22} and FiD~\cite{DBLP:conf/eacl/IzacardG21} achieve significant improvements through separate training of retrieval and generation components, end-to-end training approaches offer potential for enhanced relevance, coherence, and contextual awareness~\cite{DBLP:conf/icml/GuuLTPC20,DBLP:journals/corr/abs-2405-20680}. 
However, end-to-end training poses challenges, including high computational demands, complex data handling, and the difficulty of maintaining dynamic search indexes with accurate, up-to-date embeddings.
Techniques like batch negative sampling~\cite{DBLP:conf/emnlp/KarpukhinOMLWEC20} have improved efficiency, but require careful selection of diverse negatives.
Adapting RAG systems to new domains often necessitates simultaneous retraining of both the retriever and generator, underscoring the importance of efficient data management~\cite{DBLP:journals/tacl/SiriwardhanaWKWRN23}.
To address these challenges, methods such as \replug~\cite{shi-etal-2024-replug} use frozen language models as references to optimize retrieval without costly full-model fine-tuning.
We argue that reducing output uncertainties alone is insufficient for aligning with users’ goals when selecting tools for tasks.
Retrieval models must effectively navigate reference corpora by distinguishing between semantically similar but irrelevant options.
Inspired by techniques from LLM fine-tuning, such as \textsc{RLHF}~\cite{DBLP:conf/nips/ChristianoLBMLA17}, \textsc{DPO}~\cite{DBLP:conf/nips/RafailovSMMEF23}, and \orpo~\cite{DBLP:journals/corr/abs-2403-07691}, retrieval models can leverage comparative loss to prioritize relevant information, mirroring approaches like triplet learning.

\subsection{Task Definition}

Given an input query $q$, we aim to pair it with a candidate tool $t_i$ from a predefined set $\mathbf{T} = \{t_1, t_2, \ldots, t_{\abs{\mathbf{T}}}\}$ by maximizing their semantic alignment.
\ports is trained to prioritize the retrieval of tools' docstrings $d_{t_i}$ that most enhance tool calling accuracy when prompted to the LLM.

\subsection{\ports}

Our approach optimizes a retrieval model $\mathcal{R}$ through a dual training strategy accounting for query-docstring semantic similarity and tool support in correct answer prediction.
We shape probabilities over available data by enforcing preferences on the top-$K$ tools.
An LLM serves as an indirect ranking agent, aligning selections with tool usage patterns, resulting in a context-aware algorithm tailored to downstream tasks.

\paragraph{Goal-Directed Retrieval}
We formalize the retrieval process using an encoder $\mathcal{E}$ and a generative LLM $\mathcal{G}$.
Each instance in the dataset $\mathcal{D}$ comprises a user query $q$, the target tool call $y$, the tool required $t$ to solve the request (positive), and a set of $n$ tools irrelevant to the task (negatives). Each tool $t_i$ is associated with a description $d_{t_i}$, which includes its general characteristics, objectives, and parameters.
\ports first computes the alignment between $q$ and the tool docstrings $d_t$ using a cosine similarity function $sim(q,d_t) = \mathcal{E}_\theta(q) \cdot \mathcal{E}_\theta(d_t)$, where $\mathcal{E}: \mathbb{Z}^l \rightarrow \mathbb{R}^d$ denotes the retriever's encoding function that maps input sequences of $l$ tokens to a $d$-dimensional vector space, parameterized by weights $\theta$.
These similarities are then normalized and converted into retrieval probabilities using a softmax function with scaling factor $\gamma$:
\begin{equation}
\label{eq:retr-prob}
P_\mathcal{R}^\theta(t \vert q, d_t) = \frac{\exp \left ( \frac{sim(q,d_t)}{\gamma} \right )}{\sum\limits_{t_i \in \mathcal{T}} \exp \left ( \frac{sim(q,d_{t_i})}{\gamma} \right )}
\end{equation}
The retrieval distribution over the corpus of tools in Eq.~\ref{eq:retr-prob} is approximated by marginalizing over restricted triplet sets $\mathcal{T} = (t_i^{+}, t_{i,1}^{-},\dots, t_{i,n}^{-}) \subseteq \mathbf{T}$ for efficiency.
As in \textsc{RePlug}, we prompt each retrieved tool docstring independently with the query and then conduct $K$ separate inferences.
This enables a direct and noiseless correlation between tool selection quality and confidence in the generated output.
The output probability distribution $Q_\mathcal{G}$ of the LLM, reflecting confidence in the final prediction, is computed as in Eq.~\ref{eq:q_ppl}.
It applies a softmax function, parameterized by temperature $\beta$, to the average log-likelihood $P_\mathcal{G}$ of producing the correct tool call $y$.

\begin{equation}
\label{eq:q_ppl}
\begin{split}
&Q_\mathcal{G}(t \vert q,d_t,y) = \frac{\exp \left (  \frac{P_\mathcal{G}(y \vert q,d_t)}{\beta} \right )}{\sum\limits_{t_i \in \mathcal{T}} \exp \left ( \frac{P_\mathcal{G}(y \vert q,d_{t_i})}{\beta} \right )} \\
& P_\mathcal{G}(y\vert q,d_t) = \frac{1}{l} \log \prod_{i=1}^l P_\mathcal{G}(y_i \vert q,d_t,y_{<i})
\end{split}
\end{equation}
The retriever is trained by optimizing the Kullback-Leibler divergence between the $Q_\mathcal{G}(t \vert q,y)$ and $P_\mathcal{R}^\theta(t \vert q)$ distributions:
\begin{equation}
\begin{split}
\label{eq:loss_replug}
\mathcal{L}_\text{replug} = \E\limits_{\mathcal{T},q,y \sim \mathcal{D}} \sum_{t_i \in \mathcal{T}} 
\textsc{KL} \left( P_\mathcal{R}^\theta(t_i \vert q, d_{t_i}) \; \Vert \right. \\
\left. Q_\mathcal{G} (t_i\vert q,d_{t_i},y) \right)
\end{split}
\end{equation}
\noindent
During training, the model's perplexity---which expresses the confidence in the prediction of the correct tool call---is leveraged by $\mathcal{L}_\text{replug}$ in Eq.~\ref{eq:loss_replug} to encourage the reshape of the retrieval distribution.
As a result, the encoder model $\mathcal{E}_\theta$ learns to assign lower ranks to tools that increase the probability of generating incorrect responses.

\paragraph{Preference-Aligned Retrieval}
We introduce a contrastive loss signal that enforces a policy to favor selecting correct tools over incorrect ones (Eq.~\ref{eq:orpo-pi}).
\begin{equation}
\label{eq:orpo-pi}
\pi_\theta (t\vert q, d_t) = \frac{P_\mathcal{R}^\theta(t \vert q, d_t)}{1 - P_\mathcal{R}^\theta(t \vert q,d_t)}
\end{equation}
\noindent
For each positive-negative tool pair $(t_i^+, t^-_{i,j})$ in the input sample, we compute a ratio of their retrieval probabilities and apply the sigmoid function $\sigma$ to derive a preference score (i.e., the relative likelihood of selecting one tool over the other). 
\begin{equation}
\label{eq:orpo-loss}
\begin{split}
&\mathcal{L}_\text{po} = - \sum_{i \in {1,n}} \log \sigma \left ( \log \pi_{+,-_i} \right) \\
&\pi_{+,-_i} = \frac{\pi_\theta(t^+ \vert q, d_{t^+})}{\pi_\theta (t_i^- \vert q, d_{t_i^-})}
\end{split}
\end{equation}
The retriever incurs a penalty through the preference policies $\pi$, as indicated in Eq.~\ref{eq:orpo-loss}, when it shows an increased likelihood of selecting erroneous tools.

\paragraph{Training Objective}
\ports combines the LLM-based retrieval proxy loss ($\mathcal{L}_\text{replug}$) and the preference optimization loss ($\mathcal{L}_\text{po}$) with a weighting factor $\lambda$: $\mathcal{L}_\ports = \mathcal{L}_\text{replug} + \lambda \cdot \mathcal{L}_\text{po}$.
Our method synchronizes the encoder selections with the LLM tool-calling patterns and imposes positive-negative embedding orientation constraints similar to deep metric learning~\cite{DBLP:journals/symmetry/KayaB19}.

\paragraph{Tool Triplets and Embeddings Asynchronous Update}
The learning effectiveness in contrastive approaches is heavily influenced by the choice of negative examples~\cite{DBLP:conf/emnlp/KarpukhinOMLWEC20}.
We therefore implement a hard-negative sampling technique, choosing the $n$ tools most semantically similar to the query as negative instances, where similarity is quantified using cosine similarity between embeddings computed by the encoder itself during training.
To maintain computational efficiency while adapting to shifts in the embedding space during training, we periodically update both the tool embeddings and the selection of hard negatives every $T$ training iterations~\cite{DBLP:conf/icml/GuuLTPC20}.

\paragraph{Motivations for Contrastive Learning}
We establish that, without fine-tuning, embeddings of tool docstrings are much more concentrated in space than traditional, general-domain text documents--due to recurring data types, keywords, and concise but distinctive functional signatures (see Appendix~\ref{app:clustering}).
Without targeted supervision, these properties can lead the retriever to rely on superficial cues or converge to trivial matches. 
Our contrastive loss is designed to counteract this by steering retrieval toward semantically relevant and challenging negatives, promoting fine-grained distinctions beyond what is captured by LLM likelihoods alone. More details about the theoretical foundations of \ports are in Appendix~\ref{sec:ports_dem}.

\section{Experimental Setup}

\subsection{Datasets}

We evaluate \ports on six popular tool-augmented datasets: ToolBench~\cite{DBLP:conf/iclr/QinLYZYLLCTQZHT24}, API-Bank~\cite{li-etal-2023-api}, APIBench~\cite{DBLP:journals/tse/PengLGLWGL23}, BFCL-v2~\cite{bfcl}, ToolE~\cite{DBLP:conf/iclr/HuangSLFWZ000G024}, Octopus-v2~\cite{DBLP:journals/corr/abs-2404-01744}.
This collection offers a heterogeneous testing ground characterized by varying scales, applications, and input modalities.
When necessary, we adapt dataset instances to tool selection, which is the core task of our contributions.
Extensive dataset documentation is available in Appendix~\ref{app:software_datasets}. Key information and statistics are reported in Table~\ref{tab:datasets}.
For ToolBench, we focus on the most complex data split, G3, where queries demand the interplay of tools with dissimilar features, functions, and objectives.
Only for training, we decompose multi-tool instances from BFCL, API-Bank, and ToolBench into separate examples, each targeting a single tool.
When handling conversational inputs, we remove previous tool calls from the chat history of each fragment.
We partition test-only benchmarks (Octopus-v2, ToolE, BFCL) into train and test sets using a 70/30 ratio.
For Octopus-v2 and ToolE, which focus on single-tool selection without incorporating heterogeneous levels of complexity (e.g., difficulty levels or programming languages), we design in-domain and out-of-domain variants to evaluate generalization abilities to seen and unseen tools. Out-of-domain variants are created with an 80/20 tools split, avoiding overlap between training and test.

\begin{table*}[t]
    \centering
    \begin{adjustbox}{width=\linewidth}
    \begin{threeparttable}
    \begin{tabular}{lp{5.9cm}ccrrrrr}
    \toprule
    & & & & \multicolumn{2}{c}{\textbf{Train}} & \multicolumn{2}{c}{\textbf{Test}} & \cellcolor{tablehighlight}\textbf{All} \\
    \cmidrule(lr){5-6}
    \cmidrule(lr){7-8}
    \cmidrule(lr){9-9}
    \textbf{Dataset}\tnote{$*$} & \textbf{Description} & \textbf{Source}\tnote{$\dagger$} & \textbf{Input}\tnote{$\ddagger$} & \textbf{\# Tools} & \textbf{\# Instances} & \textbf{\# Tools} & \textbf{\# Instances} & \cellcolor{tablehighlight}\textbf{\# Tools} $\downarrow$ \\
    \hline
    \ding{182} ToolBench & REST APIs; 49 domains (e.g., Social Media, E-Commerce, Weather) & \faGear & \faQuestionCircle & 12,934 & 486,367 & 891 & 1,250 & \cellcolor{tablehighlight}12,934 \\
    \ding{183} API-Bank & General tools; 1,000 domains & \faGear & \faComments & 1,896 & 6,703 & 67 & 620 & \cellcolor{tablehighlight}1,960 \\
    \ding{184} APIBench & Tools for Java and Python programming; 90 domains & \faUser & \faQuestionCircle\hspace{0.5mm}{/}\hspace{0.5mm}\faCode & 188 & 15,218 & 152 & 188 & \cellcolor{tablehighlight}1,557 \\
    \ding{185} BFCL-v2 \faAngleDoubleRight & Python and Non-Python tools; 40 domains (e.g., Computing, Mathematics, Sports, Finance) & \faUser\hspace{0.5mm}{/}\hspace{0.5mm}\faGear & \faQuestionCircle\hspace{0.5mm}{/}\hspace{0.5mm}\faComments & 781 & 1,260 & 415 & 541 & \cellcolor{tablehighlight}1,015 \\
    \ding{186} ToolE \faAngleDoubleRight & Tools inspired to OpenAI plugins; 6 main scenarios (Software, Utilities, Finance, Home, Education, Arts) & \faGear & \faQuestionCircle & 199 & 16,491 & 199 & 4,123 & \cellcolor{tablehighlight}199 \\
    \ding{187} Octopus-v2 \faAngleDoubleRight & Android APIs (System, App, Smart Device Management) & \faGear & \faQuestionCircle & 20 & 160 & 20 & 40 & \cellcolor{tablehighlight}20 \\
    \hline
    \ding{188} \faEyeSlash\hspace{0.8mm}ToolE \faAngleDoubleRight & Same as \ding{186} & & & 160 & 16,406 & 39 & 4,208 & \cellcolor{tablehighlight}199 \\
    \ding{189} \faEyeSlash\hspace{0.8mm}Octopus-v2 \faAngleDoubleRight & Same as \ding{187} & & & 16 & 160 & 4 & 40 & \cellcolor{tablehighlight}20 \\
    \bottomrule
    \end{tabular}
    \begin{tablenotes}
    \item[$*$] \faAngleDoubleRight\hspace{0.8mm}= We split test-only benchmarks with a 70/30 ratio; \faEyeSlash\hspace{0.8mm}= Out-of-domain versions (no overlapping between train and test tools).\\
    \item[$\dagger$] \faUser\hspace{0.8mm}= Human-sourced (manual or scraped); \faGear\hspace{0.8mm}= LLM-generated (reviewed or not).
    \item[$\ddagger$] \faQuestionCircle\hspace{0.8mm}= Query, \faComments\hspace{0.8mm}= Chat, \faCode\hspace{0.8mm}= Code.
    \end{tablenotes}
    \end{threeparttable}
    \end{adjustbox}
    \caption{\textbf{Summary of tool selection datasets}, sorted by descending total tool count.}
    \label{tab:datasets}
\end{table*}

\subsection{Evaluation Metrics}

We quantify the retrieval performance using Recall@$K$~\cite{zhu2004recall} and NDCG@$K$~\cite{DBLP:journals/tois/JarvelinK02}, with $K=\{1,3,5\}$ following~\citet{DBLP:conf/iclr/QinLYZYLLCTQZHT24}.
Recall@$K$ measures the proportion of cases where the positive tool appears in the top-$K$ results, while NDCG@$K$ also considers its ranking position within the top-$K$.

\subsection{Implementation Details}

\paragraph{Models}
For the embedding function of $\mathcal{R}$, we test two prominent encoders: the 125M-parameter RoBERTa-base \cite{DBLP:journals/corr/abs-1907-11692}, a widely adopted baseline, and the 109M-parameter \textsc{BGE}-base \cite{DBLP:conf/sigir/XiaoLZMLN24}, a top performer on the MTEB leaderboard.\footnote{\href{https://huggingface.co/spaces/mteb/leaderboard}{huggingface.co/spaces/mteb/leaderboard}}
We evaluate each encoder's architectural compatibility and its impact on retrieval precision.
For $\mathcal{G}$, we examine three open-source LLMs. (1) \codestral,\footnote{\href{https://huggingface.co/mistralai/Codestral-22B-v0.1}{huggingface.co/mistralai/Codestral-22B-v0.1}} a model specialized for code generation tasks with native tool calling, motivated by \cite{nexusraven}. (2) \llamaBig \cite{dubey2024llama3herdmodels}, a foundational model, and (3) \groqLlama,\footnote{\href{https://huggingface.co/Groq/Llama-3-Groq-8B-Tool-Use}{huggingface.co/Groq/Llama-3-Groq-8B-Tool-Use}} its fine-tuned variant specifically designed for advanced tool use.
We employ 4-bit quantization for efficiency. 
Larger models were excluded due to computational limits, focusing on a selection balancing performance and efficiency.
Prompt templates are in Appendix~\ref{app:prompt_templates}.

\paragraph{Hyperparameters}
In our experiments, we used 3 negatives sampled every $T=50$ training steps upon recalculating embeddings. We adopted training and evaluation batch sizes of 2 and 4, respectively, and set maximum sequence lengths of 512 and 1024 for the encoder and LLM inputs. We applied a loss weight $\lambda$ of 0.3, with $\gamma$ and $\beta$ set to 0.5. We set the random seed to 42 for reproducibility and trained each configuration for 2 epochs using the AdamW optimizer, coupled with a cosine learning rate scheduler starting at $1e^{-5}$. We kept LLMs frozen during training. To simulate low-resource scenarios, we sampled $10$K unique instances per dataset, testing \ports' capacity to exploit few information more efficiently. Search spaces and hyperparameter insights are in Appendix~\ref{app:hyperparameters}.

\paragraph{Hardware Setup}
Each run was performed on an internal workstation using a single Nvidia GeForce RTX3090 GPU with 24GB of dedicated memory, 64GB of RAM, and an Intel® Core™ i9-10900X1080 CPU @ 3.70GHz.
The reference operating system is Ubuntu 20.04.3 LTS.

\section{Results}
We evaluate \ports against \textsc{RePlug}, our primary baseline for goal-driven retrieval. Other training methodologies are omitted from direct comparison due to divergent optimization goals and architectural assumptions.
Our core findings are in Table~\ref{tab:results}, showing metric scores for all models, datasets, and a comparison of different losses for ablation. Given space constraints, we list tool selection outcomes only for the top-performing LLM in each encoder-dataset-loss configuration. We refer the reader to Appendix~\ref{app:complete_results} for exhaustive scores.
We observe that \ports consistently elevates baseline effectiveness in all datasets (cf. the color gradients for $\Delta_{avg}$ columns linked to \lossPorts entries in Table~\ref{tab:results}).
For seen tools, \ports yields substantial gains, boosting average Recall@$\{1,2,3\}$ by up to 71.66 percentage points and average NDCG@$\{1,3,5\}$ by 70.16.
Even with unseen tools, the improvements remain remarkable, reaching +61.24 and +59.79 points in Recall and NDCG, respectively.

\begin{table*}[!t]
    \centering
    \begin{adjustbox}{width=.995\linewidth}
    \begin{threeparttable}
    \begin{tabular}{lcllcccccccc}
    \toprule
    & & & & \multicolumn{3}{c}{\textbf{Recall} (\%)} & \multicolumn{3}{c}{\textbf{NDCG} (\%)} & \multicolumn{2}{c}{\textbf{$\Delta_{avg}$ Baseline}\tnote{$\diamond$}} \\
    \cmidrule(lr){5-7}
    \cmidrule(lr){8-10}
    \cmidrule(lr){11-12}
    \textbf{Encoder} & \textbf{Dataset} & \textbf{Method} & \textbf{Best LLM} & \textbf{@1} & \textbf{@2} & \textbf{@3} & \textbf{@1} & \textbf{@3} & \textbf{@5} & \textbf{Recall} & \textbf{NDCG} \\
    \hline
    \rowcolor{robertabackgroundcolor} & & \lossPorts & \codestral & 18.23 & 21.10 & 25.60 & 18.23 & 20.28 & 24.31 & \cellcolor{robertahighlightcolor!41}20.63 & \cellcolor{robertahighlightcolor!40}19.94 \\
    \rowcolor{robertabackgroundcolor} & \multirow{-2}{*}{\ding{182}}\tnote{$\dagger$} & \lossReplug & \codestral & 12.56 & 20.11 & 24.80 & 12.56 & 19.67 & 22.19 & \cellcolor{robertahighlightcolor!38}18.14 & \cellcolor{robertahighlightcolor!37}17.14 \\
    \cline{2-12}
    \rowcolor{robertabackgroundcolor} & & \lossPorts & \groqLlama & 49.84 & 64.35 & 70.80 & 49.84 & 62.27 & 65.32 & \cellcolor{robertahighlightcolor!77}57.10 & \cellcolor{robertahighlightcolor!74}54.14 \\
    \rowcolor{robertabackgroundcolor} & \multirow{-2}{*}{\ding{183}} & \lossReplug & \groqLlama & 45.32 & 61.94 & 68.23 & 45.32 & 59.95 & 62.44 & \cellcolor{robertahighlightcolor!74}53.93 & \cellcolor{robertahighlightcolor!71}50.90 \\
    \cline{2-12}
    \rowcolor{robertabackgroundcolor} & & \lossPorts & \groqLlama & 21.50 & 27.40 & 30.53 & 21.50 & 25.22 & 26.78 & \cellcolor{robertahighlightcolor!46}25.94 & \cellcolor{robertahighlightcolor!44}23.50 \\
    \rowcolor{robertabackgroundcolor} & \multirow{-2}{*}{\ding{184}} & \lossReplug & \llamaBig & 8.74 & 12.61 & 15.35 & 8.74 & 12.55 & 14.56 & \cellcolor{robertahighlightcolor!31}10.70 & \cellcolor{robertahighlightcolor!31}10.95 \\
    \cline{2-12}
    \rowcolor{robertabackgroundcolor} & & \lossPorts & \groqLlama & 58.12 & 69.21 & 73.52 & 58.12 & 68.38 & 69.22 & \cellcolor{robertahighlightcolor!80}60.30 & \cellcolor{robertahighlightcolor!79}59.24 \\
    \rowcolor{robertabackgroundcolor} & \multirow{-2}{*}{\ding{185}} & \lossReplug & \groqLlama & 53.97 & 64.88 & 68.39 & 53.97 & 62.61 & 64.93 & \cellcolor{robertahighlightcolor!75}55.76 & \cellcolor{robertahighlightcolor!74}54.50 \\
    \cline{2-12}
    \rowcolor{robertabackgroundcolor} & & \lossPorts & \llamaBig & 60.33 & 72.45 & 77.15 & 60.33 & 70.29 & 72.34 & \cellcolor{robertahighlightcolor!77}57.51 & \cellcolor{robertahighlightcolor!75}55.65 \\
    \rowcolor{robertabackgroundcolor} & \multirow{-2}{*}{\ding{186}} & \lossReplug & \groqLlama & 49.38 & 59.84 & 64.23 & 49.38 & 58.17 & 60.79 & \cellcolor{robertahighlightcolor!65}45.35 & \cellcolor{robertahighlightcolor!64}44.11 \\
    \cline{2-12}
    \rowcolor{robertabackgroundcolor} & & \lossPorts & \llamaBig & 95.00 & 100 & 100 & 95 & 95.25 & 98.25 & \cellcolor{robertahighlightcolor!91}71.66 & \cellcolor{robertahighlightcolor!90}70.16 \\
    \rowcolor{robertabackgroundcolor} & \multirow{-2}{*}{\ding{187}} & \lossReplug & \llamaBig & 87.50 & 97.50 & 100 & 87.50 & 95.06 & 95.06 & \cellcolor{robertahighlightcolor!88.33}68.33 & \cellcolor{robertahighlightcolor!86}66.54 \\
    \cline{2-12}
    \rowcolor{robertabackgroundcolor} & & \lossPorts & \llamaBig & 74.60 & 83.90 & 86.80 & 74.60 & 81.23 & 83.55 & \cellcolor{robertahighlightcolor!81}61.24 & \cellcolor{robertahighlightcolor!79}59.79 \\
    \rowcolor{robertabackgroundcolor} & \multirow{-2}{*}{\ding{188}} & \lossReplug & \groqLlama & 58.51 & 68.54 & 74.10 & 58.51 & 67.62 & 69.78 & \cellcolor{robertahighlightcolor!66}46.53 & \cellcolor{robertahighlightcolor!65}45.30 \\
    \cline{2-12}
    \rowcolor{robertabackgroundcolor} & & \lossPorts & \llamaBig & 96.00 & 100 & 100 & 96.00 & 98.22 & 98.22\tnote{$*$} & \cellcolor{robertahighlightcolor!43}23.89 & \cellcolor{robertahighlightcolor!44}24.48 \\
    \rowcolor{robertabackgroundcolor} \multirow{-16}{*}{\shortstack[l]{RoBERTa}} & \multirow{-2}{*}{\ding{189}} & \lossReplug & \llamaBig & 80.00 & 100 & 100 & 80.00 & 92.62 & 92.62\tnote{$*$} & \cellcolor{robertahighlightcolor!39}19.16 & \cellcolor{robertahighlightcolor!35}15.41 \\    
    \hline
    \rowcolor{bgebackgroundcolor} & & \lossPorts & \codestral & 25.80 & 36.05 & 43.35 & 25.80 & 35.50 & 42.20 & \cellcolor{bgehighlightcolor!29}8.71 & \cellcolor{bgehighlightcolor!30}9.50 \\
    \rowcolor{bgebackgroundcolor} & \multirow{-2}{*}{\ding{182}}\tnote{$\dagger$} & \lossReplug & \llamaBig & 22.56 & 33.74 & 40.51 & 22.56 & 33.00 & 36.65 & \cellcolor{bgehighlightcolor!26}5.92 & \cellcolor{bgehighlightcolor!26}5.73 \\
    \cline{2-12}
    \rowcolor{bgebackgroundcolor} & & \lossPorts & \groqLlama & 59.12 & 76.80 & 81.50 & 59.12 & 75.40 & 76.10 & \cellcolor{bgehighlightcolor!35}14.94 & \cellcolor{bgehighlightcolor!34}14.21 \\
    \rowcolor{bgebackgroundcolor} & \multirow{-2}{*}{\ding{183}} & \lossReplug & \groqLlama & 56.29 & 75.00 & 80.00 & 56.29 & 70.60 & 73.32 & \cellcolor{bgehighlightcolor!33}12.90 & \cellcolor{bgehighlightcolor!31}10.73 \\
    \cline{2-12}
    \rowcolor{bgebackgroundcolor} & & \lossPorts & \groqLlama & 30.64 & 37.20 & 41.06 & 30.64 & 33.90 & 35.20 & \cellcolor{bgehighlightcolor!40}20.18 & \cellcolor{bgehighlightcolor!37}17.26 \\
    \rowcolor{bgebackgroundcolor} & \multirow{-2}{*}{\ding{184}} & \lossReplug & \llamaBig & 19.55 & 28.29 & 33.05 & 19.55 & 27.45 & 30.02 & \cellcolor{bgehighlightcolor!31}10.84 & \cellcolor{bgehighlightcolor!31}10.67 \\
    \cline{2-12}
    \rowcolor{bgebackgroundcolor} & & \lossPorts & \groqLlama & 67.20 & 73.23 & 78.10 & 67.20 & 74.60 & 73.10 & \cellcolor{bgehighlightcolor!25}5.31 & \cellcolor{bgehighlightcolor!26}5.63 \\
    \rowcolor{bgebackgroundcolor} & \multirow{-2}{*}{\ding{185}} & \lossReplug & \groqLlama & 66.17 & 73.20 & 77.82 & 66.17 & 72.92 & 74.31 & \cellcolor{bgehighlightcolor!25}4.86 & \cellcolor{bgehighlightcolor!25}5.13 \\
    \cline{2-12}
    \rowcolor{bgebackgroundcolor} & & \lossPorts & \codestral & 67.35 & 79.48 & 83.75 & 67.35 & 77.00 & 78.00 & \cellcolor{bgehighlightcolor!35}14.71 & \cellcolor{bgehighlightcolor!34}14.12 \\
    \rowcolor{bgebackgroundcolor} & \multirow{-2}{*}{\ding{186}} & \lossReplug & \groqLlama & 67.23 & 77.54 & 81.06 & 67.23 & 75.50 & 76.82 & \cellcolor{bgehighlightcolor!33}13.12 & \cellcolor{bgehighlightcolor!33}13.18 \\
    \cline{2-12}
    \rowcolor{bgebackgroundcolor} & & \lossPorts & \llamaBig & 97.50 & 100 & 100 & 97.50 & 100 & 100 & \cellcolor{bgehighlightcolor!22}1.67 & \cellcolor{bgehighlightcolor!22}2.17 \\
    \rowcolor{bgebackgroundcolor} & \multirow{-2}{*}{\ding{187}} & \lossReplug & \llamaBig & 95.00 & 100 & 100 & 95.00 & 98.00 & 98.00 & \cellcolor{bgehighlightcolor!21}0.83 & \cellcolor{bgehighlightcolor!20}0.22 \\
    \cline{2-12}
    \rowcolor{bgebackgroundcolor} & & \lossPorts & \groqLlama & 89.87 & 92.20 & 94.04 & 89.87 & 91.10 & 92.35 & \cellcolor{bgehighlightcolor!34}14.18 & \cellcolor{bgehighlightcolor!37}17.11 \\
    \rowcolor{bgebackgroundcolor} & \multirow{-2}{*}{\ding{188}} & \lossReplug & \groqLlama & 74.60 & 86.24 & 90.30 & 74.60 & 83.98 & 85.09 & \cellcolor{bgehighlightcolor!26}5.85 & \cellcolor{bgehighlightcolor!27}7.22 \\
    \cline{2-12}
    \rowcolor{bgebackgroundcolor} & & \lossPorts & \llamaBig & 97.50 & 100 & 100 & 97.50 & 100 & 100\tnote{$*$} & \cellcolor{bgehighlightcolor!21}0.84 & \cellcolor{bgehighlightcolor!21}1.17 \\
    \rowcolor{bgebackgroundcolor} \multirow{-16}{*}{\shortstack[l]{BGE}} & \multirow{-2}{*}{\ding{189}} & \lossReplug & \groqLlama & 95.00 & 100 & 100 & 95.00 & 98.00 & 98.00\tnote{$*$} & \cellcolor{bgehighlightcolor!1}0 &\cellcolor{bgehighlightcolor!1}0 \\
    \bottomrule
    \end{tabular}
    \begin{tablenotes}
    \item[$\dagger$] Results computed on the G3 split.
    \item[$*$] NDCG@4 since the out-of-domain version of Octopus-v2 has 4 tools only.\\
    \item[$\diamond$] $\Delta_{avg}$ measures the average percentage point improvement across all ranks @$K$.
    \end{tablenotes}
    \end{threeparttable}
    \end{adjustbox}
    \caption{\textbf{\texttt{PORTS} Recall@$K$ and NDCG@$K$ per encoder-dataset-loss (test set).} Reported results refer to the best LLM for each triplet. The positive gains in metric scores over the baselines are highlighted (the brighter, the better).}
    \label{tab:results}
\end{table*}

\begin{figure*}[!t]
    \centering
    \vspace{4mm}
    \begin{subfigure}[t]{\linewidth}
        \begin{tikzpicture}
            \footnotesize
            \node[draw=none, inner sep=2pt, align=center, text width=8cm] {
            \begin{tabular}{llll}
                \ref{plot:codestral} $\texttt{PORTS}_\textsc{Codestral}$ & \ref{plot:llama3groq} $\texttt{PORTS}_\textsc{LLama3-Groq}$ & \ref{plot:llama3} $\texttt{PORTS}_\textsc{Llama3}$ & \ref{plot:baseline} Baseline \\
            \end{tabular}
            };
        \end{tikzpicture}
    \end{subfigure}
    \colorbox{seenbackgroundcolor}{
    \begin{subfigure}[H]{.68\linewidth}
    \centering
    \vspace{2mm}
    \faEye\hspace{1mm}\textbf{Seen Tools}\\[-2mm]
    \hspace{-2mm}\rule{.95\linewidth}{0.4pt}\\[1mm]
    \hspace{.2cm} 
    \begin{subfigure}[H]{.32\linewidth}
        \begin{tikzpicture}
        \begin{axis}[
            width=1.1\linewidth, height=3.1cm,
            ymajorgrids=true,
            grid=both,
            grid style=dashed,
            axis background/.style={fill=plotbackground},
            ybar=0pt, 
            bar width=6pt,
            enlarge x limits=0.5,  
            xtick=data,
            symbolic x coords={
                \robertaShort,
                \bgeShort
            },
            xticklabels={},
            ymin=0,
            every tick label/.append style={font=\fontsize{8}{8}\selectfont},
            title=\textbf{ToolBench},
            title style={font=\fontsize{9}{9}\selectfont},
        ]
        
        \addplot [fill=codestralcolor, draw=none] coordinates {
            (\robertaShort, 21.87)
            (\bgeShort, 34.83)
        }; \label{plot:codestral}
        \addplot [fill=llama3groqcolor, draw=none] coordinates {
            (\robertaShort, 18.24)
            (\bgeShort, 34.13)
        }; \label{plot:llama3groq}
        \addplot [fill=llama3color, draw=none] coordinates {
            (\robertaShort, 16.49)
            (\bgeShort, 33.01)
        }; \label{plot:llama3}
        \addplot [fill=darkerbaselinecolor, draw=none] coordinates {
            (\robertaShort, 1.01)
            (\bgeShort, 26.35)
        }; \label{plot:baseline}
        
        \draw (axis cs:{[normalized]\pgfkeysvalueof{/pgfplots/xmin}},0)
            -- (axis cs:{[normalized]\pgfkeysvalueof{/pgfplots/xmax}},0);
        
        \end{axis}
        \end{tikzpicture}
    \end{subfigure}
    \hspace{-3mm}
    \begin{subfigure}[H]{.32\linewidth}
        \begin{tikzpicture}
        \begin{axis}[
            width=1.1\linewidth, height=3.1cm,
            ymajorgrids=true,
            grid=both,
            grid style=dashed,
            axis background/.style={fill=plotbackground},
            ybar=0pt, 
            bar width=6pt,
            enlarge x limits=0.5,  
            xtick=data,
            symbolic x coords={
                \robertaShort,
                \bgeShort
            },
            xticklabels={},
            ymin=0,
            every tick label/.append style={font=\fontsize{8}{8}\selectfont},
            title=\textbf{APIBench},
            title style={font=\fontsize{9}{9}\selectfont},
        ]
        
        \addplot [fill=codestralcolor, draw=none] coordinates {
            (\robertaShort, 18.06)
            (\bgeShort, 33.33)
        };
        \addplot [fill=llama3groqcolor, draw=none] coordinates {
            (\robertaShort, 26.48)
            (\bgeShort, 36.30)
        };
        \addplot [fill=llama3color, draw=none] coordinates {
            (\robertaShort, 24.11)
            (\bgeShort, 36.69)
        };
        \addplot [fill=darkerbaselinecolor, draw=none] coordinates {
            (\robertaShort, 1.53)
            (\bgeShort, 16.12)
        };
        
        \draw (axis cs:{[normalized]\pgfkeysvalueof{/pgfplots/xmin}},0)
            -- (axis cs:{[normalized]\pgfkeysvalueof{/pgfplots/xmax}},0);
        
        \end{axis}
        \end{tikzpicture}
    \end{subfigure}
    \hspace{-3mm}
    \begin{subfigure}[H]{.32\linewidth}
        \begin{tikzpicture}
        \begin{axis}[
            width=1.1\linewidth, height=3.1cm,
            ymajorgrids=true,
            grid=both,
            grid style=dashed,
            axis background/.style={fill=plotbackground},
            ybar=0pt, 
            bar width=6pt,
            enlarge x limits=0.5,  
            xtick=data,
            symbolic x coords={
                \robertaShort,
                \bgeShort
            },
            xticklabels={},
            ymin=0,
            every tick label/.append style={font=\fontsize{8}{8}\selectfont},
            title=\textbf{ToolE},
            title style={font=\fontsize{9}{9}\selectfont},
        ]
        
        \addplot [fill=codestralcolor, draw=none] coordinates {
            (\robertaShort, 67.20)
            (\bgeShort, 76.86)
        };
        \addplot [fill=llama3groqcolor, draw=none] coordinates {
            (\robertaShort, 69.57)
            (\bgeShort, 76.97)
        };
        \addplot [fill=llama3color, draw=none] coordinates {
            (\robertaShort, 69.97)
            (\bgeShort, 77.00)
        };
        \addplot [fill=darkerbaselinecolor, draw=none] coordinates {
            (\robertaShort, 12.46)
            (\bgeShort, 62.15)
        };
        
        \draw (axis cs:{[normalized]\pgfkeysvalueof{/pgfplots/xmin}},0)
            -- (axis cs:{[normalized]\pgfkeysvalueof{/pgfplots/xmax}},0);
        
        \end{axis}
        \end{tikzpicture}
    \end{subfigure}\\[-1mm]
    \hspace{.2mm} 
    \begin{subfigure}[H]{.32\linewidth}
        \begin{tikzpicture}
        \begin{axis}[
            width=1.1\linewidth, height=3.1cm,
            ymajorgrids=true,
            grid=both,
            grid style=dashed,
            axis background/.style={fill=plotbackground},
            ybar=0pt, 
            bar width=6pt,
            enlarge x limits=0.5,  
            xtick=data,
            symbolic x coords={
                \robertaShort,
                \bgeShort
            },
            ymin=0,
            every tick label/.append style={font=\fontsize{8}{8}\selectfont},
            title=\textbf{API-Bank},
            title style={font=\fontsize{9}{9}\selectfont},
        ]
        
        \addplot [fill=codestralcolor, draw=none] coordinates {
            (\robertaShort, 54.46)
            (\bgeShort, 61.67)
        };
        \addplot [fill=llama3groqcolor, draw=none] coordinates {
            (\robertaShort, 61.66)
            (\bgeShort, 72.47)
        };
        \addplot [fill=llama3color, draw=none] coordinates {
            (\robertaShort, 63.81)
            (\bgeShort, 71.82)
        };
        \addplot [fill=darkerbaselinecolor, draw=none] coordinates {
            (\robertaShort, 4.56)
            (\bgeShort, 57.53)
        };
        
        \draw (axis cs:{[normalized]\pgfkeysvalueof{/pgfplots/xmin}},0)
            -- (axis cs:{[normalized]\pgfkeysvalueof{/pgfplots/xmax}},0);
        
        \end{axis}
        \end{tikzpicture}
    \end{subfigure}
    \hspace{-3mm}
    \begin{subfigure}[H]{.32\linewidth}
        \begin{tikzpicture}
        \begin{axis}[
            width=1.1\linewidth, height=3.1cm,
            ymajorgrids=true,
            grid=both,
            grid style=dashed,
            axis background/.style={fill=plotbackground},
            ybar=0pt, 
            bar width=6pt,
            enlarge x limits=0.5,  
            xtick=data,
            symbolic x coords={
                \robertaShort,
                \bgeShort
            },
            ymin=0,
            every tick label/.append style={font=\fontsize{8}{8}\selectfont},
            title=\textbf{BFCL},
            title style={font=\fontsize{9}{9}\selectfont},
        ]
        
        \addplot [fill=codestralcolor, draw=none] coordinates {
            (\robertaShort, 49.60)
            (\bgeShort, 72.09)
        };
        \addplot [fill=llama3groqcolor, draw=none] coordinates {
            (\robertaShort, 66.95)
            (\bgeShort, 72.62)
        };
        \addplot [fill=llama3color, draw=none] coordinates {
            (\robertaShort, 65.43)
            (\bgeShort, 72.40)
        };
        \addplot [fill=darkerbaselinecolor, draw=none] coordinates {
            (\robertaShort, 6.65)
            (\bgeShort, 67.53)
        };
        
        \draw (axis cs:{[normalized]\pgfkeysvalueof{/pgfplots/xmin}},0)
            -- (axis cs:{[normalized]\pgfkeysvalueof{/pgfplots/xmax}},0);
        
        \end{axis}
        \end{tikzpicture}
    \end{subfigure}
    \hspace{-4mm}
    \begin{subfigure}[H]{.32\linewidth}
        \begin{tikzpicture}
        \begin{axis}[
            width=1.1\linewidth, height=3.1cm,
            ymajorgrids=true,
            grid=both,
            grid style=dashed,
            axis background/.style={fill=plotbackground},
            ybar=0pt, 
            bar width=6pt,
            enlarge x limits=0.5,  
            xtick=data,
            symbolic x coords={
                \robertaShort,
                \bgeShort
            },
            ymin=0,
            every tick label/.append style={font=\fontsize{8}{8}\selectfont},
            title=\textbf{Octopus-v2},
            title style={font=\fontsize{9}{9}\selectfont},
        ]
        
        \addplot [fill=codestralcolor, draw=none] coordinates {
            (\robertaShort, 98.33)
            (\bgeShort, 98.33)
        };
        \addplot [fill=llama3groqcolor, draw=none] coordinates {
            (\robertaShort, 98.33)
            (\bgeShort, 98.33)
        };
        \addplot [fill=llama3color, draw=none] coordinates {
            (\robertaShort, 98.33)
            (\bgeShort, 99.17)
        };
        \addplot [fill=darkerbaselinecolor, draw=none] coordinates {
            (\robertaShort, 26.67)
            (\bgeShort, 97.5)
        };
        
        \draw (axis cs:{[normalized]\pgfkeysvalueof{/pgfplots/xmin}},0)
            -- (axis cs:{[normalized]\pgfkeysvalueof{/pgfplots/xmax}},0);
        
        \end{axis}
        \end{tikzpicture}
    \end{subfigure}
    \end{subfigure}}
    \colorbox{unseenbackgroundcolor}{
    \begin{subfigure}[H]{.26\linewidth}
    \centering
    \vspace{2mm}
    \faEyeSlash\hspace{1mm}\textbf{Unseen Tools}\\[-2mm]
    \hspace{-2mm}\rule{.8\linewidth}{0.4pt}\\[1mm]
    \hspace{.2cm} 
    \begin{subfigure}[H]{.87\linewidth}
        \begin{tikzpicture}
        \begin{axis}[
            width=1.1\linewidth, height=3.1cm,
            ymajorgrids=true,
            grid=both,
            grid style=dashed,
            axis background/.style={fill=plotbackground},
            ybar=0pt, 
            bar width=6pt,
            enlarge x limits=0.5,  
            xtick=data,
            symbolic x coords={
                \robertaShort,
                \bgeShort
            },
            xticklabels={},
            ymin=0,
            every tick label/.append style={font=\fontsize{8}{8}\selectfont},
            title=\textbf{ToolE},
            title style={font=\fontsize{9}{9}\selectfont},
        ]
        
        \addplot [fill=codestralcolor, draw=none] coordinates {
            (\robertaShort, 79.46)
            (\bgeShort, 90.70)
        };
        \addplot [fill=llama3groqcolor, draw=none] coordinates {
            (\robertaShort, 81.03)
            (\bgeShort, 91.04)
        };
        \addplot [fill=llama3color, draw=none] coordinates {
            (\robertaShort, 81.77)
            (\bgeShort, 86.60)
        };
        \addplot [fill=darkerbaselinecolor, draw=none] coordinates {
            (\robertaShort, 20.53)
            (\bgeShort, 77.86)
        };
        
        \draw (axis cs:{[normalized]\pgfkeysvalueof{/pgfplots/xmin}},0)
            -- (axis cs:{[normalized]\pgfkeysvalueof{/pgfplots/xmax}},0);
        
        \end{axis}
        \end{tikzpicture}
    \end{subfigure}\\
    \hspace{.1cm} 
    \begin{subfigure}[H]{.87\linewidth}
        \begin{tikzpicture}
        \begin{axis}[
            width=1.1\linewidth, height=3.1cm,
            ymajorgrids=true,
            grid=both,
            grid style=dashed,
            axis background/.style={fill=plotbackground},
            ybar=0pt, 
            bar width=6pt,
            enlarge x limits=0.5,  
            xtick=data,
            symbolic x coords={
                \robertaShort,
                \bgeShort
            },
            ymin=0,
            every tick label/.append style={font=\fontsize{8}{8}\selectfont},
            title=\textbf{Octopus-v2},
            title style={font=\fontsize{9}{9}\selectfont},
        ]
        
        \addplot [fill=codestralcolor, draw=none] coordinates {
            (\robertaShort, 98.33)
            (\bgeShort, 98.67)
        };
        \addplot [fill=llama3groqcolor, draw=none] coordinates {
            (\robertaShort, 98.33)
            (\bgeShort, 99.00)
        };
        \addplot [fill=llama3color, draw=none] coordinates {
            (\robertaShort, 98.67)
            (\bgeShort, 99.17)
        };
        \addplot [fill=darkerbaselinecolor, draw=none] coordinates {
            (\robertaShort, 74.17)
            (\bgeShort, 98.33)
        };
        
        \draw (axis cs:{[normalized]\pgfkeysvalueof{/pgfplots/xmin}},0)
            -- (axis cs:{[normalized]\pgfkeysvalueof{/pgfplots/xmax}},0);
        
        \end{axis}
        \end{tikzpicture}
    \end{subfigure}
    \end{subfigure}}
    \caption{\textbf{Average Recall@$K$ for each dataset (test set).} Effectiveness of \texttt{PORTS}-tuned retrievers (\lossPorts) against frozen baselines, utilizing guidance from different LLMs. Evaluated in both in-domain and out-of-domain settings.}
    \label{fig:results}
\end{figure*}
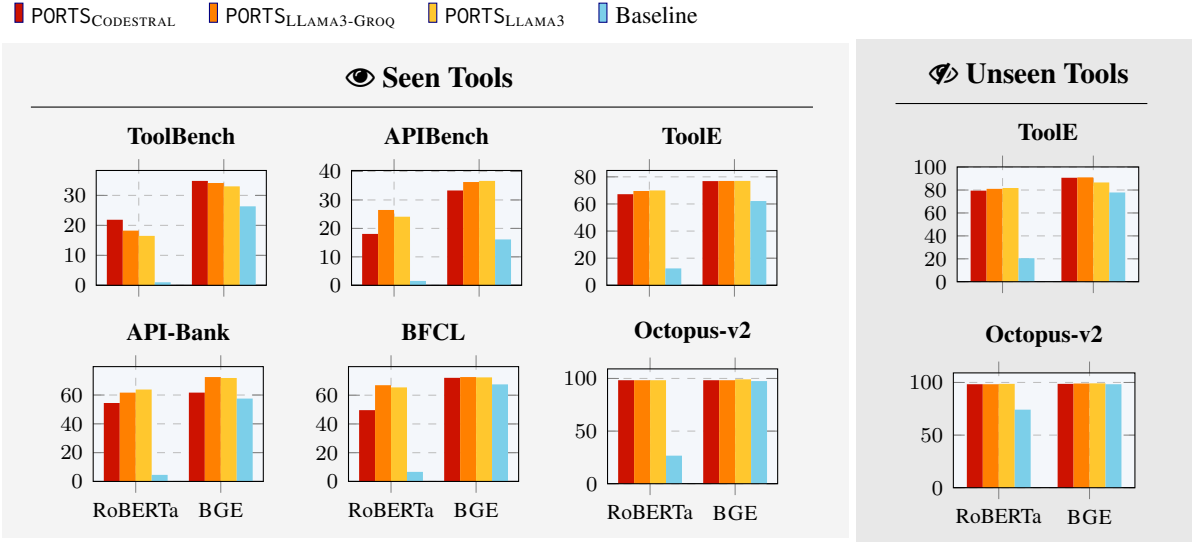

\paragraph{Baseline Impact}
Both \robertaShort and \bgeShort fine-tuned with \ports significantly outperform their respective base models, demonstrating the broad applicability of our framework.
\robertaShort exhibits a more pronounced response to \ports ($\Delta_{avg}$ Recall of 47.28 compared to \bgeShort's 10.07), indicating that simpler and less specialized retrievers are more adaptable.
Table~\ref{tab:results} clearly illustrate that \bgeShort begins with a notable disparity between positive and negative instances (average frozen Recall of 62.92 vs. \robertaShort's 18.45). This pre-existing imbalance in the retrieval distribution introduces a skew, which in turn attenuates the strength of our loss signal. Further robustness tests are reported in Appendix~\ref{app:robustness}.

\paragraph{LLM Impact}
The use of an LLM as a proxy during training requires careful scrutiny. Although the primary focus is not the direct accuracy of the downstream generative task, the distribution of perplexities between different generations raises questions about the optimal LLM to maximize learning influence. Empirical data did not elucidate a definitive model preference but instead accentuated \ports' potential to transcend its intended tool usage scope. The performance delta across all experiments ranged from +1 to +4 percentage points, with peaks attaining a +13 average recall improvement.
\llamaBig demonstrated superior performance when paired with \robertaShort, where--as previously discussed--there was markedly less resistance to adaptation.
This improvement can be attributed to the heightened uncertainty this model experiences in technical domains not anticipated during its pre-training regimen.
Although elevated perplexity is generally undesirable for downstream task optimization, it is advantageous in the \ports framework due to the resultant richer and less skewed log-likelihood distribution, as elucidated in Eq.~\ref{eq:q_ppl}.

\paragraph{Docstring Impact}
The efficacy of contrastive retrieval is heavily contingent on docstring quality.
APIBench illustrates this relationship, where vague descriptions result in smaller changes in the tool retrieval metrics.
In fact, the tool documents in the dataset reference generic features of pre-trained models from HuggingFace, which often omit parameter names and types that could better guide the retriever at training time.
Importantly, \llamaBig tends to perform better in datasets with less technical tool documentation due to its superior management of perplexity scores.
On the other hand, \groqLlama and \codestral, with their advanced tool usage capabilities, excel in datasets with more detailed docstrings (e.g., arguments, outputs, types, defaults), namely ToolBench and Octopus-v2.

\paragraph{Training Loss}
Contrastive learning and \textsc{RePlug} have known simultaneous success in various application domains.
Our assertions on the positive impact of preference-oriented learning have been substantiated through comprehensive ablations presented in Table~\ref{tab:results}.
\ports outperforms \textsc{RePlug} in all evaluated domains, with substantial disparities in recall performance, with +15.8 and +12.3 in ToolE and ApiBench, respectively.
We underline that these datasets present unique challenges: ToolE necessitates precise tool decisions among similar options in complex scenarios, while ApiBench involves code generation primarily through pre-trained neural networks, described only in broad application contexts.
Such unique characteristics are conducive to showcase the benefits of our comparative loss approach, which likely contributes to \ports's superior performance in these settings.
Zooming out, the statistics outlined in Figure \ref{fig:abstract} reflect an average Recall@$K$ improvement of 6.7 (\robertaShort) and 3.4 (\bgeShort) across all datasets and LLMs, ultimately corroborating the greater efficacy of our method over \textsc{RePlug} in the context of tool retrieval.

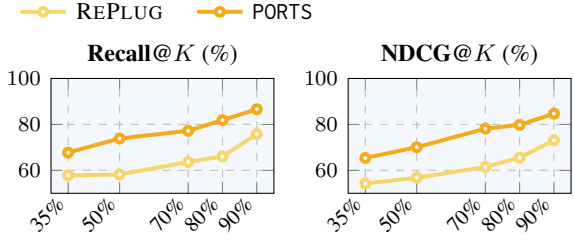
\begin{figure}[!t]
    \centering
    \begin{subfigure}[t]{\linewidth}
        \begin{tikzpicture}
            \footnotesize
            \node[draw=none, inner sep=2pt, align=left, text width=8cm] {
            \begin{tabular}{ll}
                \ref{plot:prog_replug} $\textsc{RePlug}$ & \ref{plot:prog_ports} $\texttt{PORTS}$ \\
            \end{tabular}
            };
        \end{tikzpicture}
    \end{subfigure}
    \begin{subfigure}[H]{.49\linewidth}
        \begin{tikzpicture}
        \begin{axis}[
            width=1.2\linewidth, height=3.1cm,
            ymajorgrids=true,
            grid=both,
            grid style=dashed,
            axis background/.style={fill=plotbackground},
            xmin=30, xmax=95,
            xtick={35,50,70,80,90},
            xticklabels={35\%,50\%,70\%,80\%,90\%},
            xticklabel style={rotate=45, anchor=east},
            ymin=50, ymax=100,
            every tick label/.append style={font=\fontsize{8}{8}\selectfont},
            title={\textbf{Recall@$K$} (\%)},
            title style={font=\fontsize{9}{9}\selectfont, align=center, yshift=-1ex},
        ]

        \addplot [
            color=replugcolor!97!black,
            line width=1.5pt,
            mark=*,
            mark size=1.5pt,
            mark options={fill=white},
        ] coordinates {
            (35, 57.78)
            (50, 58.18)
            (70, 63.63)
            (80, 66.09)
            (90, 75.79)
        }; \label{plot:prog_replug}
        
        \addplot [
            color=portscolor!97!black,
            line width=1.5pt,
            mark=*,
            mark size=1.5pt,
            mark options={fill=white},
        ] coordinates {
            (35, 67.73)
            (50, 73.86)
            (70, 77.22)
            (80, 81.86)
            (90, 86.61)
        }; \label{plot:prog_ports}
        
        \end{axis}
        \end{tikzpicture}
    \end{subfigure}
    \hfill
    \begin{subfigure}[H]{.49\linewidth}
        \begin{tikzpicture}
        \begin{axis}[
            width=1.2\linewidth, height=3.1cm,
            ymajorgrids=true,
            grid=both,
            grid style=dashed,
            axis background/.style={fill=plotbackground},
            xmin=30, xmax=95,
            xtick={35,50,70,80,90},
            xticklabels={35\%,50\%,70\%,80\%,90\%},
            xticklabel style={rotate=45, anchor=east},
            ymin=50, ymax=100,
            every tick label/.append style={font=\fontsize{8}{8}\selectfont},
            title={\textbf{NDCG@$K$} (\%)},
            title style={font=\fontsize{9}{9}\selectfont, align=center, yshift=-1ex},
        ]

        \addplot [
            color=replugcolor!97!black,
            line width=1.5pt,
            mark=*,
            mark size=1.5pt,
            mark options={fill=white},
        ] coordinates {
            (35, 54.23)
            (50, 56.78)
            (70, 61.45)
            (80, 65.53)
            (90, 73.07)
        };
        
        \addplot [
            color=portscolor!97!black,
            line width=1.5pt,
            mark=*,
            mark size=1.5pt,
            mark options={fill=white},
        ] coordinates {
            (35, 65.4)
            (50, 70.06)
            (70, 78.14)
            (80, 79.79)
            (90, 84.67)
        };
        
        \end{axis}
        \end{tikzpicture}
    \end{subfigure}
    \caption{\textbf{Average out-of-domain Recall@$K$ and NDGC@$K$ with a progressive number of train tools (decreasing unseen).} Reported results refer to \robertaShort on ToolE with \textsc{RePlug} and \texttt{PORTS} under \llamaBig supervision.}
    \label{fig:results_incremental_unseen}
\end{figure}

\paragraph{Out-Of-Domain Analysis}
For domains with limited resources and scarce data, retrieval systems must be capable of effectively managing unfamiliar tools. Although previous research has developed systems that can adapt to new data \cite{DBLP:conf/aaai/GaoSZF0RC0R24}, these systems are susceptible to overfitting, potentially due to biases in the distribution and usage patterns of tools. To evaluate the robustness of our method in addressing these challenges, we investigated the performance of retrieval models trained with \textsc{RePlug} and \ports when exposed to varying proportions of unseen tools from the ToolE dataset. Starting with a 90/10 ratio of seen to unseen tools, we progressively reduced the training dataset and assessed the performance of \robertaShort guided by \llamaBig on a consistent test distribution. Figure \ref{fig:results_incremental_unseen} illustrates the superior generalization capabilities of \ports. Employing preference optimization loss, our contrastive learning techniques effectively derive semantic insights into query-tool interactions without requiring extensive pairwise comparisons, thereby substantiating the enhanced low-resource capabilities of our solution and demonstrating its suitability for application areas with constantly evolving API documentation.
Figure~\ref{fig:results} reports the average recall across test datasets and encoder models for all \ports' variants fine-tuned with different LLMs.

\section{Conclusion}

We introduce \ports, a novel training method to optimize encoders for tool retrieval tasks. Our goal is twofold: to align tool selection with the preferences of the calling LLM, and to maximize the odds ratio between correct and incorrect tools.
\ports emphasizes low cost by leveraging LLMs' prior knowledge to navigate the latent space of tool document similarities, focusing on the impact of retrieved samples.
Experiments across multiple models and diverse datasets show that \ports achieves Recall@1 improvements of up to \textbf{+72.5\%} and \textbf{+58.7\%} over frozen baselines for in- and out-domain tools, with gains of \textbf{+15.24\%} and \textbf{+14.71\%} percentage points compared to \textsc{RePlug}.

To further assess retrieval effectiveness in code generation scenarios, metrics like \textsc{Pass}@$K$ could reflect the downstream impact on generative components.
Incorporating relevance signals based on actual output effects and message similarity may allow \ports to integrate discounting mechanisms for more goal-directed retrieval.
Future work could also investigate \ports in biomedical discovery workflows~\cite{DBLP:conf/acl/WangMWWJCLY25}--e.g. helping an agent decide whether to use specialized function discovery models~\cite{DBLP:journals/cmpb/DomeniconiMMP16,DBLP:conf/ic3k/DomeniconiMMP14,DBLP:journals/bmcbi/LenaDMM15}, call Gene Ontology tooling for term enrichment, query STRING for protein-protein interaction networks, or fetch data from the Unified Medical Language System--with dozens to hundreds of specialized data sources and tools with different formats, coverage, and update frequency.

\section*{Acknowledgment}
Research partially supported by: \href{https://aipact-edih.it}{AI-PACT} (CUP B47H22004450008, B47H22004460001); National Plan PNC-I.1 \href{https://www.fondazionedare.it/en/progetto-obiettivi-struttura/}{DARE} (PNC0000002, CUP B53C22006450001); PNRR Extended Partnership \href{https://fondazione-fair.it/en/}{FAIR} (PE00000013, Spoke 8); 2024 Scientific Research and High Technology Program, project ``\href{https://disi-unibo-nlp.github.io/projects/carisbo/}{AI analysis for risk assessment of empty lymph nodes in endometrial cancer surgery}'', the Fondazione Cassa di Risparmio in Bologna; Chips JU \href{https://tristan-project.eu/team/}{TRISTAN} (G.A. 101095947). LG Solution Srl for partially funding a PhD scholarship to L. Molfetta.

\section*{Limitations}

Despite its strong results, \ports has limitations that warrant further examination.
First, its effectiveness is sensitive to the quality of tool documentation, with diminished gains in domains where docstrings are vague or underspecified. 
Second, although PORTS maintains a memory efficiency comparable to \replug, it requires repeated querying of a frozen LLM to calculate the guidance signals. 
This dependence introduces additional computational overhead in both time and memory, potentially limiting scalability in resource-constrained settings. 
Future work may address these challenges by reducing the reliance on LLM inference or developing efficient approximations of guidance signals, thus improving the practicality of retrieval methods that use LLMs as proxies at training time.

\bibliography{bibliography}

@inproceedings{DBLP:conf/nips/Yao0YN22,
  author       = {Shunyu Yao and
                  Howard Chen and
                  John Yang and
                  others},
  editor       = {Sanmi Koyejo and
                  others},
  title        = {WebShop: Towards Scalable Real-World Web Interaction with Grounded
                  Language Agents},
  booktitle    = {{NeurIPS}},
  year         = {2022},
  url          = {http://papers.nips.cc/paper\_files/paper/2022/hash/82ad13ec01f9fe44c01cb91814fd7b8c-Abstract-Conference.html},
  bibsource    = {dblp computer science bibliography, https://dblp.org}
}

@article{DBLP:journals/corr/abs-2203-05115,
  author       = {Angeliki Lazaridou and
                  Elena Gribovskaya and
                  Wojciech Stokowiec and
                  others},
  title        = {Internet-augmented language models through few-shot prompting for
                  open-domain question answering},
  journal      = {CoRR},
  volume       = {abs/2203.05115},
  year         = {2022},
  url          = {https://doi.org/10.48550/arXiv.2203.05115},
  doi          = {10.48550/ARXIV.2203.05115},
  eprinttype    = {arXiv},
  eprint       = {2203.05115},
  bibsource    = {dblp computer science bibliography, https://dblp.org}
}

@article{DBLP:journals/corr/abs-2405-17935,
  author       = {Changle Qu and
                  Sunhao Dai and
                  Xiaochi Wei and
                  others},
  title        = {Tool Learning with Large Language Models: {A} Survey},
  journal      = {CoRR},
  volume       = {abs/2405.17935},
  year         = {2024},
  url          = {https://doi.org/10.48550/arXiv.2405.17935},
  doi          = {10.48550/ARXIV.2405.17935},
  eprinttype    = {arXiv},
  eprint       = {2405.17935},
  bibsource    = {dblp computer science bibliography, https://dblp.org}
}

@article{DBLP:journals/corr/abs-2311-09210,
  author       = {Wenhao Yu and
                  Hongming Zhang and
                  Xiaoman Pan and
                  others},
  title        = {Chain-of-Note: Enhancing Robustness in Retrieval-Augmented Language
                  Models},
  journal      = {CoRR},
  volume       = {abs/2311.09210},
  year         = {2023},
  url          = {https://doi.org/10.48550/arXiv.2311.09210},
  doi          = {10.48550/ARXIV.2311.09210},
  eprinttype    = {arXiv},
  eprint       = {2311.09210},
  bibsource    = {dblp computer science bibliography, https://dblp.org}
}

@inproceedings{DBLP:conf/icml/ShiCMSDCSZ23,
  author       = {Freda Shi and
                  Xinyun Chen and
                  Kanishka Misra and
                  others},
  editor       = {Andreas Krause and
                  others},
  title        = {Large Language Models Can Be Easily Distracted by Irrelevant Context},
  booktitle    = {{ICML}},
  series       = {{PMLR}},
  volume       = {202},
  pages        = {31210--31227},
  publisher    = {{PMLR}},
  year         = {2023},
  url          = {https://proceedings.mlr.press/v202/shi23a.html},
  bibsource    = {dblp computer science bibliography, https://dblp.org}
}

@article{DBLP:journals/jmlr/IzacardLLHPSDJRG23,
  author       = {Gautier Izacard and
                  Patrick S. H. Lewis and
                  Maria Lomeli and
                  others},
  title        = {Atlas: Few-shot Learning with Retrieval Augmented Language Models},
  journal      = {{JMLR}},
  volume       = {24},
  pages        = {251:1--251:43},
  year         = {2023},
  url          = {http://jmlr.org/papers/v24/23-0037.html},
  bibsource    = {dblp computer science bibliography, https://dblp.org}
}

@inproceedings{DBLP:conf/iclr/Lin0CSL00KSLZY24,
  author       = {Xi Victoria Lin and
                  Xilun Chen and
                  Mingda Chen and
                  others},
  title        = {{RA-DIT:} Retrieval-Augmented Dual Instruction Tuning},
  booktitle    = {{ICLR}},
  publisher    = {OpenReview.net},
  year         = {2024},
  url          = {https://openreview.net/forum?id=22OTbutug9},
  bibsource    = {dblp computer science bibliography, https://dblp.org}
}

@inproceedings{DBLP:conf/nips/ChengLCL0023,
  author       = {Xin Cheng and
                  Di Luo and
                  Xiuying Chen and
                  others},
  editor       = {Alice Oh and
                  others},
  title        = {Lift Yourself Up: Retrieval-augmented Text Generation with Self-Memory},
  booktitle    = {{NeurIPS}},
  year         = {2023},
  url          = {http://papers.nips.cc/paper\_files/paper/2023/hash/887262aeb3eafb01ef0fd0e3a87a8831-Abstract-Conference.html},
  bibsource    = {dblp computer science bibliography, https://dblp.org}
}

@inproceedings{DBLP:conf/iclr/QinLYZYLLCTQZHT24,
  author       = {Yujia Qin and
                  Shihao Liang and
                  Yining Ye and
                  others},
  title        = {ToolLLM: Facilitating Large Language Models to Master 16000+ Real-world
                  APIs},
  booktitle    = {{ICLR}},
  publisher    = {OpenReview.net},
  year         = {2024},
  url          = {https://openreview.net/forum?id=dHng2O0Jjr},
  bibsource    = {dblp computer science bibliography, https://dblp.org}
}

@article{DBLP:journals/corr/abs-2305-15334,
  author       = {Shishir G. Patil and
                  Tianjun Zhang and
                  Xin Wang and
                  others},
  title        = {Gorilla: Large Language Model Connected with Massive APIs},
  journal      = {CoRR},
  volume       = {abs/2305.15334},
  year         = {2023},
  url          = {https://doi.org/10.48550/arXiv.2305.15334},
  doi          = {10.48550/ARXIV.2305.15334},
  eprinttype    = {arXiv},
  eprint       = {2305.15334},
  bibsource    = {dblp computer science bibliography, https://dblp.org}
}

@article{DBLP:journals/corr/abs-2312-10332,
  author       = {Raviteja Anantha and
                  Bortik Bandyopadhyay and
                  Anirudh Kashi and
                  others},
  title        = {ProTIP: Progressive Tool Retrieval Improves Planning},
  journal      = {CoRR},
  volume       = {abs/2312.10332},
  year         = {2023},
  url          = {https://doi.org/10.48550/arXiv.2312.10332},
  doi          = {10.48550/ARXIV.2312.10332},
  eprinttype    = {arXiv},
  eprint       = {2312.10332},
  bibsource    = {dblp computer science bibliography, https://dblp.org}
}

@article{DBLP:journals/corr/abs-2405-16089,
  author       = {Changle Qu and
                  Sunhao Dai and
                  Xiaochi Wei and
                  others},
  title        = {{COLT:} Towards Completeness-Oriented Tool Retrieval for Large Language
                  Models},
  journal      = {CoRR},
  volume       = {abs/2405.16089},
  year         = {2024},
  url          = {https://doi.org/10.48550/arXiv.2405.16089},
  doi          = {10.48550/ARXIV.2405.16089},
  eprinttype    = {arXiv},
  eprint       = {2405.16089},
  bibsource    = {dblp computer science bibliography, https://dblp.org}
}

@article{DBLP:journals/corr/abs-2305-13068,
  author       = {Shuofei Qiao and
                  Honghao Gui and
                  Huajun Chen and
                  others},
  title        = {Making Language Models Better Tool Learners with Execution Feedback},
  journal      = {CoRR},
  volume       = {abs/2305.13068},
  year         = {2023},
  url          = {https://doi.org/10.48550/arXiv.2305.13068},
  doi          = {10.48550/ARXIV.2305.13068},
  eprinttype    = {arXiv},
  eprint       = {2305.13068},
  bibsource    = {dblp computer science bibliography, https://dblp.org}
}

@inproceedings{DBLP:conf/nips/YangSLZGLS23,
  author       = {Rui Yang and
                  Lin Song and
                  Yanwei Li and
                  others},
  editor       = {Alice Oh and
                  others},
  title        = {GPT4Tools: Teaching Large Language Model to Use Tools via Self-instruction},
  booktitle    = {{NeurIPS}},
  year         = {2023},
  url          = {http://papers.nips.cc/paper\_files/paper/2023/hash/e393677793767624f2821cec8bdd02f1-Abstract-Conference.html},
  bibsource    = {dblp computer science bibliography, https://dblp.org}
}

@article{DBLP:journals/corr/abs-2205-12255,
  author       = {Aaron Parisi and
                  Yao Zhao and
                  Noah Fiedel},
  title        = {{TALM:} Tool Augmented Language Models},
  journal      = {CoRR},
  volume       = {abs/2205.12255},
  year         = {2022},
  url          = {https://doi.org/10.48550/arXiv.2205.12255},
  doi          = {10.48550/ARXIV.2205.12255},
  eprinttype    = {arXiv},
  eprint       = {2205.12255},
  bibsource    = {dblp computer science bibliography, https://dblp.org}
}

@inproceedings{DBLP:conf/nips/SchickDDRLHZCS23,
  author       = {Timo Schick and
                  Jane Dwivedi{-}Yu and
                  Roberto Dess{\`{\i}} and
                  others},
  editor       = {Alice Oh and
                  others},
  title        = {Toolformer: Language Models Can Teach Themselves to Use Tools},
  booktitle    = {{NeurIPS}},
  year         = {2023},
  url          = {http://papers.nips.cc/paper\_files/paper/2023/hash/d842425e4bf79ba039352da0f658a906-Abstract-Conference.html},
  bibsource    = {dblp computer science bibliography, https://dblp.org}
}

@misc{nexusraven,
      title={NexusRaven: Surpassing the state-of-the-art in open-source function calling LLMs}, 
      author={Nexusflow.ai},
      year={2023},
      url={http://nexusflow.ai/blog}
}

@inproceedings{DBLP:conf/nips/HaoLWH23,
  author       = {Shibo Hao and
                  Tianyang Liu and
                  Zhen Wang and
                  others},
  editor       = {Alice Oh and
                  others},
  title        = {ToolkenGPT: Augmenting Frozen Language Models with Massive Tools via
                  Tool Embeddings},
  booktitle    = {{NeurIPS}},
  year         = {2023},
  url          = {http://papers.nips.cc/paper\_files/paper/2023/hash/8fd1a81c882cd45f64958da6284f4a3f-Abstract-Conference.html},
  bibsource    = {dblp computer science bibliography, https://dblp.org}
}

@article{DBLP:journals/corr/abs-2407-00121,
  author       = {Ibrahim Abdelaziz and
                  Kinjal Basu and
                  Mayank Agarwal and
                  others},
  title        = {Granite-Function Calling Model: Introducing Function Calling Abilities
                  via Multi-task Learning of Granular Tasks},
  journal      = {CoRR},
  volume       = {abs/2407.00121},
  year         = {2024},
  url          = {https://doi.org/10.48550/arXiv.2407.00121},
  doi          = {10.48550/ARXIV.2407.00121},
  eprinttype    = {arXiv},
  eprint       = {2407.00121},
  bibsource    = {dblp computer science bibliography, https://dblp.org}
}

@inproceedings{DBLP:conf/nips/0001ST00Z23,
  author       = {Yongliang Shen and
                  Kaitao Song and
                  Xu Tan and
                  others},
  editor       = {Alice Oh and
                  others},
  title        = {HuggingGPT: Solving {AI} Tasks with ChatGPT and its Friends in Hugging
                  Face},
  booktitle    = {{NeurIPS}},
  year         = {2023},
  url          = {http://papers.nips.cc/paper\_files/paper/2023/hash/77c33e6a367922d003ff102ffb92b658-Abstract-Conference.html},
  bibsource    = {dblp computer science bibliography, https://dblp.org}
}

@article{DBLP:journals/corr/abs-2306-06624,
  author       = {Yifan Song and
                  Weimin Xiong and
                  Dawei Zhu and
                  others},
  title        = {RestGPT: Connecting Large Language Models with Real-World Applications
                  via RESTful APIs},
  journal      = {CoRR},
  volume       = {abs/2306.06624},
  year         = {2023},
  url          = {https://doi.org/10.48550/arXiv.2306.06624},
  doi          = {10.48550/ARXIV.2306.06624},
  eprinttype    = {arXiv},
  eprint       = {2306.06624},
  bibsource    = {dblp computer science bibliography, https://dblp.org}
}

@inproceedings{DBLP:conf/iclr/YaoZYDSN023,
  author       = {Shunyu Yao and
                  Jeffrey Zhao and
                  Dian Yu and
                  others},
  title        = {ReAct: Synergizing Reasoning and Acting in Language Models},
  booktitle    = {{ICLR}},
  publisher    = {OpenReview.net},
  year         = {2023},
  url          = {https://openreview.net/forum?id=WE\_vluYUL-X},
  bibsource    = {dblp computer science bibliography, https://dblp.org}
}

@article{DBLP:journals/corr/abs-2303-08774,
  author       = {OpenAI},
  title        = {{GPT-4} Technical Report},
  journal      = {CoRR},
  volume       = {abs/2303.08774},
  year         = {2023},
  url          = {https://doi.org/10.48550/arXiv.2303.08774},
  doi          = {10.48550/ARXIV.2303.08774},
  eprinttype    = {arXiv},
  eprint       = {2303.08774},
  bibsource    = {dblp computer science bibliography, https://dblp.org}
}

@inproceedings{DBLP:conf/aaai/GaoSZF0RC0R24,
  author       = {Shen Gao and
                  Zhengliang Shi and
                  Minghang Zhu and
                  others},
  editor       = {Michael J. Wooldridge and
                  others},
  title        = {Confucius: Iterative Tool Learning from Introspection Feedback by
                  Easy-to-Difficult Curriculum},
  booktitle    = {{AAAI}},
  pages        = {18030--18038},
  publisher    = {{AAAI} Press},
  year         = {2024},
  url          = {https://doi.org/10.1609/aaai.v38i16.29759},
  doi          = {10.1609/AAAI.V38I16.29759},
  bibsource    = {dblp computer science bibliography, https://dblp.org}
}

@article{DBLP:journals/jd/Jones04,
  author       = {Karen Sp{\"{a}}rck Jones},
  title        = {A statistical interpretation of term specificity and its application
                  in retrieval},
  journal      = {J. Documentation},
  volume       = {60},
  number       = {5},
  pages        = {493--502},
  year         = {2004},
  url          = {https://doi.org/10.1108/00220410410560573},
  doi          = {10.1108/00220410410560573},
  bibsource    = {dblp computer science bibliography, https://dblp.org}
}

@article{DBLP:journals/ftir/RobertsonZ09,
  author       = {Stephen E. Robertson and
                  Hugo Zaragoza},
  title        = {The Probabilistic Relevance Framework: {BM25} and Beyond},
  journal      = {Found. Trends Inf. Retr.},
  volume       = {3},
  number       = {4},
  pages        = {333--389},
  year         = {2009},
  url          = {https://doi.org/10.1561/1500000019},
  doi          = {10.1561/1500000019},
  bibsource    = {dblp computer science bibliography, https://dblp.org}
}

@article{DBLP:journals/corr/abs-2311-11315,
  author       = {Yilun Kong and
                  Jingqing Ruan and
                  Yihong Chen and
                  others},
  title        = {TPTU-v2: Boosting Task Planning and Tool Usage of Large Language Model-based
                  Agents in Real-world Systems},
  journal      = {CoRR},
  volume       = {abs/2311.11315},
  year         = {2023},
  url          = {https://doi.org/10.48550/arXiv.2311.11315},
  doi          = {10.48550/ARXIV.2311.11315},
  eprinttype    = {arXiv},
  eprint       = {2311.11315},
  bibsource    = {dblp computer science bibliography, https://dblp.org}
}

@inproceedings{DBLP:conf/iclr/YuanC000J24,
  author       = {Lifan Yuan and
                  Yangyi Chen and
                  Xingyao Wang and
                  others},
  title        = {{CRAFT:} Customizing LLMs by Creating and Retrieving from Specialized
                  Toolsets},
  booktitle    = {{ICLR}},
  publisher    = {OpenReview.net},
  year         = {2024},
  url          = {https://openreview.net/forum?id=G0vdDSt9XM},
  bibsource    = {dblp computer science bibliography, https://dblp.org}
}

@inproceedings{zheng-etal-2024-toolrerank,
    title = "{T}ool{R}erank: Adaptive and Hierarchy-Aware Reranking for Tool Retrieval",
    author = "Zheng, Yuanhang  and
      Li, Peng  and
      Liu, Wei  and
      others",
    editor = "Calzolari, Nicoletta  and
      others",
    booktitle = "LREC-COLING 2024",
    month = may,
    year = "2024",
    address = "Torino, Italia",
    publisher = "ELRA and ICCL",
    url = "https://aclanthology.org/2024.lrec-main.1413",
    pages = "16263--16273",
}

@misc{bfcl,
  title={Berkeley Function Calling Leaderboard},
  author={Fanjia Yan and Huanzhi Mao and Charlie Cheng-Jie Ji and others},
  year={2024},
  howpublished={\url{https://gorilla.cs.berkeley.edu/blogs/8_berkeley_function_calling_leaderboard.html}},
}

@inproceedings{DBLP:conf/iclr/HuangSLFWZ000G024,
  author       = {Yue Huang and
                  Jiawen Shi and
                  Yuan Li and
                  others},
  title        = {MetaTool Benchmark for Large Language Models: Deciding Whether to
                  Use Tools and Which to Use},
  booktitle    = {{ICLR}},
  publisher    = {OpenReview.net},
  year         = {2024},
  url          = {https://openreview.net/forum?id=R0c2qtalgG},
  bibsource    = {dblp computer science bibliography, https://dblp.org}
}

@inproceedings{li-etal-2023-api,
    title = "{API}-Bank: A Comprehensive Benchmark for Tool-Augmented {LLM}s",
    author = "Li, Minghao  and
      Zhao, Yingxiu  and
      Yu, Bowen  and
      others",
    editor = "Bouamor, Houda  and
      others",
    booktitle = "EMNLP",
    month = dec,
    year = "2023",
    address = "Singapore",
    publisher = {ACL},
    url = "https://aclanthology.org/2023.emnlp-main.187",
    doi = "10.18653/v1/2023.emnlp-main.187",
    pages = "3102--3116",
}

@article{DBLP:journals/tse/PengLGLWGL23,
  author       = {Yun Peng and
                  Shuqing Li and
                  Wenwei Gu and
                  others},
  title        = {Revisiting, Benchmarking and Exploring {API} Recommendation: How Far
                  Are We?},
  journal      = {{IEEE} Trans. Software Eng.},
  volume       = {49},
  number       = {4},
  pages        = {1876--1897},
  year         = {2023},
  url          = {https://doi.org/10.1109/TSE.2022.3197063},
  doi          = {10.1109/TSE.2022.3197063},
  bibsource    = {dblp computer science bibliography, https://dblp.org}
}

@article{DBLP:journals/corr/abs-2404-01744,
  author       = {Wei Chen and
                  Zhiyuan Li},
  title        = {Octopus v2: On-device language model for super agent},
  journal      = {CoRR},
  volume       = {abs/2404.01744},
  year         = {2024},
  url          = {https://doi.org/10.48550/arXiv.2404.01744},
  doi          = {10.48550/ARXIV.2404.01744},
  eprinttype    = {arXiv},
  eprint       = {2404.01744},
  bibsource    = {dblp computer science bibliography, https://dblp.org}
}

@article{zhu2004recall,
  author = {Zhu, Mu},
  year = {2004},
  title = {Recall, precision and average precision},
  journal = {Department of Statistics and Actuarial Science, University of Waterloo, Waterloo},
  volume = {2},
  number = {30},
  pages = {6}
}

@article{DBLP:journals/tois/JarvelinK02,
  author       = {Kalervo J{\"{a}}rvelin and
                  Jaana Kek{\"{a}}l{\"{a}}inen},
  title        = {Cumulated gain-based evaluation of {IR} techniques},
  journal      = {{ACM} Trans. Inf. Syst.},
  volume       = {20},
  number       = {4},
  pages        = {422--446},
  year         = {2002},
  url          = {http://doi.acm.org/10.1145/582415.582418},
  doi          = {10.1145/582415.582418},
  bibsource    = {dblp computer science bibliography, https://dblp.org}
}

@inproceedings{DBLP:conf/icml/GuuLTPC20,
  author       = {Kelvin Guu and
                  Kenton Lee and
                  Zora Tung and
                  others},
  title        = {Retrieval Augmented Language Model Pre-Training},
  booktitle    = {{ICML}},
  series       = {{PMLR}},
  volume       = {119},
  pages        = {3929--3938},
  publisher    = {{PMLR}},
  year         = {2020},
  url          = {http://proceedings.mlr.press/v119/guu20a.html},
  bibsource    = {dblp computer science bibliography, https://dblp.org}
}

@inproceedings{DBLP:conf/eacl/IzacardG21,
  author       = {Gautier Izacard and
                  Edouard Grave},
  editor       = {Paola Merlo and
                  others},
  title        = {Leveraging Passage Retrieval with Generative Models for Open Domain
                  Question Answering},
  booktitle    = {{EACL}},
  pages        = {874--880},
  publisher    = {ACL},
  year         = {2021},
  url          = {https://doi.org/10.18653/v1/2021.eacl-main.74},
  doi          = {10.18653/V1/2021.EACL-MAIN.74},
  bibsource    = {dblp computer science bibliography, https://dblp.org}
}

@inproceedings{DBLP:conf/emnlp/KarpukhinOMLWEC20,
  author       = {Vladimir Karpukhin and
                  Barlas Oguz and
                  Sewon Min and
                  others},
  editor       = {Bonnie Webber and
                  others},
  title        = {Dense Passage Retrieval for Open-Domain Question Answering},
  booktitle    = {{EMNLP}},
  pages        = {6769--6781},
  publisher    = {{ACL}},
  year         = {2020},
  url          = {https://doi.org/10.18653/v1/2020.emnlp-main.550},
  doi          = {10.18653/V1/2020.EMNLP-MAIN.550},
  bibsource    = {dblp computer science bibliography, https://dblp.org}
}

@article{DBLP:journals/corr/abs-2405-20680,
  author       = {Mingda Li and
                  Xinyu Li and
                  Yifan Chen and
                  others},
  title        = {Unraveling and Mitigating Retriever Inconsistencies in Retrieval-Augmented
                  Large Language Models},
  journal      = {CoRR},
  volume       = {abs/2405.20680},
  year         = {2024},
  url          = {https://doi.org/10.48550/arXiv.2405.20680},
  doi          = {10.48550/ARXIV.2405.20680},
  eprinttype    = {arXiv},
  eprint       = {2405.20680},
  bibsource    = {dblp computer science bibliography, https://dblp.org}
}

@inproceedings{DBLP:conf/icml/BorgeaudMHCRM0L22,
  author       = {Sebastian Borgeaud and
                  Arthur Mensch and
                  Jordan Hoffmann and
                  others},
  editor       = {Kamalika Chaudhuri and
                  others},
  title        = {Improving Language Models by Retrieving from Trillions of Tokens},
  booktitle    = {{ICML}},
  series       = {{PMLR}},
  volume       = {162},
  pages        = {2206--2240},
  publisher    = {{PMLR}},
  year         = {2022},
  url          = {https://proceedings.mlr.press/v162/borgeaud22a.html},
  bibsource    = {dblp computer science bibliography, https://dblp.org}
}

@article{DBLP:journals/tacl/SiriwardhanaWKWRN23,
  author       = {Shamane Siriwardhana and
                  Rivindu Weerasekera and
                  Tharindu Kaluarachchi and
                  others},
  title        = {Improving the Domain Adaptation of Retrieval Augmented Generation
                  {(RAG)} Models for Open Domain Question Answering},
  journal      = {{TACL}},
  volume       = {11},
  pages        = {1--17},
  year         = {2023},
  url          = {https://doi.org/10.1162/tacl\_a\_00530},
  doi          = {10.1162/TACL\_A\_00530},
  bibsource    = {dblp computer science bibliography, https://dblp.org}
}

@inproceedings{shi-etal-2024-replug,
    title = "{REPLUG}: Retrieval-Augmented Black-Box Language Models",
    author = "Shi, Weijia  and
      Min, Sewon  and
      Yasunaga, Michihiro  and
      others",
    editor = "Duh, Kevin  and
      others",
    booktitle = "NAACL",
    month = jun,
    year = "2024",
    address = "Mexico City, Mexico",
    publisher = "ACL",
    url = "https://aclanthology.org/2024.naacl-long.463",
    doi = "10.18653/v1/2024.naacl-long.463",
    pages = "8371--8384",
}

@article{DBLP:journals/corr/abs-2403-07691,
  author       = {Jiwoo Hong and
                  Noah Lee and
                  James Thorne},
  title        = {{ORPO:} Monolithic Preference Optimization without Reference Model},
  journal      = {CoRR},
  volume       = {abs/2403.07691},
  year         = {2024},
  url          = {https://doi.org/10.48550/arXiv.2403.07691},
  doi          = {10.48550/ARXIV.2403.07691},
  eprinttype    = {arXiv},
  eprint       = {2403.07691},
  bibsource    = {dblp computer science bibliography, https://dblp.org}
}

@article{DBLP:journals/symmetry/KayaB19,
  author       = {Mahmut Kaya and
                  Hasan Sakir Bilge},
  title        = {Deep Metric Learning: {A} Survey},
  journal      = {Symmetry},
  volume       = {11},
  number       = {9},
  pages        = {1066},
  year         = {2019},
  url          = {https://doi.org/10.3390/sym11091066},
  doi          = {10.3390/SYM11091066},
  bibsource    = {dblp computer science bibliography, https://dblp.org}
}

@inproceedings{DBLP:conf/nips/ChristianoLBMLA17,
  author       = {Paul F. Christiano and
                  Jan Leike and
                  Tom B. Brown and
                  others},
  editor       = {Isabelle Guyon and
                  others},
  title        = {Deep Reinforcement Learning from Human Preferences},
  booktitle    = {{NeurIPS}},
  pages        = {4299--4307},
  year         = {2017},
  url          = {https://proceedings.neurips.cc/paper/2017/hash/d5e2c0adad503c91f91df240d0cd4e49-Abstract.html},
  bibsource    = {dblp computer science bibliography, https://dblp.org}
}

@inproceedings{DBLP:conf/nips/RafailovSMMEF23,
  author       = {Rafael Rafailov and
                  Archit Sharma and
                  Eric Mitchell and
                  others},
  editor       = {Alice Oh and
                  others},
  title        = {Direct Preference Optimization: Your Language Model is Secretly a
                  Reward Model},
  booktitle    = {{NeurIPS}},
  year         = {2023},
  url          = {http://papers.nips.cc/paper\_files/paper/2023/hash/a85b405ed65c6477a4fe8302b5e06ce7-Abstract-Conference.html},
  bibsource    = {dblp computer science bibliography, https://dblp.org}
}

@article{DBLP:journals/tacl/LiuLHPBPL24,
  author       = {Nelson F. Liu and
                  Kevin Lin and
                  John Hewitt and
                  Ashwin Paranjape and
                  Michele Bevilacqua and
                  Fabio Petroni and
                  Percy Liang},
  title        = {Lost in the Middle: How Language Models Use Long Contexts},
  journal      = {TACL},
  volume       = {12},
  pages        = {157--173},
  year         = {2024},
  url          = {https://doi.org/10.1162/tacl\_a\_00638},
  doi          = {10.1162/TACL\_A\_00638},
  bibsource    = {dblp computer science bibliography, https://dblp.org}
}

@inproceedings{DBLP:conf/sigir/XiaoLZMLN24,
  author       = {Shitao Xiao and
                  Zheng Liu and
                  Peitian Zhang and
                  others},
  editor       = {Grace Hui Yang and
                  others},
  title        = {C-Pack: Packed Resources For General Chinese Embeddings},
  booktitle    = {{SIGIR}},
  pages        = {641--649},
  publisher    = {{ACM}},
  year         = {2024},
  url          = {https://doi.org/10.1145/3626772.3657878},
  doi          = {10.1145/3626772.3657878},
  bibsource    = {dblp computer science bibliography, https://dblp.org}
}

@article{DBLP:journals/corr/abs-1907-11692,
  author       = {Yinhan Liu and
                  Myle Ott and
                  Naman Goyal and
                  others},
  title        = {RoBERTa: {A} Robustly Optimized {BERT} Pretraining Approach},
  journal      = {CoRR},
  volume       = {abs/1907.11692},
  year         = {2019},
  url          = {http://arxiv.org/abs/1907.11692},
  eprinttype    = {arXiv},
  eprint       = {1907.11692},
  bibsource    = {dblp computer science bibliography, https://dblp.org}
}

@misc{dubey2024llama3herdmodels,
      title={The Llama 3 Herd of Models}, 
      author={Abhimanyu Dubey and Abhinav Jauhri and Abhinav Pandey et al.},
      year={2024},
      eprint={2407.21783},
      archivePrefix={arXiv},
      primaryClass={cs.AI},
      url={https://arxiv.org/abs/2407.21783}, 
}

@article{DBLP:journals/corr/abs-2504-13181,
  author       = {Daniel Bolya and
                  Po{-}Yao Huang and
                  Peize Sun et al.},
  title        = {Perception Encoder: The best visual embeddings are not at the output of the network},
  journal      = {CoRR},
  volume       = {abs/2504.13181},
  year         = {2025},
  url          = {https://arxiv.org/abs/2504.13181},
  doi          = {10.48550/arXiv.2504.13181},
  eprinttype   = {arXiv},
  eprint       = {2504.13181},
  bibsource    = {dblp computer science bibliography, https://dblp.org}
}

@article{DBLP:journals/corr/NguyenRSGTMD16,
  author    = {Tri Nguyen and
               Mir Rosenberg and
               Xia Song and
               Jianfeng Gao and
               Saurabh Tiwary and
               Rangan Majumder and
               Li Deng},
  title     = {{MS} {MARCO:} {A} Human Generated MAchine Reading COmprehension Dataset},
  journal   = {CoRR},
  volume    = {abs/1611.09268},
  year      = {2016},
  url       = {http://arxiv.org/abs/1611.09268},
  archivePrefix = {arXiv},
  eprint    = {1611.09268},
  bibsource = {dblp computer science bibliography, https://dblp.org}
}

@inproceedings{DBLP:conf/emnlp/Yang0ZBCSM18,
  author       = {Zhilin Yang and
                  Peng Qi and
                  Saizheng Zhang and
                  Yoshua Bengio and
                  William W. Cohen and
                  Ruslan Salakhutdinov and
                  Christopher D. Manning},
  editor       = {Ellen Riloff and
                  David Chiang and
                  Julia Hockenmaier and
                  Jun'ichi Tsujii},
  title        = {HotpotQA: {A} Dataset for Diverse, Explainable Multi-hop Question
                  Answering},
  booktitle    = {EMNLP, 2018},
  pages        = {2369--2380},
  publisher    = {Association for Computational Linguistics},
  year         = {2018},
  url          = {https://doi.org/10.18653/v1/d18-1259},
  doi          = {10.18653/V1/D18-1259},
  bibsource    = {dblp computer science bibliography, https://dblp.org}
}

@inproceedings{DBLP:conf/emnlp/JinDLCL19,
  author       = {Qiao Jin and
                  Bhuwan Dhingra and
                  Zhengping Liu and
                  William W. Cohen and
                  Xinghua Lu},
  editor       = {Kentaro Inui and
                  Jing Jiang and
                  Vincent Ng and
                  Xiaojun Wan},
  title        = {PubMedQA: {A} Dataset for Biomedical Research Question Answering},
  booktitle    = {{EMNLP-IJCNLP} 2019},
  pages        = {2567--2577},
  publisher    = {Association for Computational Linguistics},
  year         = {2019},
  url          = {https://doi.org/10.18653/v1/D19-1259},
  doi          = {10.18653/V1/D19-1259},
  bibsource    = {dblp computer science bibliography, https://dblp.org}
}

@inproceedings{DBLP:conf/ic3k/DomeniconiMPS14a,
  author       = {Giacomo Domeniconi and
                  Gianluca Moro and
                  Roberto Pasolini and
                  Claudio Sartori},
  editor       = {Ana L. N. Fred and
                  Jan L. G. Dietz and
                  David Aveiro and
                  Kecheng Liu and
                  Joaquim Filipe},
  title        = {Iterative Refining of Category Profiles for Nearest Centroid Cross-Domain
                  Text Classification},
  booktitle    = {Knowledge Discovery, Knowledge Engineering and Knowledge Management, 2014},
  series       = {Communications in Computer and Information Science},
  volume       = {553},
  pages        = {50--67},
  publisher    = {Springer},
  year         = {2014},
  url          = {https://doi.org/10.1007/978-3-319-25840-9\_4},
  doi          = {10.1007/978-3-319-25840-9\_4},
  bibsource    = {dblp computer science bibliography, https://dblp.org}
}

@article{DBLP:journals/bmcbi/LenaDMM15,
  author       = {Pietro di Lena and
                  Giacomo Domeniconi and
                  Luciano Margara and
                  Gianluca Moro},
  title        = {{GOTA:} {GO} term annotation of biomedical literature},
  journal      = {{BMC} Bioinform.},
  volume       = {16},
  pages        = {346:1--346:13},
  year         = {2015},
  url          = {https://doi.org/10.1186/s12859-015-0777-8},
  doi          = {10.1186/S12859-015-0777-8},
  bibsource    = {dblp computer science bibliography, https://dblp.org}
}

@inproceedings{DBLP:conf/ic3k/DomeniconiMPS14,
  author       = {Giacomo Domeniconi and
                  Gianluca Moro and
                  Roberto Pasolini and
                  Claudio Sartori},
  editor       = {Ana L. N. Fred and
                  Joaquim Filipe},
  title        = {Cross-domain Text Classification through Iterative Refining of Target
                  Categories Representations},
  booktitle    = {{KDIR} 2014},
  pages        = {31--42},
  publisher    = {SciTePress},
  year         = {2014},
  url          = {https://doi.org/10.5220/0005069400310042},
  doi          = {10.5220/0005069400310042},
  bibsource    = {dblp computer science bibliography, https://dblp.org}
}

@inproceedings{DBLP:conf/ic3k/DomeniconiMMP14,
  author       = {Giacomo Domeniconi and
                  Marco Masseroli and
                  Gianluca Moro and
                  Pietro Pinoli},
  editor       = {Ana L. N. Fred and
                  Joaquim Filipe},
  title        = {Discovering New Gene Functionalities from Random Perturbations of
                  Known Gene Ontological Annotations},
  booktitle    = {{KDIR} 2014},
  pages        = {107--116},
  publisher    = {SciTePress},
  year         = {2014},
  url          = {https://doi.org/10.5220/0005087801070116},
  doi          = {10.5220/0005087801070116},
  bibsource    = {dblp computer science bibliography, https://dblp.org}
}

@inproceedings{DBLP:conf/ic3k/DomeniconiMPP15,
  author       = {Giacomo Domeniconi and
                  Gianluca Moro and
                  Andrea Pagliarani and
                  Roberto Pasolini},
  editor       = {Ana L. N. Fred and
                  Jan L. G. Dietz and
                  David Aveiro and
                  Kecheng Liu and
                  Joaquim Filipe},
  title        = {Markov Chain based Method for In-Domain and Cross-Domain Sentiment
                  Classification},
  booktitle    = {{KDIR} 2015},
  pages        = {127--137},
  publisher    = {SciTePress},
  year         = {2015},
  url          = {https://doi.org/10.5220/0005636001270137},
  doi          = {10.5220/0005636001270137},
  bibsource    = {dblp computer science bibliography, https://dblp.org}
}

@inproceedings{DBLP:conf/adc/LodiMS10,
  author       = {Stefano Lodi and
                  Gianluca Moro and
                  Claudio Sartori},
  editor       = {Heng Tao Shen and
                  Athman Bouguettaya},
  title        = {Distributed data clustering in multi-dimensional peer-to-peer networks},
  booktitle    = {Database Technologies 2010, Twenty-First Australasian Database Conference
                  {(ADC} 2010)},
  series       = {{CRPIT}},
  volume       = {104},
  pages        = {171--178},
  publisher    = {Australian Computer Society},
  year         = {2010},
  url          = {http://portal.acm.org/citation.cfm?id=1862264\&CFID=17470975\&CFTOKEN=71845406},
  bibsource    = {dblp computer science bibliography, https://dblp.org}
}

@inproceedings{DBLP:conf/acl/WangMWWJCLY25,
  author       = {Wenxuan Wang and
                  Zizhan Ma and
                  Zheng Wang and
                  Chenghan Wu and
                  Jiaming Ji and
                  Wenting Chen and
                  Xiang Li and
                  Yixuan Yuan},
  editor       = {Wanxiang Che and
                  Joyce Nabende and
                  Ekaterina Shutova and
                  Mohammad Taher Pilehvar},
  title        = {A Survey of LLM-based Agents in Medicine: How far are we from Baymax?},
  booktitle    = {{ACL} 2025},
  pages        = {10345--10359},
  publisher    = {Association for Computational Linguistics},
  year         = {2025},
  url          = {https://aclanthology.org/2025.findings-acl.539/},
  bibsource    = {dblp computer science bibliography, https://dblp.org}
}

@article{DBLP:journals/cmpb/DomeniconiMMP16,
  author       = {Giacomo Domeniconi and
                  Marco Masseroli and
                  Gianluca Moro and
                  Pietro Pinoli},
  title        = {Cross-organism learning method to discover new gene functionalities},
  journal      = {Comput. Methods Programs Biomed.},
  volume       = {126},
  pages        = {20--34},
  year         = {2016},
  url          = {https://doi.org/10.1016/j.cmpb.2015.12.002},
  doi          = {10.1016/J.CMPB.2015.12.002},
  bibsource    = {dblp computer science bibliography, https://dblp.org}
}

\clearpage
\newpage
\appendix

\section{In-Depth View on the Relevance of \texttt{PORTS}}
\label{sec:ports_dem}

Given an input query $q$ and a set of tools $t$ with description $d_{t_i}$, \ports combines \textsc{RePlug}'s goal-directed retrieval with preference alignment, adapted explicitly for tool selection. This approach enhances both relevance and downstream performance, effectively addressing gaps in existing methods. The foundation lies in \textsc{RePlug}'s gradient structure, derived from minimizing the KL divergence between the retriever distribution $P_\mathcal{R}(t|q,d_t)$ and the LM’s utility signal $Q_\mathcal{G}(t|q,d_t,y)$. 

\paragraph{RePlug Gradient Derivation} Starting from the KL divergence objective:
\begin{align}
    \mathcal{L}_\text{replug} &= \text{KL}(P_\mathcal{R}^\theta \parallel Q_\mathcal{G}) \nonumber \\
    &= \sum_t P_\mathcal{R}^\theta(t|q, d_t) \log \frac{P_\mathcal{R}^\theta(t|q,d_t)}{Q_\mathcal{G}(t|q,d_t,y)},
\end{align}
we decompose it into entropy and cross-entropy terms:
\begin{align}
    \text{KL} &= \underbrace{\sum_t P_\mathcal{R}^\theta \log P_\mathcal{R}^\theta}_{-H(P_\mathcal{R}^\theta)} - \underbrace{\sum_t P_\mathcal{R}^\theta \log Q_\mathcal{G}}_{-H(P_\mathcal{R}^\theta, Q_\mathcal{G})}.
\end{align}
To derive gradients with respect to retriever scores $sim(q,d_t)$, we differentiate both terms. First, for the entropy term $H(P_\mathcal{R}^\theta)$:
\begin{align}
    \begin{split}
    \frac{\partial H(P_\mathcal{R}^\theta)}{\partial sim(q,d_t)}&= \frac{1}{\gamma}P_\mathcal{R}^\theta(t|q,d_t)\; \cdot \\
    &\cdot\; \Bigl( 1 + \log P_\mathcal{R}^\theta(t|q,d_t) - H(P_\mathcal{R}^\theta) \Bigr),
    \end{split}
\end{align}
where $\gamma$ is the retriever’s temperature parameter. For the cross-entropy term $H(P_\mathcal{R}^\theta, Q_\mathcal{G})$:
\begin{align}
    \begin{split}
    &\frac{\partial H(P_\mathcal{R}^\theta, Q_\mathcal{G})}{\partial sim(q,d_t)} = \frac{1}{\gamma}P_\mathcal{R}^\theta(t|q,d_t)\cdot \\& \hspace{1cm} \cdot \left(\log Q_\mathcal{G}(t|q,d_t,y) - \mathbb{E}_{P_\mathcal{R}^\theta}[\log Q_\mathcal{G}]\right).
    \end{split}
\end{align}
Subtracting these gradients yields:
\begin{align}
    \begin{split}
        &\frac{\partial \mathcal{L}_\text{replug}}{\partial sim(q,d_t)} = \frac{P_\mathcal{R}^\theta(t|q,d_t)}{\gamma} \cdot \\ &\;\;\;\;\;\; \cdot \left( \log\frac{P_\mathcal{R}^\theta(t|q,d_t)}{Q_\mathcal{G}(t|q,d_t,y)}
        \quad - \text{KL}(P_\mathcal{R}^\theta \parallel Q_\mathcal{G}) \right).
    \end{split}
\end{align}
Here, $\log\frac{P_\mathcal{R}^\theta}{Q_\mathcal{G}}$ provides per-document alignment signals, while the $\text{KL}$ term stabilizes training by serving as a baseline for global distribution shifts.

\paragraph{PORTS Gradient Extension} Our analysis of the full \ports loss gradient highlights the limitations of \textsc{RePlug} alone:
\begin{align}
    \begin{split}
    &\nabla_\theta \mathcal{L}_\text{PORTS} \propto     \nabla_\theta \mathcal{L}_\text{replug} + \nabla_\theta \mathcal{L}_\text{po} \\
   &\nabla \mathcal{L}_\text{replug} =  P_\mathcal{R}^\theta(t|q,d_t) \Biggl( \log\frac{P_R(t|q,d_t)}{Q_\mathcal{G}(t|q,d_t,y)} +\\&\hspace{4cm} - \text{KL}(P_\mathcal{R}^\theta \parallel Q_\mathcal{G}) \Biggr) \\
    & \nabla_\theta \mathcal{L}_\text{po} = \left(1 + \frac{\pi_\theta(t|q,d_{t^+})}{\pi_\theta(t|q,d_{t^-})}\right)^{-1}.
    \end{split}
\end{align}
While $P_\mathcal{R}^\theta/Q$ compares the retriever confidence against the LLM's docstring utility assessment, the KL term acts as a global stabilizer that prevents over-adjustments to individual docstrings. Without our preference optimization component, the system simply minimizes differences between retrieval probabilities $P_\mathcal{R}^\theta$ and downstream log-likelihoods $\mathcal{Q}_\mathcal{G}$, overlooking LLM handling of imperfect descriptions. Specifically, pure \textsc{RePlug} gradients lack awareness of \textit{relative tool utility} – a critical shortfall when multiple tools have overlapping or ambiguous descriptions.

The preference optimization contrastive term addresses this by enforcing refinement of tool selection through pairwise comparisons. Unlike the KL penalty, which operates globally, the term
\[
\left( 1 + \frac{\pi_\theta(t|q,d_{t^+})}{\pi_\theta(t|q,d_{t^-})} \right)^{-1}
\]  
explicitly rewards the retriever for distinguishing between \textit{semantically similar but functionally distinct} tools ($d_{t^+}$ and $d_{t^-}$). This creates implicit links between tools based on their downstream task performance rather than surface-level similarity. For example, when two API descriptions share terminology but differ in required parameters, the contrastive component amplifies gradients for the tool whose documentation better resolves this ambiguity in practice.

In so doing, \ports benefits from negative examples not only through the downstream task signal (which can be noisy due to LLM approximation errors) but also through structured semantic comparisons. The retriever learns to associate subtle linguistic cues in tool descriptions with their \textit{functional outcomes}, even when $Q_\mathcal{G}$ provides imperfect supervision. This dual mechanism proves critical in real-world scenarios where tool documentation quality varies widely – the contrastive term compensates for sparse or ambiguous $\mathcal{Q}_\mathcal{G}$ signals by reinforcing discriminative features across the toolset.
\ports' triplet formulation also enables flexible, state-independent negative sampling, unlike \textsc{RePlug}'s iterative sampling, which may introduce bias.
A qualitative example of \ports' retrieval and disambiguation capabilities is shown in Appendix~\ref{app:qual_examples}.

\section{Prompt Templates}
\label{app:prompt_templates}

The prompt template for Large Language Models (LLMs) is typically divided into two distinct components: a system message and a user instruction.
The system message serves to establish the model's role and behavioral parameters.
In contrast, the user instruction delineates the specific task or query to be addressed.
In cases where a model's chat template does not inherently accommodate a discrete system message, this information is instead prepended to the user instruction.
This ensures that the model is primed with all necessary contextual and behavioral guidelines before processing the task at hand. Within the \ports framework, a frozen LLM is prompted with the input query and the docstring of a retrieved tool to gauge the probability of predicting the target call. During training, we mask the input up to the ``Answer'' tag and compute the next-token probability of the gold answer using the system and instruction sections as input with a causal attention approach. Our prompt templates are reported in Listing~\ref{lst:prompt_template_1} and Listing~\ref{lst:prompt_template_2}.
To better recall the prior knowledge of the model, we describe tools as API functions.
We use each model's specific chat template, omitting special tokens in the listings for clarity.
We use a different prompt when working with \llamaBig to better adhere to the chat template on which it was trained, as suggested in the HuggingFace model card.\footnote{\href{https://huggingface.co/Groq/Llama-3-Groq-8B-Tool-Use}{huggingface.co/Groq/Llama-3-Groq-8B-Tool-Use}}

\section{Software and Datasets: Details, Intended Use, and Impact}
\label{app:software_datasets}

Despite the popularity of \textsc{RePlug}, no implementation code was publicly available. \textbf{We dedicated significant effort to reconstructing the method \underline{from scratch}}, carefully clarifying its methodological choices. To benefit the broader research community, \textbf{we release our complete implementation--including \ports--as fully open-source under a permissive MIT license}. This ensures full reproducibility and establishes the first open-source solution for goal-directed encoder fine-tuning.

Queries, tools, and docstrings can vary greatly depending on the dataset. Table~\ref{tab:sample_toolbench}, Table~\ref{tab:sample_octopus}, Table~\ref{tab:sample_apibank}, Table~\ref{tab:sample_apibench}, Table~\ref{tab:sample_bfcl}, and Table~\ref{tab:sample_toole} show representative input-output examples sampled from the test set of each dataset.
Each dataset has been pre-processed to ensure compatibility with our tool-selection task by decoupling instances that require the use of multiple tools. Tool descriptions have been enhanced with detailed information about input and output parameter types, formats, and purposes, following Python-style docstrings to better define the scope of each tool and facilitate the retrieval process.
Given the conversational and multi-tool nature of the API-Bank dataset, we have distinguished between inputs for the retriever and generative models. The retriever's input excludes previous tool calls to avoid biases and inconsistencies in the similarity-based search, while generative models receive the full conversation history. This approach enhances tool selection accuracy while allowing generative models to leverage complete contextual information.

\paragraph{Licenses of Used Datasets}
All datasets used in our experiments are publicly available and released under permissive open-source licenses. Specifically, \textsc{ToolBench}, \textsc{ApiBench}, \textsc{Octopus}, and \textsc{BFCLv2} are distributed under the Apache 2.0 License, while \textsc{ToolE} and \textsc{ApiBank} are released under the MIT License. These licenses allow for both academic and commercial use, ensuring full compliance with open-source standards and enabling reproducibility of our experiments.

\section{Hyperparameter Space}
\label{app:hyperparameters}

Table~\ref{tab:hyperparams} presents a comprehensive overview of the hyperparameters explored in our study. This extensive search space was designed to optimize the model's performance across various dimensions, from basic configuration settings to more nuanced training parameters.

\textcolor{purple}{}\textcolor{purple}{}\textcolor{purple}{}\textcolor{purple}{}\textcolor{purple}{}\textcolor{purple}{}

\begin{figure}[!ht]
    \centering
    \footnotesize
    \begin{tcolorbox}[colframe=black, colback=gray!5, title=Prompt Template for Instruct LLMs]
\textbf{\textcolor{teal}{\#\# System Message}}

\medskip
\texttt{You are a function caller. You are given a user query and the description (docstring) of a single API function.}

\medskip
\texttt{You must generate a function call using the exact name and parameters of the provided API function. You are not allowed to use any other function besides the one given.}

\medskip
\texttt{Return only the function call, using single quotes for strings and separating parameters with commas.}

\medskip
\texttt{You are not permitted to deviate from the given API function in any way. You must use the exact function name and parameter types specified, even if you think another function would be more appropriate for the user's request.}

\medskip
\texttt{Example:\\
==============\\
Docstring:\\
def add\_reminder(text: str, date: str, time: str):\\
\hspace*{2em}"""\\
\hspace*{2em}Description:\\
\hspace*{2em}Set a reminder for a task on a specified date and time.\\
\hspace*{2em}\\
\hspace*{2em}Arguments:\\
\hspace*{2em}---------\\
\hspace*{2em}- text : str\\
\hspace*{4em}The description or name of the task for which the reminder is\\
\hspace*{4em}being set.\\
\hspace*{2em}- date : str\\
\hspace*{4em}The date on which the reminder should be scheduled.\\
\hspace*{2em}- time : str\\
\hspace*{4em}The time at which the reminder should be scheduled.\\
\hspace*{2em}"""}

\medskip
\texttt{Query:\\
"Add a reminder to buy groceries tomorrow at 2 PM"\\
Answer:\\
add\_reminder(\\
\hspace*{1em}text='Buy groceries',\\
\hspace*{1em}date='tomorrow',\\ 
\hspace*{1em}time='2 PM'\\
)\\
==============}

\medskip
\textbf{\textcolor{teal}{\#\# Instruction Message}}

\texttt{Docstring: \textcolor{purple}{\textbf{\$\{Docstring\}}}\\
Query: \textcolor{purple}{\textbf{\$\{Query\}}}\\
Answer: \textcolor{purple}{\textbf{\$\{Answer\}}}}
    \end{tcolorbox}
    \caption{Prompt template for call generation with the retrieved tool for \gemma, \qwen, \llamaSmall, \llamaBig, and \codestral.}
    \label{lst:prompt_template_1}
\end{figure}

\begin{figure}[!ht]
    \centering
    \footnotesize
    \vspace{-1.45cm}
    \begin{tcolorbox}[colframe=black, colback=gray!5, title=Prompt Template for Tool LLMs]
\textbf{\textcolor{teal}{\#\# System Message}}

\medskip
\texttt{You are a function caller. You are given a user query and the definition of a single tool function within <tools></tools> XML tags.}

\medskip
\texttt{You must generate a function call using the exact name and parameters of the provided tool. You are not allowed to use any other function besides the one given.}

\medskip
\texttt{Return only the function call, using single quotes for strings and separating parameters with commas.}

\medskip
\texttt{You are not permitted to deviate from the given API function in any way. You must use the exact function name and parameter types specified, even if you think another function would be more appropriate for the user's request.}

\medskip
\texttt{Example:\\
==============\\
Docstring:\\
def add\_reminder(text: str, date: str, time: str):\\
\hspace*{2em}"""\\
\hspace*{2em}Description:\\
\hspace*{2em}Set a reminder for a task on a specified date and time.\\
\hspace*{2em}\\
\hspace*{2em}Arguments:\\
\hspace*{2em}---------\\
\hspace*{2em}- text : str\\
\hspace*{4em}The description or name of the task for which the reminder is\\
\hspace*{4em}being set.\\
\hspace*{2em}- date : str\\
\hspace*{4em}The date on which the reminder should be scheduled.\\
\hspace*{2em}- time : str\\
\hspace*{4em}The time at which the reminder should be scheduled.\\
\hspace*{2em}"""}

\medskip
\texttt{Query:\\
"Add a reminder to buy groceries tomorrow at 2 PM"\\
Answer:\\
add\_reminder(\\
\hspace*{1em}text='Buy groceries',\\
\hspace*{1em}date='tomorrow',\\ 
\hspace*{1em}time='2 PM'\\
)\\
==============}

\medskip
\textbf{\textcolor{teal}{\#\# Instruction Message}}

\texttt{Docstring: \textcolor{purple}{\textbf{\$\{Docstring\}}}\\
Query: \textcolor{purple}{\textbf{\$\{Query\}}}\\
Answer: \textcolor{purple}{\textbf{\$\{Answer\}}}}
    \end{tcolorbox}
    \caption{Prompt template for call generation with the retrieved tool for \groqLlama.}
    \label{lst:prompt_template_2}
\end{figure}

\newpage

\begin{table*}[!t]
    \centering
    \fontsize{9}{9}\selectfont
    \begin{tabularx}{\linewidth}{>{\hsize=1.5cm}XX}
        \toprule
        \arrayrulecolor{black}
        \textbf{Field} & \textbf{Text} \\
        \midrule
        \arrayrulecolor[rgb]{0.6,0.6,0.6}
        \rowcolor{yellow!10} Query & \begin{allintypewriter}
        Please provide me with the user information for the user with the username 'michaelbrown'. Also, fetch the order details for order ID 31415 and get the inventory status of the store.
        \end{allintypewriter} \\
        \midrule
        \rowcolor{gray!5} Gold Tool's Docstring & \begin{allintypewriter}
        def petstore\_blitz.getUserByName():\newline
    	\hspace*{10pt}"""\newline
    	\hspace*{10pt}Description:\newline
        \hspace*{10pt}Fetch user by name.\newline\newline
        
    	\hspace*{10pt}Arguments:\newline
    	\hspace*{10pt}---------\newline
    	\hspace*{10pt}- username : STRING (required)\newline
    	  \hspace*{10pt}Description: The name that needs to be fetched. Use user1 for testing.\newline
    	\hspace*{10pt}"""
        \end{allintypewriter} \\
        \midrule
        \rowcolor{yellow!10} Answer & \begin{allintypewriter}
        petstore\_blitz.getUserByName(username="michaelbrown")
        \end{allintypewriter} \\
        \arrayrulecolor{black}
        \bottomrule
    \end{tabularx}
    \caption{ToolBench dataset sample \ding{182}.}
    \label{tab:sample_toolbench}
\end{table*}

\begin{table*}[!t]
    \centering
    \fontsize{9}{9}\selectfont
    \begin{tabularx}{\linewidth}{>{\hsize=1.5cm}XX}
        \toprule
        \arrayrulecolor{black}
        \textbf{Field} & \textbf{Text} \\
        \midrule
        \arrayrulecolor[rgb]{0.6,0.6,0.6}
        \rowcolor{yellow!10} Query & \begin{allintypewriter}
        What's the weather like in New York City for the next three days?
        \end{allintypewriter} \\
        \midrule
        \rowcolor{gray!5} Gold Tool's Docstring & \begin{allintypewriter}
        def get\_weather\_forecast():\newline
        \hspace*{10pt}"""\newline
        \hspace*{10pt}Provides a weather forecast for a specified location over a given number \newline\hspace*{10pt}of days. Each day's forecast includes a brief description of the expected \newline\hspace*{10pt}weather conditions.\newline\newline
    
        \hspace*{10pt}Arguments:\newline
        \hspace*{10pt}----------\newline
        \hspace*{10pt}- location (str): The location for which the weather forecast is desired. \newline
        \hspace*{10pt}Can be a city name, ZIP code, or other location identifiers.\newline\newline
    
        \hspace*{10pt}Output:\newline
    	\hspace*{10pt}---------\newline
        \hspace*{10pt}- list[str]: A list of strings, each representing the weather \newline\hspace*{10pt}forecast for one day. \newline
        \hspace*{10pt}Each string includes the date and a brief description of the weather \newline\hspace*{10pt}conditions. Formatted in 'YYYY-MM-DD: Description' format.\newline
        \hspace*{10pt}"""
        \end{allintypewriter} \\
        \midrule
        \rowcolor{yellow!10} Answer & \begin{allintypewriter}
        get\_weather\_forecast('New York City')
        \end{allintypewriter} \\
        \arrayrulecolor{black}
        \bottomrule
    \end{tabularx}
    \caption{Octopus dataset sample \ding{187}/\ding{189}.}
    \label{tab:sample_octopus}
\end{table*}

\begin{table*}[!t]
    \centering
    \fontsize{7}{7}\selectfont
    \begin{tabularx}{\linewidth}{>{\hsize=1.2cm}XX}
        \toprule
        \arrayrulecolor{black}
        \textbf{Field} & \textbf{Text} \\
        \midrule
        \arrayrulecolor[rgb]{0.6,0.6,0.6}
        \rowcolor{yellow!10} Query & \begin{allintypewriter}
        \textbf{User:} Can you give me a list of all the stress reduction technique categories you offer?\newline 
        \textbf{API-Request:} [getCategories()]->["Breathing Techniques", "Progressive Muscle Relaxation", "Meditation", "Yoga", "Biofeedback"]\newline 
        \textbf{AI:} Our available stress reduction technique categories are Breathing Techniques, Progressive Muscle Relaxation, Meditation, Yoga, and Biofeedback.\newline 
        \textbf{User:} Can you suggest some stress-reducing techniques that are easy to learn?\newline
        \textbf{API-Request:} [getTechniques(category=\'None\', difficulty=\'easy\')]->[{"name": "Breathing and Grounding Exercise", "description": "This simple technique involves taking a deep breath, exhaling, and grounding yourself by imagining roots growing out of your feet into the ground. This exercise is an excellent way to fall asleep quickly or to calm down when feeling anxious or stressed.", "category": "Breathing Techniques", "difficulty": "easy"}, {"name": "Progressive Muscle Relaxation", "description": "A technique in which you slowly tense and then relax each muscle group of the body. It is particularly useful for relaxation and reducing anxiety.", "category": "Progressive Muscle Relaxation", "difficulty": "easy"}, {"name": "Body Scan Meditation", "description": "This technique requires focusing your attention on different parts of your body and deliberately relaxing them, which can help reduce stress and anxiety.", "category": "Meditation", "difficulty": "easy"}]\newline
        \textbf{AI:} I recommend the following techniques for easy stress reduction: Breathing and Grounding Exercise, Progressive Muscle Relaxation, and Body Scan Meditation.\newline
        \textbf{User:} Can you add the \'Visualization\' technique to your stress-reducing technique list?\newline
        \textbf{Generate API Request:}    
        \end{allintypewriter} \\
        \midrule
        \rowcolor{gray!5} Gold Tool's Docstring & \begin{allintypewriter}
        \textbf{User:} Can you give me a list of all the stress reduction technique categories you offer?\newline
        \textbf{AI:} Our available stress reduction technique categories are Breathing Techniques, Progressive Muscle Relaxation, Meditation, Yoga, and Biofeedback.\newline
        \textbf{User:} Can you suggest some stress-reducing techniques that are easy to learn?\newline
        \textbf{AI:} I recommend the following techniques for easy stress reduction: Breathing and Grounding Exercise, Progressive Muscle Relaxation, and Body Scan Meditation.\newline
        \textbf{User:} Can you add the 'Visualization' technique to your stress-reducing technique list?\newline
        \textbf{Generate API Request:}
        \end{allintypewriter} \\
        \midrule
        \rowcolor{yellow!10} Answer & \begin{allintypewriter}
        def addTechnique():\newline
    	\hspace*{10pt}"""\newline
    	\hspace*{10pt}Description:\newline
    	\hspace*{10pt}Add a new stress reduction technique\newline\newline
    	
    	\hspace*{10pt}Arguments:\newline
    	\hspace*{10pt}---------\newline
    	\hspace*{10pt}- name : string (optional)\newline
    	  \hspace*{10pt}Description: The name of the new stress reduction technique\newline
    	  \hspace*{10pt}Format: Not specified\newline
    	\hspace*{10pt}- description : string (optional)\newline
    	  \hspace*{10pt}Description: A description of the new stress reduction technique\newline
    	  \hspace*{10pt}Format: Not specified\newline
    	\hspace*{10pt}- category : string (optional)\newline
    	  \hspace*{10pt}Description: The category of the new stress reduction technique\newline
    	  \hspace*{10pt}Format: Not specified\newline
    	\hspace*{10pt}- difficulty : string (optional)\newline
    	  \hspace*{10pt}Description: The difficulty level of the new stress reduction technique\newline
    	  \hspace*{10pt}Format: Not specified\newline\newline
    	
    	\hspace*{10pt}Output:\newline
    	\hspace*{10pt}---------\newline
    	\hspace*{10pt}- data : object (optional)\newline
    	  \hspace*{10pt}Description: The newly added stress reduction technique\newline
    	  \hspace*{10pt}Format: Not specified\newline
    	  \hspace*{10pt}Properties:\newline
    	\hspace*{10pt}    - name : string (optional)\newline
    	  \hspace*{10pt}    Description: The name of the newly added stress reduction technique\newline
    	  \hspace*{10pt}    Format: Not specified\newline
    	\hspace*{10pt}    - description : string (optional)\newline
    	  \hspace*{10pt}    Description: A description of the newly added stress reduction \hspace*{10pt}technique\newline
    	  \hspace*{10pt}    Format: Not specified\newline
    	\hspace*{10pt}    - category : string (optional)\newline
    	  \hspace*{10pt}    Description: The category of the newly added stress reduction \hspace*{10pt}technique\newline
    	  \hspace*{10pt}    Format: Not specified\newline
    	\hspace*{10pt}    - difficulty : string (optional)\newline
    	  \hspace*{10pt}    Description: The difficulty level of the newly added stress reduction \hspace*{10pt}technique\newline
    	  \hspace*{10pt}    Format: Not specified\newline
    	\hspace*{10pt}"""\newline
        \end{allintypewriter} \\
        \rowcolor{gray!5} Answer & \begin{allintypewriter}
        addTechnique(name='Visualization', description='a relaxation exercise in which you create a peaceful mental image of a place or situation', category='Meditation', difficulty='easy')
        \end{allintypewriter} \\
        \arrayrulecolor{black}
        \bottomrule
    \end{tabularx}
    \caption{API-Bank dataset sample \ding{183}.}
    \label{tab:sample_apibank}
\end{table*}

\begin{table*}[!t]
    \centering
    \fontsize{9}{9}\selectfont
    \begin{tabularx}{\linewidth}{>{\hsize=1.5cm}XX}
        \toprule
        \arrayrulecolor{black}
        \textbf{Field} & \textbf{Text} \\
        \midrule
        \arrayrulecolor[rgb]{0.6,0.6,0.6}
        \rowcolor{yellow!10} Query & \begin{allintypewriter}
        Users want to engage in a conversation with a fictional character based on their persona. This conversation will be used as part of a script for an animation series.
        \end{allintypewriter} \\
        \midrule
        \rowcolor{gray!5} Gold Tool's Docstring & \begin{allintypewriter}
       def AutoModelForCausalLM.from\_pretrained('pygmalion-6b'):\newline
    	\hspace*{10pt}"""\newline
    	\hspace*{10pt}Description:\newline\newline
    	\hspace*{10pt}Pygmalion 6B is a proof-of-concept dialogue model based on EleutherAI's \newline\hspace*{10pt}GPT-J-6B. The fine-tuning dataset consisted of 56MB of dialogue data\newline\hspace*{10pt}gathered from multiple sources, which includes both real and partially \newline\hspace*{10pt}machine-generated conversations. The model was initialized from the uft-6b \newline\hspace*{10pt}ConvoGPT model and fine-tuned on ~48.5 million tokens for ~5k steps \newline\hspace*{10pt}on 4 NVIDIA A40s using DeepSpeed.\newline
    	\hspace*{10pt}"""
        \end{allintypewriter} \\
        \midrule
        \rowcolor{yellow!10} Answer & \begin{allintypewriter}
        AutoModelForCausalLM.from\_pretrained('pygmalion-6b')
        \end{allintypewriter} \\
        \arrayrulecolor{black}
        \bottomrule
    \end{tabularx}
    \caption{APIBench dataset sample \ding{184}.}
    \label{tab:sample_apibench}
\end{table*}

\begin{table*}[!t]
    \centering
    \fontsize{9}{9}\selectfont
    \begin{tabularx}{\linewidth}{>{\hsize=1.5cm}XX}
        \toprule
        \arrayrulecolor{black}
        \textbf{Field} & \textbf{Text} \\
        \midrule
        \arrayrulecolor[rgb]{0.6,0.6,0.6}
        \rowcolor{yellow!10} Query & \begin{allintypewriter}
        Search for a Chicken Noodle Soup recipe and a Vegan Salad recipe.
        \end{allintypewriter} \\
        \midrule
        \rowcolor{gray!5} Gold Tool's Docstring & \begin{allintypewriter}
        def recipe\_search.find():\newline
    	\hspace*{10pt}"""\newline
    	\hspace*{10pt}Description:\newline
    	\hspace*{10pt}Locate recipes based on the type of dish.\newline\newline
    	
    	\hspace*{10pt}Arguments:\newline
    	\hspace*{10pt}---------\newline
    	\hspace*{10pt}- dish : string = None (required) The name of the dish to search for.\newline
    	\hspace*{10pt}- diet : string = Keto (optional) Dietary preference.\newline
        \hspace*{10pt}"""
        \end{allintypewriter} \\
        \midrule
        \rowcolor{yellow!10} Answer & \begin{allintypewriter}
        recipe\_search.find(dish="Chicken Noodle Soup", diet="Vegan")
        \end{allintypewriter} \\
        \arrayrulecolor{black}
        \bottomrule
    \end{tabularx}
    \caption{BFCL dataset sample \ding{185}.}
    \label{tab:sample_bfcl}
\end{table*}

\begin{table*}[!t]
    \centering
    \fontsize{9}{9}\selectfont
    \begin{tabularx}{\linewidth}{>{\hsize=1.5cm}XX} 
        \toprule
        \arrayrulecolor{black}
        \textbf{Field} & \textbf{Text} \\
        \midrule
        \arrayrulecolor[rgb]{0.6,0.6,0.6}
        \rowcolor{yellow!10} Query & \begin{allintypewriter}
        Help me with a quick d20 roll, I've got a crucial decision to make in my game.
        \end{allintypewriter} \\
        \midrule
        \rowcolor{gray!5} Gold Tool's Docstring & \begin{allintypewriter}
        def diceroller():\newline
        \hspace*{10pt}"""\newline
        \hspace*{10pt}Description:\newline
        \hspace*{10pt}App for rolling dice using the d20 or Fate/Fudge systems.\newline
        \hspace*{10pt}"""
        \end{allintypewriter} \\
        \midrule
        \rowcolor{yellow!10} Answer & \begin{allintypewriter}
        diceroller()
        \end{allintypewriter} \\
        \arrayrulecolor{black}
        \bottomrule
    \end{tabularx}
    \caption{ToolE dataset sample \ding{186}/\ding{188}.}
    \label{tab:sample_toole}
    \vspace{5pt}
\end{table*}

\newpage
\clearpage

\begin{table}[!htb]
    \centering
    \begin{adjustbox}{width=\linewidth}
        \begin{tabular}{lr}
        \toprule
        \textbf{Hyperparameter} & \textbf{Search space} \\
        \midrule
        \rowcolor{gray!15} Random seed & \{0, 42$\ast$, 100\}\\
        $|$Negatives$|$ & \{1, 2, 3$\ast$\}\\
        \rowcolor{gray!15} Negatives selection & \begin{tabular}[t]{r}Sampling every $T=50$\\training steps\end{tabular}\\
        Max sequence length (encoder) & 512\\
        \rowcolor{gray!15} Max sequence length (LLM) & 1024\\
        Loss weighting factor $\lambda$ & \{0.1, 0.3$\ast$, 0.5, 0.7, 0.9\}\\
        \rowcolor{gray!15} Retriever likelihood temperature $\gamma$ & \{0.3, 0.5$\ast$, 0.7, 1\}\\
        LLM likelihood temperature $\beta$ & \{0.3, 0.5$\ast$, 0.7, 1\}\\
        \rowcolor{gray!15} $|$Epochs$|$ & 2\\
        Fine-tuning optimizer & \begin{tabular}[t]{r}AdamW (0.9 $\beta_1$, 0.999 $\beta_2$,\\0.01 w. decay)\end{tabular}\\
        \rowcolor{gray!15} Training batch size & 2\\
        Test batch size & 4\\
        \rowcolor{gray!15} Cosine learning rate & \begin{tabular}[t]{r}\{$1e^{-6}$, $1e^{-5}\ast$, $5e^{-5}$,\\$1e^{-4}$, $2e^{-4}$\}\end{tabular}\\
        \bottomrule
        \end{tabular}
    \end{adjustbox}
    \caption{Explored hyperparameters along with their empirical search grid. $\ast$ marks the final picked values.}
    \label{tab:hyperparams}
\end{table}

\section{Computational Budget}
\label{app:computational_budget}

All experiments were performed on machines equipped with NVIDIA RTX 3090 GPUs (24GB VRAM). The total compute time required for training and evaluation across all \ports' variants amounted to approximately 500 GPU-hours. This includes finetuning on multiple datasets, ablative experiments, running inference with large language models, and conducting retrieval evaluations.

\section{Robustness}
\label{app:robustness}

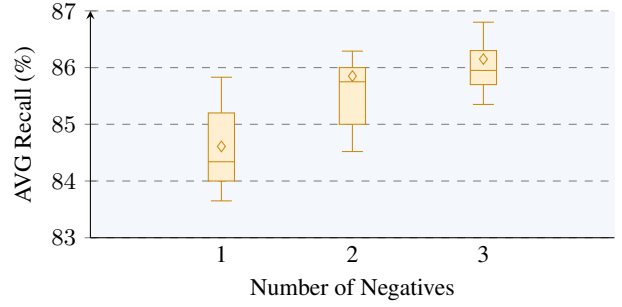
\begin{figure}[!t]
    \centering
    \begin{tikzpicture}[
        node font=\small,
        farbe/.style={draw=#1!80!black,fill=#1!20}
    ] 
  
    \begin{axis}[
        height=3cm, width=0.9\columnwidth,
        major grid style={gray},
        minor grid style={gray},
        axis background/.style={fill=plotbackground},
        axis x line*=bottom,
        axis y line*=left,
        scale only axis,
        ymin=83, ymax=87,  
        ylabel={AVG Recall (\%)},
        xlabel={Number of Negatives},
        boxplot/draw direction=y,
        grid style=dashed,
        xmin=0, xmax=4,
        xtick={1,2,3},
        xticklabels={1,2,3},
        xtick style={draw=none},
        x tick label style={font=\footnotesize,align=center,},
        compat=newest,
        axis y line=left,
        ymajorgrids=true, 
    ]

    \addplot[
        farbe=portscolor, 
        boxplot prepared={
            draw position=1, 
            lower whisker=83.65, 
            lower quartile=84.00, 
            median=84.34, 
            average=84.61, 
            upper quartile=85.20, 
            upper whisker=85.83, 
            box extend=0.2,
        },
    ] coordinates {};
    
    \addplot[
        farbe=portscolor,
        boxplot prepared={
            draw position=2, 
            lower whisker=84.52, 
            lower quartile=85.00, 
            median=85.75, 
            average=85.85, 
            upper quartile=86.00, 
            upper whisker=86.29, 
            box extend=0.2,
        },
    ] coordinates {};
    
    \addplot[
        farbe=portscolor, 
        boxplot prepared={
            draw position=3, 
            lower whisker=85.35, 
            lower quartile=85.70, 
            median=85.95, 
            average=86.15, 
            upper quartile=86.30, 
            upper whisker=86.8, 
            box extend=0.2,
        },
    ] coordinates {};

    \end{axis}
    \end{tikzpicture}
    \caption{Average Recall@$K$ across different numbers of negatives and seeds on the ToolE~\ding{186} dataset using \modernBertShort and \llamaBig.}
    \label{fig:negatives_recall}
\end{figure}

The efficacy of \ports was evaluated through ablation studies to determine optimal parameter configuration and assess robustness across configuration variations, using the ToolE~\ding{186} dataset with \robertaShort as encoder and \llamaBig as generative models.
To demonstrate the effectiveness of the contrastive loss, we examined the impact of using different numbers of negative examples in the learning process, with results in Figure~\ref{fig:negatives_recall} illustrating advantages of incorporating larger numbers of examples which better guide preference optimization.
We investigated the effects of varying weighting factors $\beta$ and $\gamma$, with Figure~\ref{fig:conf_matrix} showing higher $\beta$ and lower $\gamma$ values yield improved results, optimal when both are set to 0.5.
Additionally, we examined the influence of random seeds on our method, focusing on their impact on input data distribution and dropout layer behavior, with results in Table~\ref{tab:avg_recall_seed} demonstrating the robustness and effectiveness of \ports and its low variance in response to such configuration changes.

\def\myConfMat{{
{83.64, 84.78, 85.22, 84.61},
{83.78, 85.78, 85.29, 84.52},
{83.86, 84.42, 85.05, 84.08},
{84.61, 84.52, 84.30, 84.61},
}}

\def\classNames{{"0.3","0.5","0.7","1"}}
\def\numClasses{4}
\def\myScale{1.35} 

\begin{figure}[!t]
    \centering
    \begin{tikzpicture}[
        scale = \myScale,
        font={\footnotesize},
    ]
    
    \tikzset{vertical label/.style={rotate=90,anchor=east}}
    \tikzset{diagonal label/.style={rotate=45,anchor=north east}}
    
    \foreach \y in {1,...,\numClasses}
    {
        \node [anchor=east] at (0.4,-\y) {\pgfmathparse{\classNames[\y-1]}\pgfmathresult}; 
        
        \foreach \x in {1,...,\numClasses}
        {
            \pgfmathparse{\myConfMat[\y-1][\x-1]}
            \let\mVal\pgfmathresult

            \pgfmathsetmacro{\minval}{83.64}
            \pgfmathsetmacro{\maxval}{85.78}
            \pgfmathsetmacro{\normalizedValue}{(\mVal-\minval)/(\maxval-\minval)}

            \pgfmathsetmacro{\r}{0.95 - 0.25*\normalizedValue}  
            \pgfmathsetmacro{\g}{0.85 - 0.35*\normalizedValue}  
            \pgfmathsetmacro{\b}{0.75 - 0.45*\normalizedValue}  

            \definecolor{cellcolor}{rgb}{\r,\g,\b}

            \def\txtcol{black}

            \node[
                draw,
                fill=cellcolor,
                text=\txtcol,
                align=center,
                minimum size=\myScale*10mm,
                inner sep=0,
            ] at (\x,-\y) {\mVal};

            \ifnum\y=\numClasses
                \node [] at (\x,-\y-0.75)
                {\pgfmathparse{\classNames[\x-1]}\pgfmathresult};
            \fi
        }
    }

    \coordinate (yaxis) at (-0.2, -\numClasses/2-0.35);
    \coordinate (xaxis) at (\numClasses/2+0.5, -\numClasses-1);
    
    \node [] at (yaxis) {$\boldsymbol{\gamma}$};
    \node [] at (xaxis) {$\boldsymbol{\beta}$};
    
    \end{tikzpicture}
    \caption{Confusion Matrix for Average Recall@$K$ on the ToolE~\ding{186} dataset using \modernBertShort and \llamaBig, with varying $\beta$ and $\gamma$ hyperparameters.}
    \label{fig:conf_matrix}
\end{figure}

\begin{table}[!hb]
    \centering
    \begin{adjustbox}{width=.85\linewidth}
    \begin{threeparttable}
    \begin{tabular}{llll}
        \toprule
        \textbf{Dataset} & \textbf{AVG Recall} & \textbf{Seed} & \textbf{$\sigma^2$} \\
        \hline
        & \textbf{17.45} & \textbf{0} & \\
        & 14.38 & 42 & \\
        \multirow{-3}{*}{ToolBench \ding{182}} & 
        \underline{16.84} & \underline{100} & \multirow{-3}{*}{1.76} \\
        \hline
        & \underline{56.93} & \underline{0} & \\
        & \textbf{60.38} & \textbf{42} & \\
        \multirow{-3}{*}{API-Bank \ding{183}} & 46.73 & 100 & \multirow{-3}{*}{33.59} \\
        \hline
        & 26.10 & 0 & \\
        & \textbf{27.75} & \textbf{42} & \\
        \multirow{-3}{*}{APIBench \ding{184}} & \underline{27.10} & \underline{100} & \multirow{-3}{*}{1.20} \\
        \hline
        & 55.45 & 0 & \\
        & \underline{60.51} & \underline{42} & \\
        \multirow{-3}{*}{BFCL \ding{185}} & 
        \textbf{61.98} & \textbf{100} & \multirow{-3}{*}{7.82} \\
        \hline
        & 86.86 & 0 & \\
        & \textbf{88.32} & \textbf{42} & \\
        \multirow{-3}{*}{ToolE \ding{186}} & 
        \underline{86.37} & \underline{100} & \multirow{-3}{*}{0.69} \\
        \hline
        & \underline{95.00} & \underline{0} & \\
        & \textbf{96.66} & \textbf{42} & \\
        \multirow{-3}{*}{Octopus \ding{187}} & 86.60 & 100 & \multirow{-3}{*}{19.39} \\
        \bottomrule
    \end{tabular}
    \begin{tablenotes}
    \end{tablenotes}
    \end{threeparttable}
    \end{adjustbox}
    \caption{Per-dataset variance ($\sigma^2$) of the average Recall@$K$ with training runs using different random seeds. Bold and underline denote the best and second-best runs for each dataset.}
    \label{tab:avg_recall_seed}
\end{table}

\clearpage
\newpage

\section{Clustering Properties of Tool Embeddings}
\label{app:clustering}
\begin{figure*}[!t]
    \centering
    \begin{subfigure}[t]{\linewidth}
        \centering
        \begin{tikzpicture}
            \footnotesize
            \node[draw=none, inner sep=2pt, align=center, text width=8cm] {
            \begin{tabular}{llll}
                \ref{plot:agglomerative} $\texttt{Agglomerative}$ & \ref{plot:dbscan} $\texttt{DBScan}$ & \ref{plot:kmeans} $\texttt{KMeans}$ & \ref{plot:avg} $\texttt{Average}$ \\
            \end{tabular}
            };
        \end{tikzpicture}
    \end{subfigure}
    \begin{subfigure}[t]{.96\linewidth}
        \begin{tikzpicture}
            \begin{axis}[
                width=\linewidth, height=22cm,
                ymajorgrids=true,
                grid=both,
                grid style=dashed,
                axis background/.style={fill=plotbackground},
                boxplot/draw direction=x, 
                boxplot={box extend=0.15, draw position=1},
                enlarge y limits=0.05,  
                xmin=-0.1,
                xmax=0.6,
                xlabel={\textbf{Silhouette}},
                ytick={1,2,3,5,6,7,8,9,10},
                yticklabels={HotpotQA,PubMedQA,MSMarco,\ding{182} ToolBench,\ding{183} API-Bank,\ding{184} APIBench,\ding{185} BFCL-v2,\ding{186} ToolE,\ding{187} Octopus-v2},
                y dir=reverse,
                every tick label/.append style={font=\fontsize{8}{8}\selectfont},
                xlabel style={font=\fontsize{8}{8}\selectfont},
                ylabel style={font=\fontsize{8}{8}\selectfont},
            ]
    
            \addplot+[
                boxplot prepared={
                    average=0.035,
                    median=0.013,
                    upper whisker=0.11,
                    lower whisker=-0.003
                },
                draw=agglomerativecolor,
                fill=agglomerativecolor,
                solid,
                boxplot/draw position=0.7
            ] coordinates {};
            \addplot+[const plot, thick, draw=agglomerativecolor, mark=none, forget plot] coordinates {(-1,0) (-1,0)};
            \label{plot:agglomerative}
            \addplot+[
                boxplot prepared={
                    average=0.009,
                    median=0.000,
                    upper whisker=0.034,
                    lower whisker=0.001
                },
                draw=dbscancolor,
                fill=dbscancolor,
                solid,
                boxplot/draw position=0.9
            ] coordinates {};
            \addplot+[const plot, thick, draw=dbscancolor, mark=none, forget plot] coordinates {(-1,0) (-1,0)};
            \label{plot:dbscan}
            \addplot+[
                boxplot prepared={
                    average=0.042,
                    median=0.021,
                    upper whisker=0.104,
                    lower whisker=0.007
                },
                draw=kmeanscolor,
                fill=kmeanscolor,
                solid,
                boxplot/draw position=1.1
            ] coordinates {};
            \addplot+[const plot, thick, draw=kmeanscolor, mark=none, forget plot] coordinates {(-1,0) (-1,0)};
            \label{plot:kmeans}
            \addplot+[
                boxplot prepared={
                    average=0.028,
                    median=0.035,
                    upper whisker=0.042,
                    lower whisker=0.008
                },
                draw=avgcolor,
                fill=avgcolor,
                solid,
                boxplot/draw position=1.3
            ] coordinates {};
            \addplot+[const plot, thick, draw=avgcolor, mark=none, forget plot] coordinates {(-1,0) (-1,0)};
            \label{plot:avg}

            \addplot+[
                boxplot prepared={
                    average=0.035,
                    median=-0.003,
                    upper whisker=0.147,
                    lower whisker=-0.007
                },
                draw=agglomerativecolor,
                fill=agglomerativecolor,
                solid,
                boxplot/draw position=1.7
            ] coordinates {};
            \addplot+[
                boxplot prepared={
                    average=0.000,
                    median=0.000,
                    upper whisker=0.000,
                    lower whisker=-0.000
                },
                draw=dbscancolor,
                fill=dbscancolor,
                solid,
                boxplot/draw position=1.9
            ] coordinates {};
            \addplot+[
                boxplot prepared={
                    average=0.061,
                    median=0.021,
                    upper whisker=0.172,
                    lower whisker=0.015
                },
                draw=kmeanscolor,
                fill=kmeanscolor,
                solid,
                boxplot/draw position=2.1
            ] coordinates {};
            \addplot+[
                boxplot prepared={
                    average=0.032,
                    median=0.035,
                    upper whisker=0.061,
                    lower whisker=0.000
                },
                draw=avgcolor,
                fill=avgcolor,
                solid,
                boxplot/draw position=2.3
            ] coordinates {};

            \addplot+[
                boxplot prepared={
                    average=0.033,
                    median=0.017,
                    upper whisker=0.085,
                    lower whisker=0.012
                },
                draw=agglomerativecolor,
                fill=agglomerativecolor,
                solid,
                boxplot/draw position=2.7
            ] coordinates {};
            \addplot+[
                boxplot prepared={
                    average=0.038,
                    median=0.000,
                    upper whisker=0.152,
                    lower whisker=0.000
                },
                draw=dbscancolor,
                fill=dbscancolor,
                solid,
                boxplot/draw position=2.9
            ] coordinates {};
            \addplot+[
                boxplot prepared={
                    average=0.036,
                    median=0.023,
                    upper whisker=0.075,
                    lower whisker=0.022
                },
                draw=kmeanscolor,
                fill=kmeanscolor,
                solid,
                boxplot/draw position=3.1
            ] coordinates {};
            \addplot+[
                boxplot prepared={
                    average=0.036,
                    median=0.036,
                    upper whisker=0.039,
                    lower whisker=0.033
                },
                draw=avgcolor,
                fill=avgcolor,
                solid,
                boxplot/draw position=3.3
            ] coordinates {};

            \addplot+[
                boxplot prepared={
                    average=0.041,
                    median=0.023,
                    upper whisker=0.090,
                    lower whisker=0.015
                },
                draw=agglomerativecolor,
                fill=agglomerativecolor,
                solid,
                boxplot/draw position=4.7
            ] coordinates {};
            \addplot+[
                boxplot prepared={
                    average=0.103,
                    median=0.000,
                    upper whisker=0.413,
                    lower whisker=0.000
                },
                draw=dbscancolor,
                fill=dbscancolor,
                solid,
                boxplot/draw position=4.9
            ] coordinates {};
            \addplot+[
                boxplot prepared={
                    average=0.051,
                    median=0.029,
                    upper whisker=0.099,
                    lower whisker=0.035
                },
                draw=kmeanscolor,
                fill=kmeanscolor,
                solid,
                boxplot/draw position=5.1
            ] coordinates {};
            \addplot+[
                boxplot prepared={
                    average=0.065,
                    median=0.051,
                    upper whisker=0.103,
                    lower whisker=0.041
                },
                draw=avgcolor,
                fill=avgcolor,
                solid,
                boxplot/draw position=5.3
            ] coordinates {};

            \addplot+[
                boxplot prepared={
                    average=0.068,
                    median=0.042,
                    upper whisker=0.128,
                    lower whisker=0.044
                },
                draw=agglomerativecolor,
                fill=agglomerativecolor,
                solid,
                boxplot/draw position=5.7
            ] coordinates {};
            \addplot+[
                boxplot prepared={
                    average=0.112,
                    median=0.066,
                    upper whisker=0.315,
                    lower whisker=0.000
                },
                draw=dbscancolor,
                fill=dbscancolor,
                solid,
                boxplot/draw position=5.9
            ] coordinates {};
            \addplot+[
                boxplot prepared={
                    average=0.080,
                    median=0.051,
                    upper whisker=0.153,
                    lower whisker=0.048
                },
                draw=kmeanscolor,
                fill=kmeanscolor,
                solid,
                boxplot/draw position=6.1
            ] coordinates {};
            \addplot+[
                boxplot prepared={
                    average=0.086,
                    median=0.080,
                    upper whisker=0.111,
                    lower whisker=0.068
                },
                draw=avgcolor,
                fill=avgcolor,
                solid,
                boxplot/draw position=6.3
            ] coordinates {};

            \addplot+[
                boxplot prepared={
                    average=0.077,
                    median=0.070,
                    upper whisker=0.108,
                    lower whisker=0.062
                },
                draw=agglomerativecolor,
                fill=agglomerativecolor,
                solid,
                boxplot/draw position=6.7
            ] coordinates {};
            \addplot+[
                boxplot prepared={
                    average=0.159,
                    median=0.127,
                    upper whisker=0.383,
                    lower whisker=0.000
                },
                draw=dbscancolor,
                fill=dbscancolor,
                solid,
                boxplot/draw position=6.9
            ] coordinates {};
            \addplot+[
                boxplot prepared={
                    average=0.086,
                    median=0.079,
                    upper whisker=0.120,
                    lower whisker=0.070
                },
                draw=kmeanscolor,
                fill=kmeanscolor,
                solid,
                boxplot/draw position=7.1
            ] coordinates {};
            \addplot+[
                boxplot prepared={
                    average=0.107,
                    median=0.086,
                    upper whisker=0.159,
                    lower whisker=0.077
                },
                draw=avgcolor,
                fill=avgcolor,
                solid,
                boxplot/draw position=7.3
            ] coordinates {};

            \addplot+[
                boxplot prepared={
                    average=0.049,
                    median=0.028,
                    upper whisker=0.121,
                    lower whisker=0.016
                },
                draw=agglomerativecolor,
                fill=agglomerativecolor,
                solid,
                boxplot/draw position=7.7
            ] coordinates {};
            \addplot+[
                boxplot prepared={
                    average=0.230,
                    median=0.231,
                    upper whisker=0.492,
                    lower whisker=-0.033
                },
                draw=dbscancolor,
                fill=dbscancolor,
                solid,
                boxplot/draw position=7.9
            ] coordinates {};
            \addplot+[
                boxplot prepared={
                    average=0.068,
                    median=0.037,
                    upper whisker=0.152,
                    lower whisker=0.038
                },
                draw=kmeanscolor,
                fill=kmeanscolor,
                solid,
                boxplot/draw position=8.1
            ] coordinates {};
            \addplot+[
                boxplot prepared={
                    average=0.116,
                    median=0.068,
                    upper whisker=0.230,
                    lower whisker=0.050
                },
                draw=avgcolor,
                fill=avgcolor,
                solid,
                boxplot/draw position=8.3
            ] coordinates {};

            \addplot+[
                boxplot prepared={
                    average=0.033,
                    median=0.025,
                    upper whisker=0.064,
                    lower whisker=0.020
                },
                draw=agglomerativecolor,
                fill=agglomerativecolor,
                solid,
                boxplot/draw position=8.7
            ] coordinates {};
            \addplot+[
                boxplot prepared={
                    average=-0.007,
                    median=0.000,
                    upper whisker=0.000,
                    lower whisker=-0.028
                },
                draw=dbscancolor,
                fill=dbscancolor,
                solid,
                boxplot/draw position=8.9
            ] coordinates {};
            \addplot+[
                boxplot prepared={
                    average=0.036,
                    median=0.028,
                    upper whisker=0.054,
                    lower whisker=0.028
                },
                draw=kmeanscolor,
                fill=kmeanscolor,
                solid,
                boxplot/draw position=9.1
            ] coordinates {};
            \addplot+[
                boxplot prepared={
                    average=0.021,
                    median=0.033,
                    upper whisker=0.036,
                    lower whisker=-0.007
                },
                draw=avgcolor,
                fill=avgcolor,
                solid,
                boxplot/draw position=9.3
            ] coordinates {};

            \addplot+[
                boxplot prepared={
                    average=0.109,
                    median=0.096,
                    upper whisker=0.164,
                    lower whisker=0.087
                },
                draw=agglomerativecolor,
                fill=agglomerativecolor,
                solid,
                boxplot/draw position=9.7
            ] coordinates {};
            \addplot+[
                boxplot prepared={
                    average=0.041,
                    median=0.000,
                    upper whisker=0.163,
                    lower whisker=0.000
                },
                draw=dbscancolor,
                fill=dbscancolor,
                solid,
                boxplot/draw position=9.9
            ] coordinates {};
            \addplot+[
                boxplot prepared={
                    average=0.092,
                    median=0.080,
                    upper whisker=0.144,
                    lower whisker=0.071
                },
                draw=kmeanscolor,
                fill=kmeanscolor,
                solid,
                boxplot/draw position=10.1
            ] coordinates {};
            \addplot+[
                boxplot prepared={
                    average=0.081,
                    median=0.092,
                    upper whisker=0.110,
                    lower whisker=0.041
                },
                draw=avgcolor,
                fill=avgcolor,
                solid,
                boxplot/draw position=10.3
            ] coordinates {};

            legend style={
                font=\fontsize{7}{7}\selectfont,
                at={(0.5,-0.05)},
                anchor=north,
                legend columns=4,
                /tikz/every even column/.append style={column sep=0.4cm}
            }
            
            \end{axis}
        \end{tikzpicture}
    \end{subfigure}
    \caption{Average Silhouette scores across datasets (top--general-domain, bottom--tool-specific) and clustering algorithms, computed over 20{,}000 sampled embeddings per dataset. Results are aggregated across multiple encoder models. Higher scores indicate more compact and well-separated clusters, reflecting stronger semantic structure in the embedding space.}
    \label{fig:clustering}
\end{figure*}
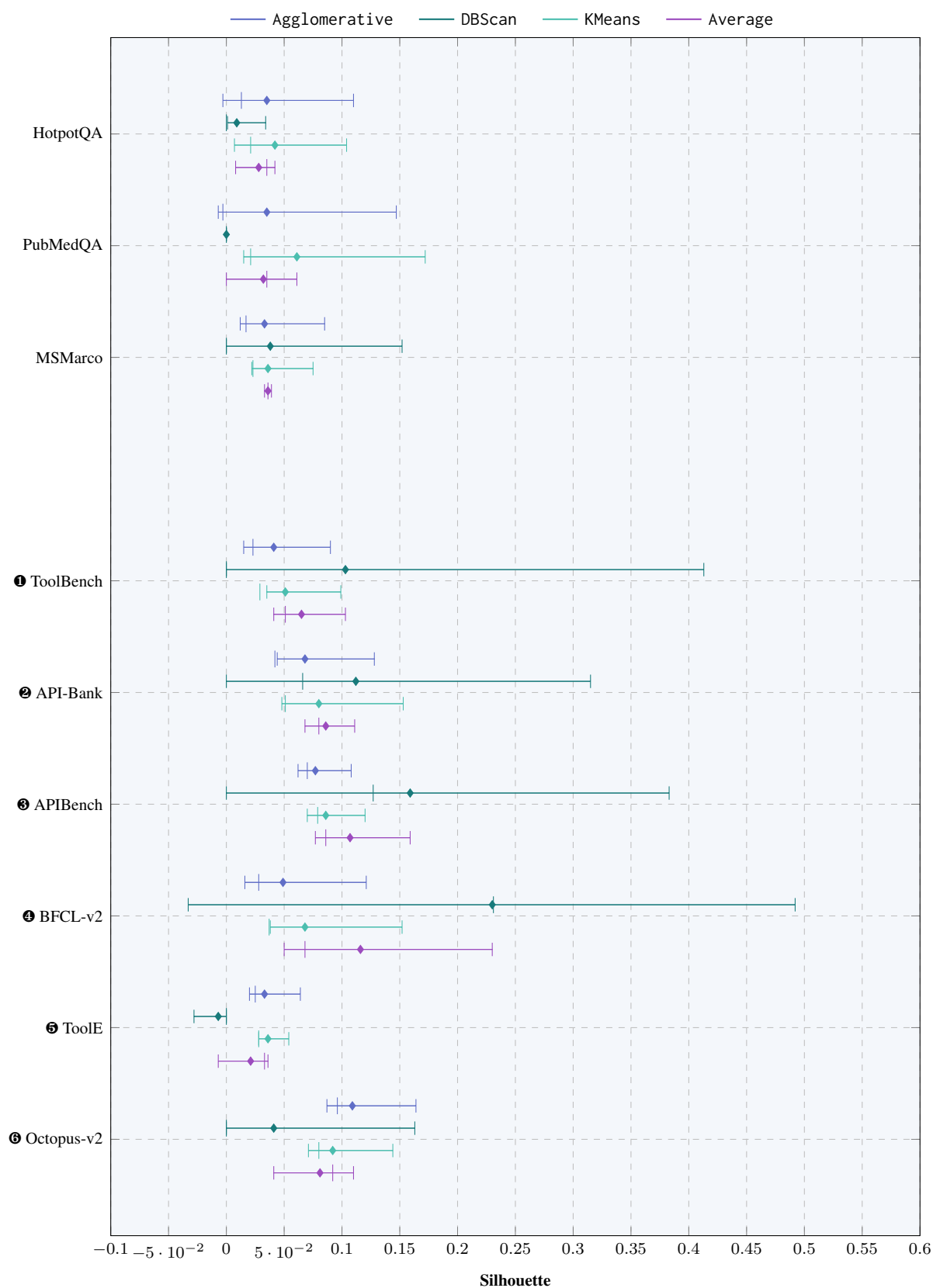

API docstrings exhibit skewed token distributions, dominated by recurring elements such as data types and keywords.
Compared to general-domain retrieval corpora, tool-related datasets form well-separated semantic clusters, owing to their concise yet distinctive functional signatures. 
Without targeted supervision, these structural properties can lead retrieval models to rely on superficial lexical cues or converge toward trivial matches. To characterize the clustering tendency of these representations, we apply standard unsupervised algorithms--including K-Means (with $K \in [2, 16]$), Agglomerative Clustering (Ward linkage), and DBSCAN (with $\epsilon = 0.3$)—on 20{,}000 randomly sampled embedding vectors per dataset. 
We use four top-performing models from the MTEB leaderboard\footnote{\href{https://huggingface.co/spaces/mteb/leaderboard}{huggingface.co/spaces/mteb/leaderboard}} to extract these representations: \texttt{BAAI/bge-m3},\footnote{\href{https://huggingface.co/BAAI/bge-m3}{huggingface.co/BAAI/bge-m3}} \texttt{intfloat/multilingual-e5-large-instruct},\footnote{\href{https://huggingface.co/intfloat/multilingual-e5-large-instruct}{huggingface.co/intfloat/multilingual-e5-large-instruct}} \texttt{answerdotai/ModernBERT-base},\footnote{\href{https://huggingface.co/answerdotai/ModernBERT-base}{huggingface.co/answerdotai/ModernBERT-base}} and \texttt{Alibaba-NLP/gte-multilingual-base}.\footnote{\href{https://huggingface.co/Alibaba-NLP/gte-multilingual-base}{huggingface.co/Alibaba-NLP/gte-multilingual-base}} The silhouette coefficient is computed to quantify clustering quality, capturing both intra-cluster cohesion and inter-cluster separation.
As summarized in Figure~\ref{fig:clustering}, tool-centric datasets consistently achieve higher silhouette scores than general-domain corpora, including MSMarco~\cite{DBLP:journals/corr/NguyenRSGTMD16}, HotpotQA~\cite{DBLP:conf/emnlp/Yang0ZBCSM18}, and PubMedQA~\cite{DBLP:conf/emnlp/JinDLCL19}. 
This high clustering tendency presents a fundamental limitation where semantically similar tools concentrate within the same dense clusters, making contrastive supervision essential to differentiate tools that share similar descriptive features yet possess distinct functionalities and produce varying effects on LLM behavior, necessitating targeted intra-cluster contrastive learning to prevent training from optimizing merely for tool relevance rather than functional effectiveness.
This gap highlights the stronger intrinsic structure of tool embeddings and further motivates our contrastive supervision strategy to promote fine-grained, functionally meaningful distinctions beyond those induced by LLM likelihoods alone.

\begin{table*}[t]
    \centering
    \begin{adjustbox}{width=\linewidth}
    \begin{threeparttable}
    \begin{tabular}{lcllcccccccc}
    \toprule
    & & & & \multicolumn{3}{c}{\textbf{Recall} (\%)} & \multicolumn{3}{c}{\textbf{NDCG} (\%)} & \multicolumn{2}{c}{\textbf{$\Delta$ Baseline}} \\
    \cmidrule(lr){5-7}
    \cmidrule(lr){8-10}
    \cmidrule(lr){11-12}
    \textbf{Encoder} & \textbf{Dataset} & \textbf{Method} & \textbf{LLM} & \textbf{@1} & \textbf{@2} & \textbf{@3} & \textbf{@1} & \textbf{@3} & \textbf{@5} & \textbf{Recall} & \textbf{NDCG} \\
    \hline
    \rowcolor{robertabackgroundcolor} & & \lossPorts & \llamaBig & 11.89 & 17.13 & 20.45 & 11.89 & 17.10 & 19.10 & \cellcolor{robertahighlightcolor!35}15.48 & \cellcolor{robertahighlightcolor!35}15.18\\
    \rowcolor{robertabackgroundcolor} & & \lossReplug & \llamaBig & 10.34 & 16.50 & 20.58 & 10.34 & 16.11 & 18.08 & \cellcolor{robertahighlightcolor!35}14.80 & \cellcolor{robertahighlightcolor!34}14.09\\
    \cline{3-12}
    \rowcolor{robertabackgroundcolor} & & \lossPorts & \groqLlama & 12.11 & 19.20 & 23.40 & 12.11 & 19.30 & 23.10 & \cellcolor{robertahighlightcolor!37}17.23 & \cellcolor{robertahighlightcolor!38}18.13\\
    \rowcolor{robertabackgroundcolor} & & \lossReplug & \groqLlama & 11.53 & 18.52 & 22.80 & 11.53 & 18.05 & 21.19 & \cellcolor{robertahighlightcolor!37}16.61 & \cellcolor{robertahighlightcolor!36}16.11\\
    \cline{3-12}
    \rowcolor{robertabackgroundcolor} & & \lossPorts & \codestral & 18.23 & 21.10 & 25.60 & 18.23 & 20.28 & 24.31 & \cellcolor{robertahighlightcolor!41}20.63 & \cellcolor{robertahighlightcolor!40}19.94 \\
    \rowcolor{robertabackgroundcolor} & \multirow{-6}{*}{\ding{182}} & \lossReplug & \codestral & 12.56 & 20.11 & 24.80 & 12.56 & 19.67 & 22.19 & \cellcolor{robertahighlightcolor!38}18.14 & \cellcolor{robertahighlightcolor!37}17.14 \\    
    \cline{2-12}
    \rowcolor{robertabackgroundcolor} & & \lossPorts & \llamaBig & 49.70 & 62.78 & 68.06 & 49.70 & 61.29 & 63.38 & \cellcolor{robertahighlightcolor!77}56.47 & \cellcolor{robertahighlightcolor!74}53.75\\
    \rowcolor{robertabackgroundcolor} & & \lossReplug & \llamaBig & 4.00 & 5.81 & 8.87 & 4.00 & 6.30 & 8.00 & \cellcolor{robertahighlightcolor!22}1.67 & \cellcolor{robertahighlightcolor!21}1.12\\
    \cline{3-12}
    \rowcolor{robertabackgroundcolor} & & \lossPorts & \groqLlama & 49.84 & 64.35 & 70.80 & 49.84 & 62.27 & 65.32 & \cellcolor{robertahighlightcolor!77}57.10 & \cellcolor{robertahighlightcolor!74}54.14 \\
    \rowcolor{robertabackgroundcolor} & & \lossReplug & \groqLlama & 45.32 & 61.94 & 68.23 & 45.32 & 59.95 & 62.44 & \cellcolor{robertahighlightcolor!74}53.93 & \cellcolor{robertahighlightcolor!71}50.90 \\
    \cline{3-12}
    \rowcolor{robertabackgroundcolor} & & \lossPorts & \codestral & 43.87 & 54.67 & 64.84 & 43.87 & 58.10 & 61.20 & \cellcolor{robertahighlightcolor!70}49.90 & \cellcolor{robertahighlightcolor!70}50.23\\
    \rowcolor{robertabackgroundcolor} & \multirow{-6}{*}{\ding{183}} & \lossReplug & \codestral & 10.00 & 14.19 & 18.06 & 10.00 & 15.17 & 18.16 & \cellcolor{robertahighlightcolor!30}9.52 & \cellcolor{robertahighlightcolor!29}9.13\\
    \cline{2-12}
    \rowcolor{robertabackgroundcolor} & & \lossPorts & \llamaBig & 18.94 & 25.27 & 28.12 & 18.94 & 22.00 & 23.80 & \cellcolor{robertahighlightcolor!43}22.58 & \cellcolor{robertahighlightcolor!40}20.34\\
    \rowcolor{robertabackgroundcolor} & & \lossReplug & \llamaBig & 8.74 & 12.61 & 15.35 & 8.74 & 12.55 & 14.56 & \cellcolor{robertahighlightcolor!31}10.70 & \cellcolor{robertahighlightcolor!31}10.95 \\
    \cline{3-12}
    \rowcolor{robertabackgroundcolor} & & \lossPorts & \groqLlama & 21.50 & 27.40 & 30.53 & 21.50 & 25.22 & 26.78 & \cellcolor{robertahighlightcolor!46}25.94 & \cellcolor{robertahighlightcolor!44}23.50 \\
    \rowcolor{robertabackgroundcolor} & & \lossReplug & \groqLlama & 5.66 & 8.46 & 9.86 & 5.66 & 8.21 & 9.23 & \cellcolor{robertahighlightcolor!26}6.46 & \cellcolor{robertahighlightcolor!26}6.22\\
    \cline{3-12}
    \rowcolor{robertabackgroundcolor} & & \lossPorts & \codestral & 13.45 & 18.93 & 21.80 & 13.45 & 16.90 & 18.00 & \cellcolor{robertahighlightcolor!37}16.53 & \cellcolor{robertahighlightcolor!35}15.06\\
    \rowcolor{robertabackgroundcolor} & \multirow{-6}{*}{\ding{184}} & \lossReplug & \codestral & 6.78 & 9.52 & 11.99 & 6.78 & 10.01 & 11.03 & \cellcolor{robertahighlightcolor!28}7.90 & \cellcolor{robertahighlightcolor!28}8.02\\
    \cline{2-12}
    \rowcolor{robertabackgroundcolor} & & \lossPorts & \llamaBig & 57.85 & 67.65 & 70.79 & 57.85 & 66.11 & 67.12 & \cellcolor{robertahighlightcolor!79}58.78 & \cellcolor{robertahighlightcolor!77}57.09\\
    \rowcolor{robertabackgroundcolor} & & \lossReplug & \llamaBig & 48.43 & 58.60 & 64.14 & 48.43 & 58.18 & 60.09 & \cellcolor{robertahighlightcolor!70}50.40 & \cellcolor{robertahighlightcolor!69}49.06\\
    \cline{3-12}
    \rowcolor{robertabackgroundcolor} & & \lossPorts & \groqLlama & 58.12 & 69.21 & 73.52 & 58.12 & 68.38 & 69.22 & \cellcolor{robertahighlightcolor!80}60.30 & \cellcolor{robertahighlightcolor!79}59.24 \\
    \rowcolor{robertabackgroundcolor} & & \lossReplug & \groqLlama & 53.97 & 64.88 & 68.39 & 53.97 & 62.61 & 64.93 & \cellcolor{robertahighlightcolor!76}55.76 & \cellcolor{robertahighlightcolor!75}54.50 \\
    \cline{3-12}
    \rowcolor{robertabackgroundcolor} & & \lossPorts & \codestral & 39.74 & 52.14 & 56.93 & 39.74 & 49.60 & 52.50 & \cellcolor{robertahighlightcolor!63}42.95 & \cellcolor{robertahighlightcolor!61}41.34\\
    \rowcolor{robertabackgroundcolor} & \multirow{-6}{*}{\ding{185}} & \lossReplug & \codestral & 38.26 & 47.87 & 53.05 & 38.26 & 47.11 & 49.18 & \cellcolor{robertahighlightcolor!60}39.74 & \cellcolor{robertahighlightcolor!58}38.12\\
    \cline{2-12}
    \rowcolor{robertabackgroundcolor} & & \lossPorts & \llamaBig & 60.33 & 72.45 & 77.15 & 60.33 & 70.29 & 72.34 & \cellcolor{robertahighlightcolor!78}57.51 & \cellcolor{robertahighlightcolor!76}55.65 \\
    \rowcolor{robertabackgroundcolor} & & \lossReplug & \llamaBig & 10.53 & 13.92 & 16.18 & 10.53 & 14.21 & 15.23 & \cellcolor{robertahighlightcolor!21}1.08 & \cellcolor{robertahighlightcolor!21}1.21\\
    \cline{3-12}
    \rowcolor{robertabackgroundcolor} & & \lossPorts & \groqLlama & 59.90 & 71.90 & 76.90 & 59.90 & 70.04 & 72.06 & \cellcolor{robertahighlightcolor!77}57.10 & \cellcolor{robertahighlightcolor!75}55.02\\
    \rowcolor{robertabackgroundcolor} & & \lossReplug & \groqLlama & 49.38 & 59.84 & 64.23 & 49.38 & 58.17 & 60.79 & \cellcolor{robertahighlightcolor!65}45.35 & \cellcolor{robertahighlightcolor!64}44.11 \\
    \cline{3-12}
    \rowcolor{robertabackgroundcolor} & & \lossPorts & \codestral & 56.70 & 70.00 & 74.89 & 56.70 & 67.11 & 69.16 & \cellcolor{robertahighlightcolor!75}54.73 & \cellcolor{robertahighlightcolor!72}52.12\\
    \rowcolor{robertabackgroundcolor} & \multirow{-6}{*}{\ding{186}} & \lossReplug & \codestral & 40.65 & 51.54 & 57.53 & 40.65 & 51.03 & 53.04 & \cellcolor{robertahighlightcolor!57}37.44 & \cellcolor{robertahighlightcolor!56}36.02\\
    \cline{2-12}
    \rowcolor{robertabackgroundcolor} & & \lossPorts & \llamaBig & 95.00 & 100 & 100 & 95.00 & 95.25 & 98.25 & \cellcolor{robertahighlightcolor!92}71.66 & \cellcolor{robertahighlightcolor!90}70.16 \\
    \rowcolor{robertabackgroundcolor} & & \lossReplug & \llamaBig & 87.50 & 97.50 & 100 & 87.50 & 95.06 & 95.06 & \cellcolor{robertahighlightcolor!88}68.33 & \cellcolor{robertahighlightcolor!87}66.54 \\
    \cline{3-12}
    \rowcolor{robertabackgroundcolor} & & \lossPorts & \groqLlama & 95.00 & 100 & 100 & 95.00 & 95.20 & 98.20 & \cellcolor{robertahighlightcolor!92}71.53 & \cellcolor{robertahighlightcolor!90}70.03\\
    \rowcolor{robertabackgroundcolor} & & \lossReplug & \groqLlama & 75.00 & 90.00 & 97.50 & 75.00 & 88.00 & 89.00 & \cellcolor{robertahighlightcolor!81}60.83 & \cellcolor{robertahighlightcolor!78}58.00\\
    \cline{3-12}
    \rowcolor{robertabackgroundcolor} & & \lossPorts & \codestral & 95.00 & 100 & 100 & 95.00 & 95.20 & 98.20 & \cellcolor{robertahighlightcolor!92}71.53 & \cellcolor{robertahighlightcolor!90}70.03\\
    \rowcolor{robertabackgroundcolor} & \multirow{-6}{*}{\ding{187}} & \lossReplug & \codestral & 85.00 & 90.00 & 97.50 & 85.00 & 92.00 & 93.00 & \cellcolor{robertahighlightcolor!84}64.17 & \cellcolor{robertahighlightcolor!84}64.00\\
    \cline{2-12}
    \rowcolor{robertabackgroundcolor} & & \lossPorts & \llamaBig & 74.60 & 83.90 & 86.80 & 74.60 & 81.23 & 83.55 & \cellcolor{robertahighlightcolor!81}61.24 & \cellcolor{robertahighlightcolor!80}59.79 \\
    \rowcolor{robertabackgroundcolor} & & \lossReplug & \llamaBig & 56.82 & 66.99 & 72.55 & 56.82 & 66.02 & 68.14 & \cellcolor{robertahighlightcolor!65}44.94 & \cellcolor{robertahighlightcolor!64}44.07\\
    \cline{3-12}
    \rowcolor{robertabackgroundcolor} & & \lossPorts & \groqLlama & 72.98 & 83.60 & 86.50 & 72.98 & 81.21 & 83.01 & \cellcolor{robertahighlightcolor!81}60.51 & \cellcolor{robertahighlightcolor!79}59.11\\
    \rowcolor{robertabackgroundcolor} & & \lossReplug & \groqLlama & 58.51 & 68.54 & 74.10 & 58.51 & 67.62 & 69.78 & \cellcolor{robertahighlightcolor!67}46.53 & \cellcolor{robertahighlightcolor!65}45.30 \\
    \cline{3-12}
    \rowcolor{robertabackgroundcolor} & & \lossPorts & \codestral & 71.10 & 81.70 & 85.57 & 71.10 & 79.13 & 81.08 & \cellcolor{robertahighlightcolor!79}58.94 & \cellcolor{robertahighlightcolor!77}57.09\\
    \rowcolor{robertabackgroundcolor} & \multirow{-6}{*}{\ding{188}} & \lossReplug & \codestral & 53.90 & 64.00 & 69.08 & 53.90 & 63.11 & 66.02 & \cellcolor{robertahighlightcolor!62}41.81 & \cellcolor{robertahighlightcolor!61}41.00\\
    \cline{2-12}
    \rowcolor{robertabackgroundcolor} & & \lossPorts & \llamaBig & 96.00 & 100 & 100 & 96.00 & 98.22 & 98.22\tnote{$*$} & \cellcolor{robertahighlightcolor!44}23.89 & \cellcolor{robertahighlightcolor!44}24.48 \\
    \rowcolor{robertabackgroundcolor} & & \lossReplug & \llamaBig & 80.00 & 100 & 100 & 80.00 & 92.62 & 92.62\tnote{$*$} & \cellcolor{robertahighlightcolor!39}19.16 & \cellcolor{robertahighlightcolor!35}15.41 \\
    \cline{3-12}
    \rowcolor{robertabackgroundcolor} & & \lossPorts & \groqLlama & 95.00 & 100 & 100 & 95.00 & 98.00 & 98.20\tnote{$*$} & \cellcolor{robertahighlightcolor!44}23.57 & \cellcolor{robertahighlightcolor!44}24.00\\
    \rowcolor{robertabackgroundcolor} & & \lossReplug & \groqLlama & 77.50 & 100 & 100 & 77.50 & 92.00 & 92.00\tnote{$*$} & \cellcolor{robertahighlightcolor!38}18.33 & \cellcolor{robertahighlightcolor!35}15.00\\
    \cline{3-12}
    \rowcolor{robertabackgroundcolor} & & \lossPorts & \codestral & 95.00 & 100 & 100 & 95.00 & 98.00 & 98.20\tnote{$*$} & \cellcolor{robertahighlightcolor!44}23.57 & \cellcolor{robertahighlightcolor!44}24.00\\
    \rowcolor{robertabackgroundcolor} \multirow{-48}{*}{\modernBert} & \multirow{-6}{*}{\ding{189}} & \lossReplug & \codestral & 77.50 & 100 & 100 & 77.50 & 92.00 & 92.00\tnote{$*$} & \cellcolor{robertahighlightcolor!38}18.33 & \cellcolor{robertahighlightcolor!35}15.00\\
    \bottomrule
    \end{tabular}
    \begin{tablenotes}
    \item[$*$] NDCG@4 since the out-of-domain version of Octopus-v2 has 4 tools only.
    \end{tablenotes}
    \end{threeparttable}
    \end{adjustbox}
    \caption{\textbf{\texttt{PORTS} Recall@$K$ and NDCG@$K$ per dataset-loss-generator (test set) with \modernBertShort as base encoder model.} The positive gains in metric scores over the baselines are highlighted (the brighter, the better).}
    \label{tab:results_roberta}
\end{table*}

\begin{table*}[!t]
    \centering
    \begin{adjustbox}{width=\linewidth}
    \begin{threeparttable}
    \begin{tabular}{lcllcccccccc}
    \toprule
    & & & & \multicolumn{3}{c}{\textbf{Recall} (\%)} & \multicolumn{3}{c}{\textbf{NDCG} (\%)} & \multicolumn{2}{c}{\textbf{$\Delta$ Baseline}} \\
    \cmidrule(lr){5-7}
    \cmidrule(lr){8-10}
    \cmidrule(lr){11-12}
    \textbf{Encoder} & \textbf{Dataset} & \textbf{Method} & \textbf{LLM} & \textbf{@1} & \textbf{@2} & \textbf{@3} & \textbf{@1} & \textbf{@3} & \textbf{@5} & \textbf{Recall} & \textbf{NDCG} \\
    \hline
    \rowcolor{bgebackgroundcolor} & & \lossPorts & \llamaBig & 23.20 & 34.60 & 41.24 & 23.20 & 36.20 & 38.70 & \cellcolor{bgehighlightcolor!27}6.66 & \cellcolor{bgehighlightcolor!28}8.00\\
    \rowcolor{bgebackgroundcolor} & & \lossReplug & \llamaBig & 22.56 & 33.74 & 40.51 & 22.56 & 33.00 & 36.65 & \cellcolor{bgehighlightcolor!26}5.92 & \cellcolor{bgehighlightcolor!26}5.73 \\
    \cline{3-12}
    \rowcolor{bgebackgroundcolor} & & \lossPorts & \groqLlama & 24.64 & 35.30 & 42.46 & 24.64 & 35.22 & 41.60 & \cellcolor{bgehighlightcolor!28}7.78 & \cellcolor{bgehighlightcolor!31}11.13\\
    \rowcolor{bgebackgroundcolor} & & \lossReplug & \groqLlama & 22.32 & 33.58 & 40.43 & 22.32 & 32.83 & 36.13 & \cellcolor{bgehighlightcolor!26}5.84 & \cellcolor{bgehighlightcolor!26}5.67\\
    \cline{3-12}
    \rowcolor{bgebackgroundcolor} & & \lossPorts & \codestral & 25.80 & 36.05 & 43.35 & 25.80 & 35.50 & 42.20 & \cellcolor{bgehighlightcolor!29}8.71 & \cellcolor{bgehighlightcolor!30}9.50 \\
    \rowcolor{bgebackgroundcolor} & \multirow{-6}{*}{\ding{182}} & \lossReplug & \codestral & 21.51 & 33.38 & 40.60 & 21.51 & 33.04 & 37.07 & \cellcolor{bgehighlightcolor!25}5.48 & \cellcolor{bgehighlightcolor!26}6.05\\
    \cline{2-12}
    \rowcolor{bgebackgroundcolor} & & \lossPorts & \llamaBig & 59.00 & 75.65 & 80.80 & 59.00 & 72.21 & 74.23 & \cellcolor{bgehighlightcolor!34}14.29 & \cellcolor{bgehighlightcolor!32}12.22\\
    \rowcolor{bgebackgroundcolor} & & \lossReplug & \llamaBig & 55.32 & 71.94 & 78.06 & 55.32 & 69.01 & 72.05 & \cellcolor{bgehighlightcolor!31}10.91 & \cellcolor{bgehighlightcolor!29}9.03\\
    \cline{3-12}
    \rowcolor{bgebackgroundcolor} & & \lossPorts & \groqLlama & 59.12 & 76.80 & 81.50 & 59.12 & 75.40 & 76.10 & \cellcolor{bgehighlightcolor!35}14.94 & \cellcolor{bgehighlightcolor!34}14.21 \\
    \rowcolor{bgebackgroundcolor} & & \lossReplug & \groqLlama & 56.29 & 75.00 & 80.00 & 56.29 & 70.60 & 73.32 & \cellcolor{bgehighlightcolor!33}12.90 & \cellcolor{bgehighlightcolor!31}10.73 \\
    \cline{3-12}
    \rowcolor{bgebackgroundcolor} & & \lossPorts & \codestral & 49.52 & 64.03 & 71.45 & 49.52 & 62.30 & 66.10 & \cellcolor{bgehighlightcolor!24}4.14 & \cellcolor{bgehighlightcolor!23}3.00\\
    \rowcolor{bgebackgroundcolor} & \multirow{-6}{*}{\ding{183}} & \lossReplug & \codestral & 45.40 & 60.15 & 65.97 & 45.40 & 57.20 & 59.38 & \cellcolor{bgehighlightcolor!24}3.98 & \cellcolor{bgehighlightcolor!23}2.79\\
    \cline{2-12}
    \rowcolor{bgebackgroundcolor} & & \lossPorts & \llamaBig & 30.42 & 37.11 & 40.68 & 30.42 & 33.50 & 34.70 & \cellcolor{bgehighlightcolor!40}20.10 & \cellcolor{bgehighlightcolor!37}17.21\\
    \rowcolor{bgebackgroundcolor} & & \lossReplug & \llamaBig & 19.55 & 28.29 & 33.05 & 19.55 & 27.45 & 30.02 & \cellcolor{bgehighlightcolor!31}10.84 & \cellcolor{bgehighlightcolor!31}10.67 \\
    \cline{3-12}
    \rowcolor{bgebackgroundcolor} & & \lossPorts & \groqLlama & 30.64 & 37.20 & 41.06 & 30.64 & 33.90 & 35.20 & \cellcolor{bgehighlightcolor!40}20.18 & \cellcolor{bgehighlightcolor!37}17.26 \\
    \rowcolor{bgebackgroundcolor} & & \lossReplug & \groqLlama & 19.10 & 28.57 & 33.05 & 19.10 & 27.05 & 30.04 & \cellcolor{bgehighlightcolor!31}10.79 & \cellcolor{bgehighlightcolor!30}10.03\\
    \cline{3-12}
    \rowcolor{bgebackgroundcolor} & & \lossPorts & \codestral & 25.83 & 34.45 & 39.72 & 25.83 & 30.90 & 32.90 & \cellcolor{bgehighlightcolor!37}17.22 & \cellcolor{bgehighlightcolor!34}14.06\\
    \rowcolor{bgebackgroundcolor} & \multirow{-6}{*}{\ding{184}} & \lossReplug & \codestral & 18.88 & 28.42 & 32.89 & 18.88 & 26.80 & 29.80 & \cellcolor{bgehighlightcolor!28}7.53 & \cellcolor{bgehighlightcolor!29}9.07\\
    \cline{2-12}
    \rowcolor{bgebackgroundcolor} & & \lossPorts & \llamaBig & 65.25 & 73.75 & 78.19 & 65.25 & 72.70 & 74.60 & \cellcolor{bgehighlightcolor!25}4.87 & \cellcolor{bgehighlightcolor!25}5.13\\
    \rowcolor{bgebackgroundcolor} & & \lossReplug & \llamaBig & 65.06 & 73.57 & 77.82 & 65.06 & 73.22 & 74.24 & \cellcolor{bgehighlightcolor!25}4.62 & \cellcolor{bgehighlightcolor!25}5.11\\
    \cline{3-12}
    \rowcolor{bgebackgroundcolor} & & \lossPorts & \groqLlama & 67.20 & 73.23 & 78.10 & 67.20 & 74.60 & 73.10 & \cellcolor{bgehighlightcolor!25}5.31 & \cellcolor{bgehighlightcolor!26}5.63 \\
    \rowcolor{bgebackgroundcolor} & & \lossReplug & \groqLlama & 66.17 & 73.20 & 77.82 & 66.17 & 72.92 & 74.31 & \cellcolor{bgehighlightcolor!25}4.86 & \cellcolor{bgehighlightcolor!25}5.13 \\
    \cline{3-12}
    \rowcolor{bgebackgroundcolor} & & \lossPorts & \codestral & 64.00 & 74.00 & 78.00 & 64.00 & 72.02 & 73.30 & \cellcolor{bgehighlightcolor!25}4.56 & \cellcolor{bgehighlightcolor!24}4.03\\
    \rowcolor{bgebackgroundcolor} & \multirow{-6}{*}{\ding{185}} & \lossReplug & \codestral & 59.35 & 65.40 & 73.06 & 59.35 & 67.26 & 71.18 & \cellcolor{bgehighlightcolor!24}3.56 & \cellcolor{bgehighlightcolor!24}3.78\\
    \cline{2-12}
    \rowcolor{bgebackgroundcolor} & & \lossPorts & \llamaBig & 66.60 & 78.10 & 82.29 & 66.60 & 76.03 & 77.05 & \cellcolor{bgehighlightcolor!35}14.59 & \cellcolor{bgehighlightcolor!34}14.06\\
    \rowcolor{bgebackgroundcolor} & & \lossReplug & \llamaBig & 66.65 & 76.91 & 80.84 & 66.65 & 75.04 & 77.07 & \cellcolor{bgehighlightcolor!33}12.65 & \cellcolor{bgehighlightcolor!32}12.06\\
    \cline{3-12}
    \rowcolor{bgebackgroundcolor} & & \lossPorts & \groqLlama & 66.59 & 78.02 & 82.29 & 66.59 & 76.16 & 77.17 & \cellcolor{bgehighlightcolor!35}14.61 & \cellcolor{bgehighlightcolor!34}14.10\\
    \rowcolor{bgebackgroundcolor} & & \lossReplug & \groqLlama & 67.23 & 77.54 & 81.06 & 67.23 & 75.50 & 76.82 & \cellcolor{bgehighlightcolor!33}13.12 & \cellcolor{bgehighlightcolor!33}13.18 \\
    \cline{3-12}
    \rowcolor{bgebackgroundcolor} & & \lossPorts & \codestral & 67.35 & 79.48 & 83.75 & 67.35 & 77.00 & 78.00 & \cellcolor{bgehighlightcolor!35}14.71 & \cellcolor{bgehighlightcolor!34}14.12 \\
    \rowcolor{bgebackgroundcolor} & \multirow{-6}{*}{\ding{186}} & \lossReplug & \codestral & 65.10 & 77.01 & 80.95 & 65.10 & 73.78 & 76.67 & \cellcolor{bgehighlightcolor!32}12.37 & \cellcolor{bgehighlightcolor!33}13.00\\
    \cline{2-12}
    \rowcolor{bgebackgroundcolor} & & \lossPorts & \llamaBig & 97.50 & 100 & 100 & 97.50 & 100 & 100 & \cellcolor{bgehighlightcolor!22}1.67 & \cellcolor{bgehighlightcolor!22}2.17 \\
    \rowcolor{bgebackgroundcolor} & & \lossReplug & \llamaBig & 95.00 & 100 & 100 & 95.00 & 98.00 & 98.00 & \cellcolor{bgehighlightcolor!21}0.83 & \cellcolor{bgehighlightcolor!20}0.22 \\
    \cline{3-12}
    \rowcolor{bgebackgroundcolor} & & \lossPorts & \groqLlama & 95.00 & 100 & 100 & 95.00 & 98.00 & 98.00 & \cellcolor{bgehighlightcolor!21}0.83 & \cellcolor{bgehighlightcolor!20}0.22\\
    \rowcolor{bgebackgroundcolor} & & \lossReplug & \groqLlama & 95.00 & 97.50 & 100 & 95.00 & 98.00 & 98.00 & \cellcolor{bgehighlightcolor!1}0 & \cellcolor{bgehighlightcolor!1}0\\
    \cline{3-12}
    \rowcolor{bgebackgroundcolor} & & \lossPorts & \codestral & 95.00 & 100 & 100 & 95.00 & 98.00 & 98.00 & \cellcolor{bgehighlightcolor!21}0.83 & \cellcolor{bgehighlightcolor!20}0.22\\
    \rowcolor{bgebackgroundcolor} & \multirow{-6}{*}{\ding{187}} & \lossReplug & \codestral & 95.00 & 97.50 & 100 & 95.00 & 98.00 & 98.00 & \cellcolor{bgehighlightcolor!1}0 & \cellcolor{bgehighlightcolor!1}0\\
    \cline{2-12}
    \rowcolor{bgebackgroundcolor} & & \lossPorts & \llamaBig & 78.85 & 89.07 & 91.89 & 78.85 & 86.21 & 87.19 & \cellcolor{bgehighlightcolor!29}8.74 & \cellcolor{bgehighlightcolor!29}9.17\\
    \rowcolor{bgebackgroundcolor} & & \lossReplug & \llamaBig & 73.93 & 85.19 & 89.14 & 73.93 & 83.11 & 84.13 & \cellcolor{bgehighlightcolor!25}4.89 & \cellcolor{bgehighlightcolor!26}6.12\\
    \cline{3-12}
    \rowcolor{bgebackgroundcolor} & & \lossPorts & \groqLlama & 89.87 & 92.20 & 94.04 & 89.87 & 91.10 & 92.35 & \cellcolor{bgehighlightcolor!34}14.18 & \cellcolor{bgehighlightcolor!37}17.11 \\
    \rowcolor{bgebackgroundcolor} & & \lossReplug & \groqLlama & 74.60 & 86.24 & 90.30 & 74.60 & 83.98 & 85.09 & \cellcolor{bgehighlightcolor!26}5.85 & \cellcolor{bgehighlightcolor!27}7.22 \\
    \cline{3-12}
    \rowcolor{bgebackgroundcolor} & & \lossPorts & \codestral & 89.30 & 92.10 & 92.90 & 80.00 & 87.00 & 88.00 & \cellcolor{bgehighlightcolor!33}12.84 & \cellcolor{bgehighlightcolor!31}11.00\\
    \rowcolor{bgebackgroundcolor} & \multirow{-6}{*}{\ding{188}} & \lossReplug & \codestral & 74.60 & 86.24 & 90.30 & 74.60 & 84.12 & 85.17 & \cellcolor{bgehighlightcolor!26}5.85 & \cellcolor{bgehighlightcolor!27}7.11\\
    \cline{2-12}
    \rowcolor{bgebackgroundcolor} & & \lossPorts & \llamaBig & 97.50 & 100 & 100 & 97.50 & 100 & 100\tnote{$*$} & \cellcolor{bgehighlightcolor!21}0.84 & \cellcolor{bgehighlightcolor!21}1.17\\
    \rowcolor{bgebackgroundcolor} & & \lossReplug & \llamaBig & 95.00 & 100 & 100 & 95.00 & 98.00 & 98.00\tnote{$*$} & \cellcolor{bgehighlightcolor!1}0 & \cellcolor{bgehighlightcolor!1}0\\
    \cline{3-12}
    \rowcolor{bgebackgroundcolor} & & \lossPorts & \groqLlama & 97.00 & 100 & 100 & 96.00 & 100 & 100\tnote{$*$} & \cellcolor{bgehighlightcolor!21}0.67 & \cellcolor{bgehighlightcolor!20}0.36\\
    \rowcolor{bgebackgroundcolor} & & \lossReplug & \groqLlama & 95.00 & 100 & 100 & 95.00 & 98.00 & 98.00\tnote{$*$} & \cellcolor{bgehighlightcolor!1}0 &\cellcolor{bgehighlightcolor!1}0\\
    \cline{3-12}
    \rowcolor{bgebackgroundcolor} & & \lossPorts & \codestral & 96.00 & 100 & 100 & 95.00 & 100 & 100\tnote{$*$} & \cellcolor{bgehighlightcolor!20}0.33 & \cellcolor{bgehighlightcolor!20}0.12\\
    \rowcolor{bgebackgroundcolor} \multirow{-48}{*}{\bge} & \multirow{-6}{*}{\ding{189}} & \lossReplug & \codestral & 95.00 & 100 & 100 & 95.00 & 98.00 & 98.00\tnote{$*$} & \cellcolor{bgehighlightcolor!1}0 & \cellcolor{bgehighlightcolor!1}0\\
    \bottomrule
    \end{tabular}
    \begin{tablenotes}
    \item[$*$] NDCG@4 since the out-of-domain version of Octopus-v2 has 4 tools only.
    \end{tablenotes}
    \end{threeparttable}
    \end{adjustbox}
    \caption{\textbf{\texttt{PORTS} Recall@$K$ and NDCG@$K$ per dataset-loss-generator (test set) with \bgeShort as base encoder model.} The positive gains in metric scores over the baselines are highlighted (the brighter, the better).}
    \label{tab:results_bge}
\end{table*}

\section{Complete Results}
\label{app:complete_results}

Table~\ref{tab:results_roberta} and Table~\ref{tab:results_bge} complement the results of the main paper, listing the retrieval scores achieved by training encoders under the supervision signal of each LLM explored.

\section{Qualitative Example}
\label{app:qual_examples}

Table~\ref{tab:input_output_examples} presents a specific query from the ToolE test set, demonstrating the contrast between the top-3 tools retrieved by BGE with and without \ports tuning. The results clearly illustrate that our alignment process not only successfully positions the correct tool at the top rank, but also generates a significantly sharpened preference distribution.

\begin{table*}[!htb]
    \centering
    \fontsize{9}{9}\selectfont
    \begin{adjustbox}{width=\linewidth}
    \begin{tabular}{cc}
        \multicolumn{2}{c}{\textbf{Query:} I'm looking for a hotel in Sapporo.} \\[4mm]
        \hspace{2mm}
        \begin{tikzpicture}
            \centering
            \begin{axis}[
                width=6cm, height=4cm,
                ymajorgrids=true,
                grid=both,
                grid style=dashed,
                axis background/.style={fill=plotbackground},
                ybar=0pt, 
                bar width=35pt,
                enlarge x limits=0.28,  
                xtick=data,
                symbolic x coords={
                    {TripTool\\\faCheck},
                    SmartTicket,
                    Local
                },
                x tick label style={text width=2cm, align=center},
                ymin=0.0, ymax=1.0,
                every tick label/.append style={font=\fontsize{8}{8}\selectfont},
                title=\textbf{\ports Retrieved Tools},
                title style={font=\fontsize{9}{9}\selectfont},
            ]
                \addplot [fill=bluebar, draw=none] coordinates {
                    ({TripTool\\\faCheck}, 0.96)
                    (SmartTicket, 0.57)
                    (Local, 0.44)
                };
            \end{axis}
        \end{tikzpicture}
        &
        \begin{tikzpicture}
            \centering
            \begin{axis}[
                width=6cm, height=4cm,
                ymajorgrids=true,
                grid=both,
                grid style=dashed,
                axis background/.style={fill=plotbackground},
                ybar=0pt, 
                bar width=35pt,
                enlarge x limits=0.28,  
                xtick=data,
                symbolic x coords={
                    Sakenowa,
                    {TripTool\\\faCheck},
                    Local
                },
                x tick label style={text width=2cm, align=center},
                ymin=0.0, ymax=1.0,
                every tick label/.append style={font=\fontsize{8}{8}\selectfont},
                title=\textbf{\bge Retrieved Tools},
                title style={font=\fontsize{9}{9}\selectfont},
            ]
                \addplot [fill=bluebar, draw=none] coordinates {
                    (Sakenowa, 0.55)
                    ({TripTool\\\faCheck}, 0.52)
                    (Local, 0.50)
                };
            \end{axis}
        \end{tikzpicture} \\[4mm]
        \multicolumn{2}{c}{
            \parbox{0.7\linewidth}{
                \textbf{Gold Tool Docstring:} Offer discounted hotel and accommodation bookings, along with personalized hotel and product searches, travel planning, image editing, and more, helping users easily plan their trips and find accommodation and transportation options.
            }
        }
    \end{tabular}
    \end{adjustbox}
    \caption{\textbf{Input-output tool selection example from the ToolE test set.} Cosine similarity comparison between \ports-tuned BGE (left) and baseline, frozen BGE (right).}
    \label{tab:input_output_examples}
\end{table*}

\end{document}